\documentclass[twocolumn]{emulateapj}  
\usepackage{epstopdf}
\usepackage{ulem}
\usepackage{amssymb}
\usepackage{times}
\usepackage{graphicx}
\usepackage[usenames,dvipsnames]{pstricks}
\usepackage{psfrag}
\usepackage{lipsum}
\usepackage{natbib}
\usepackage[colorlinks=true,linkcolor=blue,citecolor=blue]{hyperref}
\def\aj{AJ}
\def\apj{ApJ}
\def\apjl{ApJ}
\def\apjs{ApJS}
\def\aap{A\&A}
\def\aaps{A\&AS}
\def\mnras{MNRAS}
\def\nat{Nature}
\def\physrep{Phys.~Rep.}

\newcommand{\be}{\begin{equation}}
\newcommand{\ee}{\end{equation}}
\newcommand{\bary}{\begin{eqnarray}}
\newcommand{\eary}{\end{eqnarray}}

\begin{document}



\title{Analysis and Modelling of the multi-wavelength observations\\ of the luminous GRB 190114C}

\author{N.~ Fraija\altaffilmark{1$\dagger$}, S. Dichiara\altaffilmark{2,3}, A.C. Caligula do E. S. Pedreira\altaffilmark{1}, A.~ Galvan-Gamez  \altaffilmark{1}, R. L. Becerra\altaffilmark{1}, R. Barniol Duran\altaffilmark{4} and B. B. Zhang\altaffilmark{6,7}
}
\affil{$^1$Instituto de Astronom\'ia, Universidad Nacional Aut\'{o}noma de M\'{e}xico, Apdo. Postal 70-264, Cd. Universitaria, Ciudad de M\'{e}xico 04510}
\affil{$^2$ Department of Astronomy, University of Maryland, College Park, MD 20742-4111, USA}
\affil{$^3$ Astrophysics Science Division, NASA Goddard Space Flight Center, 8800 Greenbelt Rd, Greenbelt, MD 20771, USA}
\affil{$^4$ Department of Physics and Astronomy, California State University, Sacramento, 6000 J Street, Sacramento, CA 95819-6041, USA}
\affil{$^5$ School of Astronomy and Space Science, Nanjing University, Nanjing 210093, China}
\affil{$^6$ Key Laboratory of Modern Astronomy and Astrophysics (Nanjing University), Ministry of Education, China}
\email[$\dagger$ ]{nifraija@astro.unam.mx}
%
\begin{abstract}
Very-high-energy (VHE; $\geq 10$ GeV) photons are expected from the nearest and brightest Gamma-ray bursts (GRBs). VHE photons, at energies higher than 300 GeV, were recently reported by the MAGIC collaboration for this burst.   Immediately, GRB 190114C was followed up  by a massive observational campaign covering a large fraction of the electromagnetic spectrum.   In this paper,  we obtain the LAT light curve  of GRB 190114C and  show that it exhibits similar features to other bright LAT-detected bursts; the first high-energy photon ($\geq$ 100 MeV) is delayed with the onset of the prompt phase and the flux light curve exhibits a long-lived emission (lasting much longer than the prompt phase) and a short-lasting bright peak (located at the beginning of long-lived emission).  Analyzing the multi-wavelength observations, we show that the short-lasting LAT and GBM bright peaks are consistent with the synchrotron self-Compton reverse-shock model and the long-lived observations  with the standard synchrotron forward-shock model that evolves from a stratified stellar-wind like medium to a uniform ISM-like medium. Given the best-fit values, a bright optical flash produced by synchrotron reverse-shock emission is expected.   From our analysis we infer that the high-energy photons are produced in the deceleration phase of the outflow and some additional processes to synchrotron in the forward shocks  should be considered to properly describe the LAT photons with energies beyond the synchrotron limit. Moreover, we claim that an outflow  endowed with magnetic fields could describe the polarization and properties exhibited in the light curve of GRB 190114C.
\end{abstract}
%
\keywords{Gamma-rays bursts: individual (GRB 190114C)  --- Physical data and processes: acceleration of particles  --- Physical data and processes: radiation mechanism: nonthermal --- ISM: general - magnetic fields}

\section{Introduction}
%
Gamma-ray bursts (GRBs), the most luminous gamma-ray transient events in the universe, are non-repeating flashes usually associated to core collapse of massive stars  when the duration of the prompt emission is longer than $\gtrsim 2$ s or to the merger of compact object binaries when the duration is less than  $\lesssim  2$ s \cite[e.g., see][for reviews]{2004IJMPA..19.2385Z, 2015PhR...561....1K}.  Irrespective of the progenitor associated to the prompt emission, a long-lived  afterglow emission is generated via the deceleration of  the outflow in the cirbumburst medium.  The transition between the prompt and afterglow phase is recognized by early  signatures observed in multi-wavelength light curves and broadband spectral energy distributions (SEDs).  These signatures are associated with abrupt changes in the spectral features \citep{1999ApJ...524L..47G}, the sudden decrease in the density flux interpreted as high-latitude emission \citep{2000ApJ...541L..51K, 2019ApJ...871..123F}, rapid variations in the evolution of the polarimetric observations \citep{2017Natur.547..425T, 2003ApJ...596L..17G, 2017ApJ...848...94F}  and an outstanding multi-frequency peak  generated by the reverse shock \citep{2007ApJ...655..973K, 2016ApJ...818..190F, 2018ApJ...859...70F,  2019arXiv190405987B}.\\
The detection of very-high-energy (VHE; $\gtrsim $ 10 GeV) photons and their arrival times provides a crucial piece of information to quantify the baryonic composition of the outflow, the particle acceleration efficiency, the emitting region and the radiation processes, among others \cite[e.g., see][for reviews]{2004IJMPA..19.2385Z, 2015PhR...561....1K}.  The Large Area Telescope (LAT) on-board the Fermi satellite has detected more than  100 GRBs which exhibited photons greater than $\geq$ 100 MeV and only one dozen of  bursts with VHE ($\geq 10$ GeV) photons.  The most powerful bursts have displayed that the energetic photons arrive late with respect to the onset of the prompt emission and the LAT light curves display two distinct components: one that lasts much longer than the prompt emission called long-lived emission and another short-lasting bright peak located at the beginning of the long-lived  emission. Using multi-wavelength observations at lower energies for these powerful events, several authors modelled the long-lived emission with the standard synchrotron forward-shock model \citep{2009MNRAS.400L..75K, 2010MNRAS.409..226K, 2010MNRAS.403..926G, 2014MNRAS.443.3578N, 2009MNRAS.396.1163Z, 2017ApJ...837..116B, 2019arXiv190513572F} and the short-lasting bright peak with synchrotron self-Compton reverse-shock model \citep{2015ApJ...804..105F, 2016ApJ...818..190F, 2017ApJ...848...94F}, indicating  that the LAT fluxes were generated during the external shocks. However, this is not the case for VHE photons, which cannot be interpreted in the framework of the synchrotron forward-shock model. The maximum photon energy generated by this radiative process is $\sim 10~{\rm GeV}~\left(\frac{\Gamma}{100}\right)\left(1+z\right)^{-1}$, where $\Gamma$ is the bulk Lorentz factor and $z$ the redshift \citep{2010ApJ...718L..63P, 2009ApJ...706L.138A, 2011MNRAS.412..522B}. Given that the bulk Lorentz factor evolves during the afterglow as $\propto t^{-\frac38}$ and $\propto t^{-\frac14}$ for a uniform ISM-like medium and a stratified stellar-wind like medium, respectively, VHE photons from synchrotron radiation are not expected at the end of this phase. Therefore, we want to emphasize that the LAT photons below the maximum synchrotron energy can be explained well by synchrotron forward shock and beyond the synchrotron limit  some additional mechanisms must be invoked to explain the VHE LAT photons.\\
\\
The BAT (Burst Area Telescope) instrument on-board the Swift satellite triggered on GRB 190114C on January 14, 2019 at 20:57:06.012 UTC (trigger 883832) \citep{2019GCN.23688....1G}. GRB 190114C was also detected by the two instruments on-board Fermi satellite; Gamma-Ray Burst Monitor \citep[GBM;][]{2019GCN.23709....1D} and LAT \citep{2019GCN.23709....1D}. Immediately after the detection, counterparts were observed by the X-ray Telescope \citep[XRT;][]{2019GCN.23688....1G, 2019GCN.23704....1O} and Ultraviolet/Optical Telescope \citep[UVOT;][]{2019GCN.23688....1G, 2019GCN.23725....1S} on-board Swift satellite, by SPI-ACS instrument on-board INTEGRAL \citep{2019GCN.23714....1M},  by Mini-CALorimeter (MCAL) instrument on-board the AGILE satellite \citep{2019GCN.23712....1U}, by Hard X-ray Modulation Telescope (HXMT) instrument on-board the Insight satellite \citep{2019GCN.23716....1X}, by Konus-Wind \citep{2019GCN.23737....1F}, by the Atacama Large Millimeter/submillimeter Array (ALMA),  by Very Large Array (VLA) \citep{2019arXiv190407261L} and by a massive campaign of optical instruments and telescopes   \citep{2019GCN.23690....1T, 2019GCN.23693....1L, 2019GCN.23695....1S, 2019GCN.23699....1L, 2019GCN.23701....1M, 2019GCN.23702....1B, 2019GCN.23717....1I, 2019GCN.23726....1K, 2019GCN.23729....1D, 2019GCN.23732....1K, 2019GCN.23733....1K, 2019GCN.23734....1K, 2019GCN.23740....1I, 2019GCN.23741....1M}. For the first time an excess of gamma-ray events with a significance of $\>$20 $\sigma$ was detected during the first 20 minutes and photons with  energies above 300 GeV were reported by MAGIC collaboration from GRB 190114C  \citep{2019GCN.23701....1M}.\\
In this paper,  we analyze the LAT light curve obtained at the position for GRB 190114C and show that it exhibits similar features of other LAT-detected bursts.  Analyzing the multi-wavelength observations, we show that the short-lasting LAT and GBM bright peaks are consistent with synchrotron self-Compton reverse-shock model and the long-lived LAT, GBM, X-ray, optical and radio emissions with the synchrotron forward-shock model that evolves from a stratified stellar-wind like medium to a uniform ISM-like medium.  The paper is arranged as follows. In Section 2 we present multi-wavelength observations and/or data reduction. In Section 3 we describe the multi-wavelength observations through the synchrotron forward-shock model and the SSC reverse-shock model in a stratified  stellar-wind like and uniform ISM-like medium. In Section 4, the discussion and results of the analysis done using the multi-wavelength data are presented.   Finally,  in Section 5 we give a brief summary.  The convention $Q_{\rm x}=Q/10^{\rm x}$  in cgs units and  the universal constants c=$\hbar$=1 in natural units will be adopted through this paper.  
\vspace{1cm}
\section{GRB 190114C: Multi-wavelength Observations and/or data Reduction}\label{sec:observations}
\subsection{Fermi LAT observations and Data reduction}
\begin{figure}[h!]
{ \centering
\resizebox*{0.9\textwidth}{0.4\textheight}
{\includegraphics{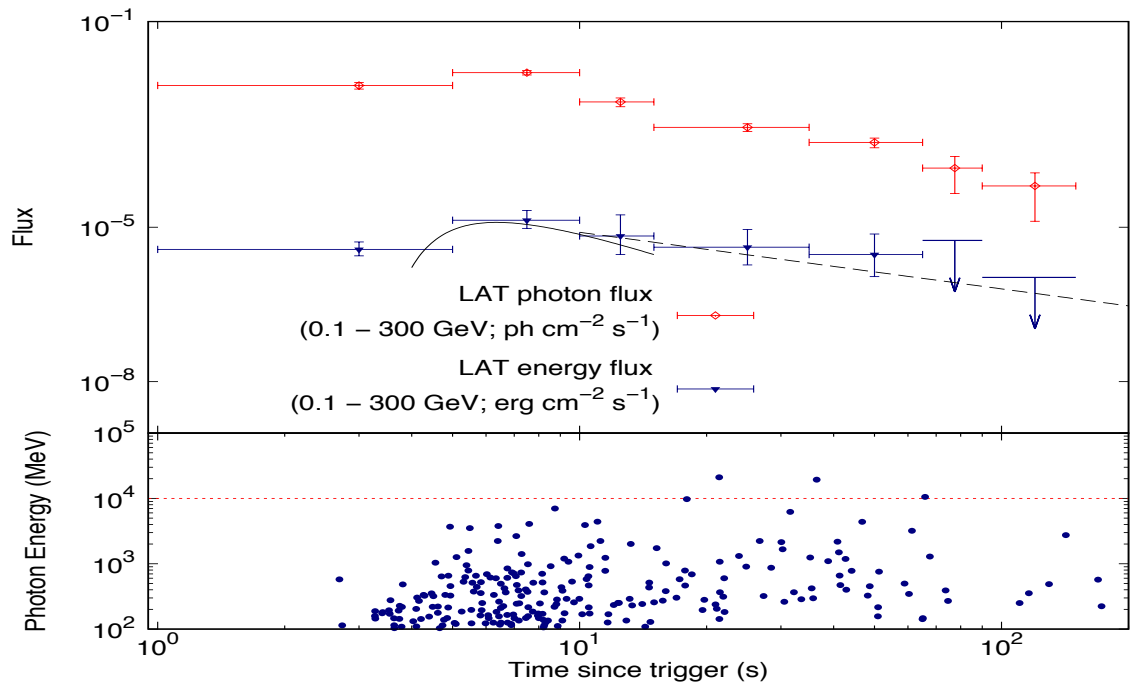}}\\
}
\caption{Upper panel:  Fermi-LAT energy flux (blue) and photon flux (red) light curves obtained between 0.1 and 300 GeV. The solid black line represents the best-fit curve found using our model. Lower panel:  All the photons with energies $> 100$ MeV and probabilities $>90$\% of being associated with GRB 190114C.  Fermi-LAT data  were reduced using the public database at the Fermi website.}
\label{LAT_lc}
\end{figure}

 The Fermi LAT instrument detected VHE emission from GRB 190114C. LAT data exhibited a representative increase in the event rate.  The preliminary photon index above 100 MeV was $\Gamma_{\rm LAT}=\beta_{\rm LAT} + 1 =1.98 \pm 0.06$, with an estimated  energy flux of  $(2.06\pm 0.14)\times 10^{-6}\, {\rm erg\,cm^{-2}\,s^{-1}}$. Later,   \cite{2019arXiv190107505W} analyzed the LAT spectrum in two time intervals, $\sim$ 6 - 7 s and 11 - 14 s, reporting PL indexes of  $\Gamma_{\rm LAT}=\beta_{\rm LAT}+1=2.06\pm0.30$ and $2.10\pm0.31$, respectively.\\
\\
Fermi LAT event data files are retrieved from the online data repository\footnote{http://fermi.gsfc.nasa.gov/ssc/data} starting few seconds before the GBM trigger time, 20:57:02.63 UT \citep{2019GCN.23707....1H}. These data are analyzed using Fermi Science tools\footnote{https://fermi.gsfc.nasa.gov/ssc/data/analysis/software/} version v11r06p03 and reprocessed with Pass 8 extended, spacecraft data, and the instrument response  functions  "P8R3\_TRANSIENT020\_V2".  Transient events are selected using \textit{gtselect} (evtclass=16) in the energy range between 100 MeV and 300 GeV, within 15$^\circ$ of the reported GRB position and with a maximum zenith angle of 100$^\circ$.
After taking into account of a model for the source and diffuse components (galactic and extragalactic) using \textit{gtdiffrsp}, we generate the spectra and related response files using \textit{gtbin} and \textit{gtrspgen}, respectively. Data are binned in 7 time bins: [1-5, 5-10, 10-15, 15-35, 35-65, 65-90, 90-150] s from the trigger. This binning pattern results from a trade-off aimed to preserve the time profile of the signal and the minimum statistical significance needed to analyze the spectrum.
We derive the spectrum for each bin and fit it with a simple power law (SPL) using the software XSPECv12.10.1 \citep{1996ASPC..101...17A}. The resulting fluxes are computed after the fit with 90\% confidence errors in each time bin. The light curves with the flux above 100 MeV are shown in the upper panel of Figure \ref{LAT_lc}.\\  
Figure \ref{LAT_lc} shows the Fermi LAT energy flux (blue) and photon flux (red) light curves obtained between 0.1 and 300 GeV (upper panel) and the energies of all the photons ($\geq 100$ MeV) with probabilities $>90$\% of being associated with this burst (lower panel).    In the upper panel we can observe that the energy flux and the photon flux light curves exhibit a bright peak at $\sim$ 6 - 7 s followed by a monotonic decreasing emission extended for $\sim$ 70 s.\\
In order to model the Fermi LAT data, the function \citep{2006Natur.442..172V}
\be\label{funcion_lat}
F(t)= A \left(\frac{t-t_0}{t_0}\right)^{-\alpha_{\rm \gamma,pk}}\,e^{-\frac{\tau}{t-t_0}}\,,
\ee
and a SPL ($\propto t^{-\alpha_{\rm LAT}}$) are used to describe the short-lasting bright peak and  the long-lived emission, respectively.  Here,  $t_0$ is the  starting time,  $A$ is the amplitude, $\tau$ is the timescale of the flux rise and $\alpha_{\rm \gamma,pk}$ is the temporal decay index of the peak.   The  energy flux light curve together with the best-fit curve  is shown in this upper panel.   The best-fit values found are $t_0=2.61\pm0.15$ s,  $\tau=8.11\pm1.22$ s,  $\alpha_{\rm \gamma, pk}=2.65\pm0.19$ and $\alpha_{\rm LAT}=1.10\pm0.15$ ($\chi^2$ = 0.86).\\
The lower panel in Figure  \ref{LAT_lc} displays several features: i) the first high-energy photon of 571.4 MeV was observed at 2.7 s after the GBM trigger,   ii) this burst exhibited 238 photons with energies larger than 100 MeV, 41 with energies larger than 1 GeV and 5 photons with energies larger than 10 GeV and  iii) the highest-energy photon exhibited in the LAT observations was 21.42 GeV detected at 21 s after the GBM trigger\footnote{It is worth noting that photons at energy higher than $\sim$ 300 GeV were reported by MAGIC collaboration.}.
\subsection{GBM observations}
\begin{figure}[h!]
{ \centering
\resizebox*{0.45\textwidth}{0.33\textheight}
{\includegraphics{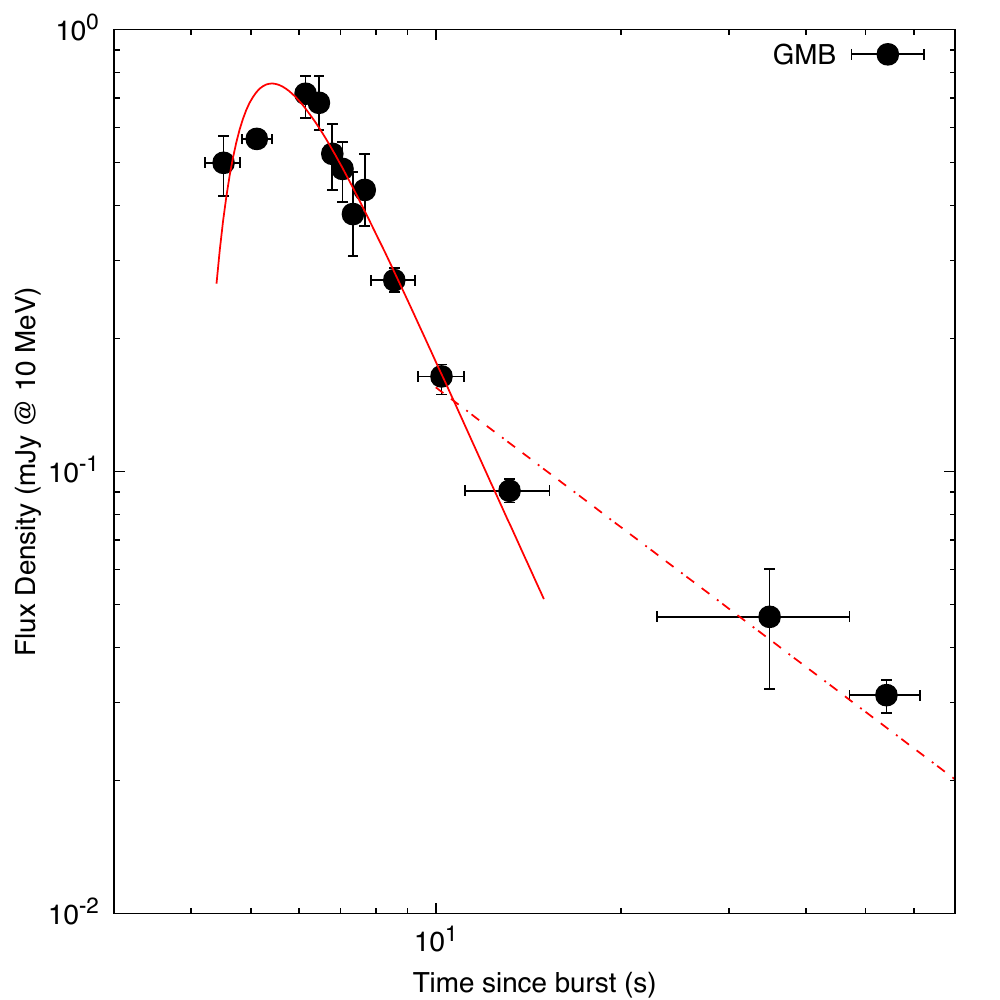}}
\resizebox*{0.45\textwidth}{0.33\textheight}
{\includegraphics{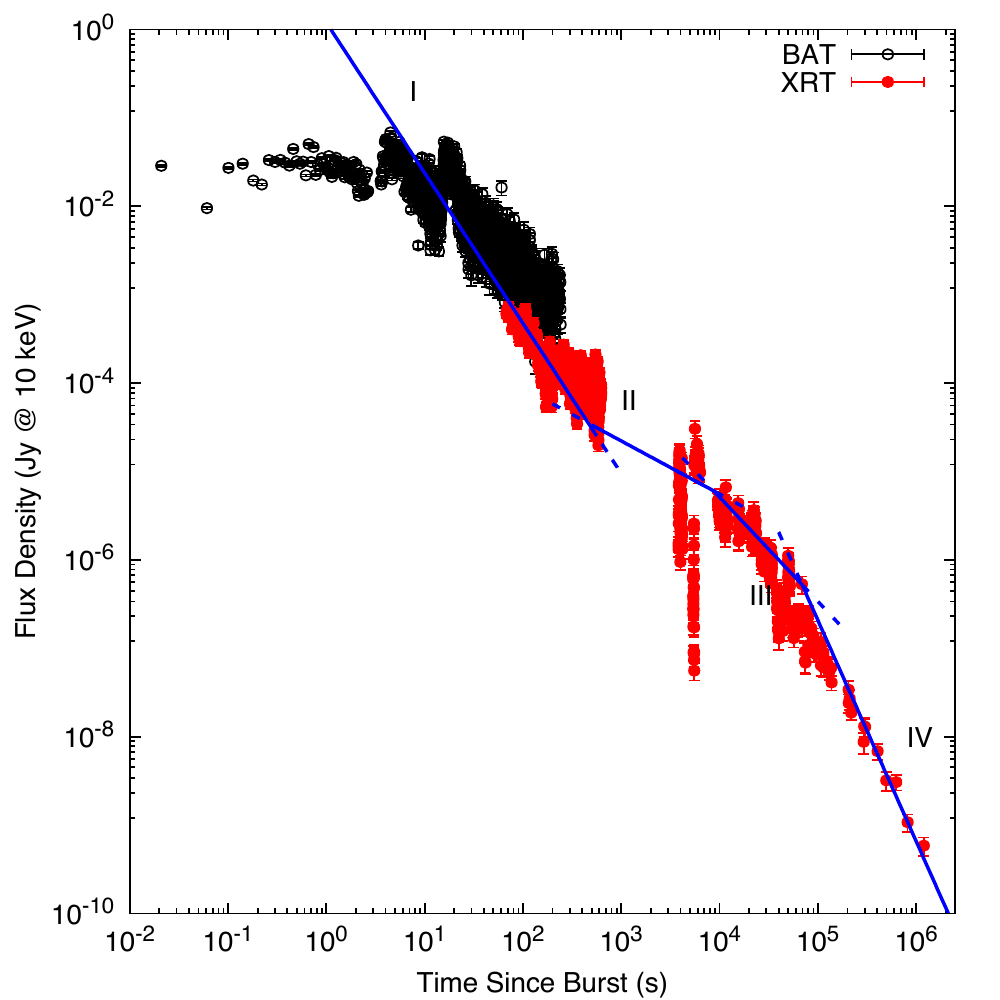}}\\
\resizebox*{0.45\textwidth}{0.33\textheight}
{\includegraphics{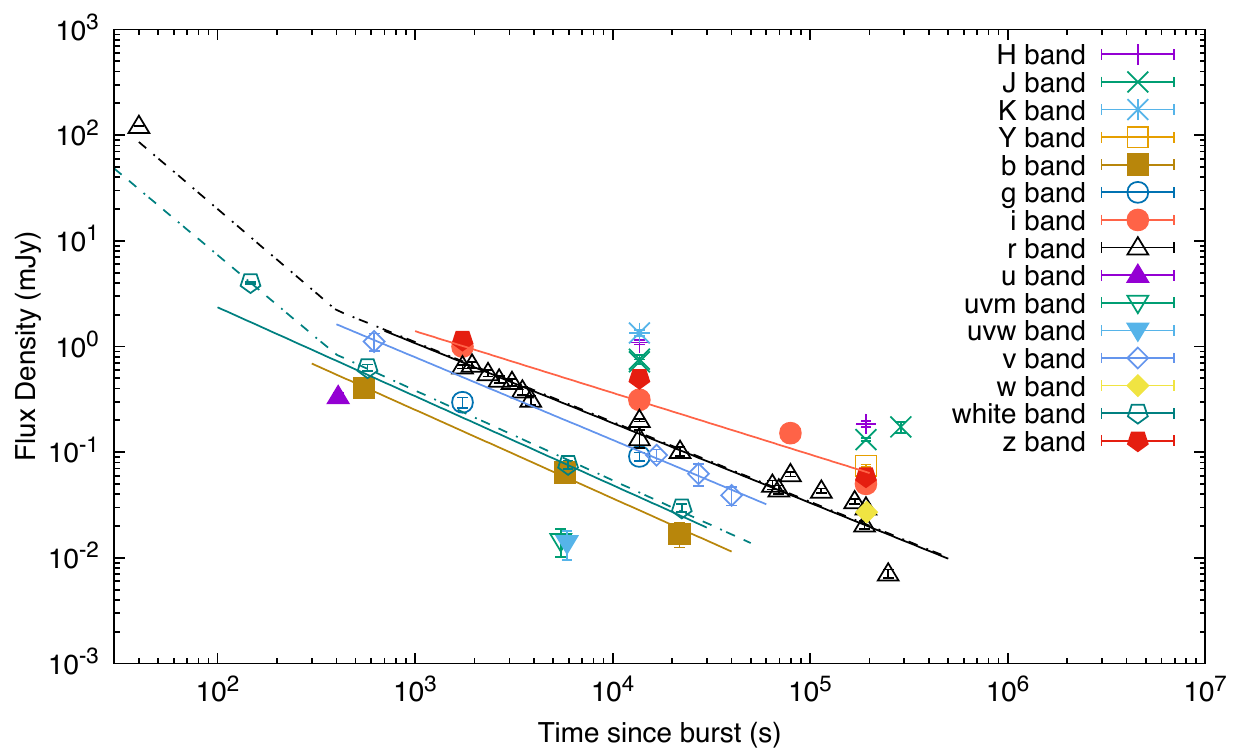}}
\resizebox*{0.45\textwidth}{0.33\textheight}
{\includegraphics{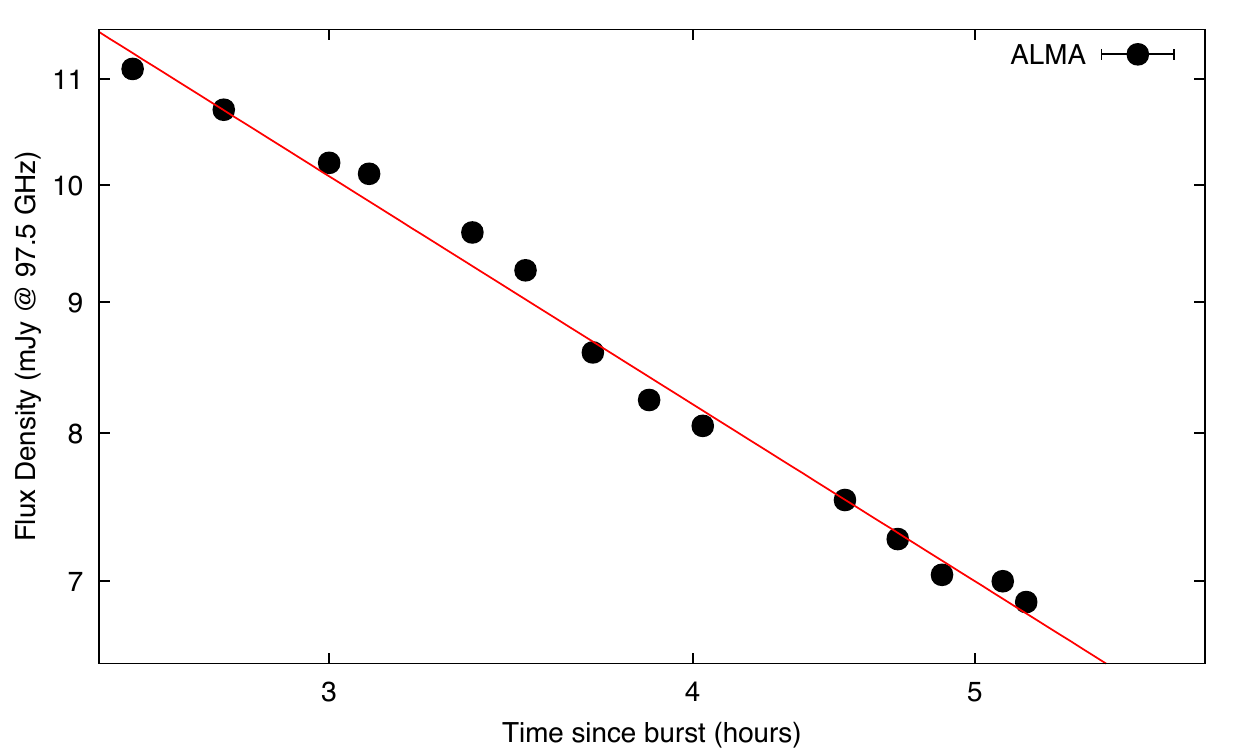}}\\
}
\caption{The upper left-hand panel shows the GBM light curve at 10 MeV. The continuous red line corresponds to the best-fit curve using the eq. (\ref{LAT_lc}) and the dashed red line correspond to a SPL.  Data were taken from \cite{2019arXiv190201861R}. The upper right-hand panel shows the X-ray light curve obtained with Swift BAT (black) and XRT (red) instruments at 10 keV. Blue lines correspond to the best-fit curves using SPL functions. The Swift data were obtained using the  publicly available database at the official Swift web site.  The lower left-hand panel
shows the optical light curves of GRB 190114C in different filters with the best-fit functions.  The continuous line corresponds to the best-fit curve using a  SPL function and the dotted-dashed line using a  BPL function. Optical data were collected from \cite{2019GCN.23699....1L, 2019GCN.23701....1M, 2019GCN.23702....1B, 2019GCN.23717....1I, 2019GCN.23726....1K, 2019GCN.23729....1D, 2019GCN.23732....1K, 2019GCN.23733....1K, 2019GCN.23734....1K, 2019GCN.23740....1I, 2019GCN.23741....1M}. The lower right-hand panel shows the radio light curve obtained with ALMA at 97.5 GHz. The red line corresponds to the best-fit curve using a  SPL function. Radio data were taken from \cite{2019arXiv190407261L}.}
\label{X-ray-optical_lc}
\end{figure}

The Fermi GBM instrument triggered and localized GRB 190114C at 2019 January 14 20:57:02.63 UTC. During the first 15 s after the trigger, the GBM light curve showed a very bright, multi-peaked pulse followed by a weaker pulse occurring between $15$ to $25$ s.  In addition, a fainter emission with a duration of $200$ s after the trigger was detected.  The GBM team reported a  duration of the main emission of $T_{90}$=116 s (50-300 keV).    This burst presented an equivalent  isotropic energy of $ 3\times 10^{53}$ erg in the energy range of 1 keV - 10 MeV \citep{2019GCN.23707....1H}.\\
Recently, \cite{2019arXiv190201861R} analyzed the GBM data finding two different spectral components: a smoothly broken power law (SBPL) and a power law (PL).  Authors showed that the EPL component in the energy range of 10 keV - 40 MeV reached the maximum flux (at the peak) of $(1.7\pm 0.2)\times 10^{-5}\,{\rm erg\,cm^{-2}\,s}$ in the time interval of $\sim$ 6 - 7 s. After the peak, this component decreased with a temporal index of 2.8 up to 15 sec and finally, with 1. They reported a spectral PL index for the GBM spectrum of $\Gamma_{\rm GBM}=\beta_{\rm GBM}+1=1.81\pm0.08$\\
Given the similarity between the LAT (see Figure  \ref{LAT_lc}) and GBM \cite[see Figure 1 in][]{2019arXiv190201861R} light curves, we take the Fermi GBM data reported in \cite{2019arXiv190201861R} and model the EPL component.  Again, the function described by eq. \ref{funcion_lat} and a SPL ($\propto t^{-\alpha_{\rm GBM}}$) are used  to describe the short-lasting peak and the long-lived emission, respectively.  In this case, the best-fit values found are $t_0=3.09\pm0.23$ s,  $\tau=7.29\pm0.46$ s,  $\alpha_{\rm \gamma, pk}=2.96\pm 0.19$ and $\alpha_{\rm GBM}=1.05\pm0.13$ ($\chi^2$ = 1.33). The values obtained with our model are very similar to those ones reported by \cite{2019arXiv190201861R}.\\
The upper left-hand panel in Figure \ref{X-ray-optical_lc} shows the GBM light curve of the EPL component at 10 MeV. The continuous and dashed red lines correspond to the best-fit curves.  Data were taken from \cite{2019arXiv190201861R}.
\subsection{X-ray observations and data reduction}
The Swift BAT instrument triggered on GRB 190114C at 2019 January 14 20:57:06.012 UTC  \citep{2019GCN.23688....1G}.   During the first 25 s, the BAT light curve exhibited a very bright multi-peaked structure.  The main brightest emission consist of two complex pulses, ending at about $50$ s after the trigger time.  Afterwards, the X-ray flux appeared to decay exponentially out to beyond $720$ s after the trigger, when the burst went out of the BAT field of view. GRB 190114C returned to the field of view of BAT at $\sim$ $3800$ s after the trigger, although no significant flux was detected at that time \citep{2019GCN.23724....1K}.\\ 
The Swift XRT instrument began observing GRB 190114C at 64 s after the trigger time.  This instrument found a bright, uncatalogued X-ray source from 03:38:01.20  to  26:56:47.6 (J2000) with a 90\% uncertainty radius of 1.4 arcsec \citep{2019GCN.23688....1G, 2019GCN.23704....1O}.\\ 
The upper right-hand panel in Figure \ref{X-ray-optical_lc} shows the Swift X-ray light curve obtained with Swift BAT (black) and XRT (red) instruments at 10 keV. Blue lines correspond to the best-fit curves using SPL functions. Swift data were obtained using the  public available database at the official Swift web site. Four PL segments are identified in the X-ray light curve: (I) an initial PL segment with a temporal index of $1.59\pm0.12$. This value clearly is not related with the typical decay slope, which is explained in terms of the high-latitude emission of the prompt GRB \citep[the emission has abruptly ceased;][]{2000ApJ...541L..51K}, (II) a PL segment with a temporal index of $0.57\pm0.09$. This value is consistent with shallow ``plateau" decay segment \citep{2006ApJ...642..354Z, 2018ApJ...869..155S}, (III)  a PL segment with a temporal index of $1.09\pm0.11$ \citep{2006ApJ...642..354Z}. This value is consistent with the normal decay segment and (IV) a late steeper decay with a temporal PL index of $2.54\pm0.14$. This value is consistent with the jet break \citep{2006ApJ...638..920V}. The best-fit values of the X-ray data are reported in Table \ref{table1}.\\ 
\subsection{Optical observations and data reduction}
The Swift UVOT began observing a candidate afterglow of GRB 190114C at 73 s after the trigger trigger \citep{2019GCN.23688....1G}. The observations using the near-ultraviolet (NUV) filters of the first few orbits indicated  that the afterglow faded rapidly \citep{2019GCN.23725....1S}.\\
Using the MASTER-IAC telescope, \cite{2019GCN.23690....1T} pointed to GRB 190114C 25 s after notice time and 47 s after trigger time. On their first set they found one optical transient within Swift error-box (RA=54.5042,  DEC=-26.9383) brighter than 16.54 magnitude. Furthermore, MASTER-SAAO with MASTER-IAC telescopes reported a polarization photometry in 4-position angles \citep{2019GCN.23693....1L}.   \cite{2019GCN.23692....1U} detected a source in Pan-STARRS archival in the field of GRB 190114C suggesting this source as the possible host galaxy of GRB 190114C.  This was confirmed by NOT \citep{2019GCN.23695....1S}, which derived a redshift of  $z$ = 0.42.   Additional photometry was reported in \cite{2019GCN.23699....1L, 2019GCN.23701....1M, 2019GCN.23702....1B, 2019GCN.23717....1I, 2019GCN.23726....1K, 2019GCN.23729....1D, 2019GCN.23732....1K, 2019GCN.23733....1K, 2019GCN.23734....1K, 2019GCN.23740....1I, 2019GCN.23741....1M}.\\
The lower left-hand panel Figure  \ref{X-ray-optical_lc} shows the optical light curves of GRB 190114C in different filters with the best-fit functions. The continuous line corresponds to the best-fit curve using a  SPL function and the dotted-dashed line using a  BPL function. SPL functions are used for the ${\rm i}$, ${\rm r}$, ${\rm v}$,  ${\rm white}$ and ${\rm b}$ bands (solid lines) and BPL functions for ${\rm r}$ and ${\rm white}$ bands (dotted-dashed lines).   Optical data were collected from  several instruments and taken from the GCN circulars showed above.  The optical fluxes and their corresponding uncertainties used in this work were calculated using the standard conversion for AB magnitudes shown in \cite{1996AJ....111.1748F}.  The optical data were corrected by the galactic extinction using the relation derived in \cite{2019arXiv190106051B}. The values of $\beta_{\rm O}=0.83$ for optical filters and a reddening of $E_{B-V}=0.01$ \citep{2019GCN.23702....1B} were used.\\
The best-fit values of the temporal PL indexes with their respective $\chi^2/ndf$ are reported in Table \ref{table2}.  This table shows that optical fluxes present two distinct decays separated by a break at $\sim$ 400 s. Before this break,  the temporal PL indexes are  stepper ($\alpha_{\rm O}= 1.593\pm0.012$ for r-band and $1.567\pm0.097$  and after they lie in the range of $0.6\lesssim \alpha_{\rm O} \lesssim 0.9$.  Due to the large amount of optical data collected in the r-band, the multi-wavelength analysis is done  considering the optical r-band data points. The r-band optical observation collected the 9$^{\rm th}$ day after the burst trigger was removed due to the contamination by the host galaxy and Supernova associated to this burst \citep{2019GCN.23766....1R, 2019GCN.23983....1M}.\\
\vspace{0.2cm}
\subsection{Radio observations}
The Atacama Large Millimeter/submillimeter Array (ALMA; at 97.5 GHz) and the Karl G. Jansky Very Large Array (VLA; at 5 - 38 GHz) began observing the afterglow of GRB 190114C at 2.2 and 4.7  hours after the burst trigger, respectively \citep{2019arXiv190407261L}. The ALMA and VLA observations were extended up to 5.2 and 6.3 hours after the burst trigger, respectively.  Authors described the SED of the radio data at 0.2 days; VLA at radio cm-band and ALMA at mm-band.  Using a BPL model  they found a spectral index  of $\beta_{\rm R}=0.3\pm 0.2$ below the break of $24\pm 4$ GHz. In addition, \cite{2019arXiv190407261L} found that the GROND K-band and ALMA observations were consistent with a SPL at 0.16 days.\\
The lower right-hand panel in Figure  \ref{X-ray-optical_lc} shows the radio light curves of the ALMA observations with the best-fit curve using a SPL function. The best-fit value of the temporal index of $0.71\pm0.01$ is reported in Table \ref{table3}. Radio data were taken from \cite{2019arXiv190407261L}.
\subsection{VHE observations}
MAGIC telescopes detected VHE gamma-ray emission from GRB 190114C. Their data showed a clear excess of gamma-ray events with the significance $\>$20$\sigma$  in the first 20 min (starting at $T+50$ s) for photon energies around 300 GeV.  
Other TeV gamma-ray observatories such as the High altitude water Cherenkov (HAWC) and H.E.S.S. neither reported VHE detection nor upper limits in the directions of GRB 190114C.
\section{Description of the multi-wavelength observations}
\subsection{Multi-wavelength analysis of observations}
\begin{figure}[h!]
{\centering
\resizebox*{0.77\textwidth}{0.4\textheight}
{\includegraphics{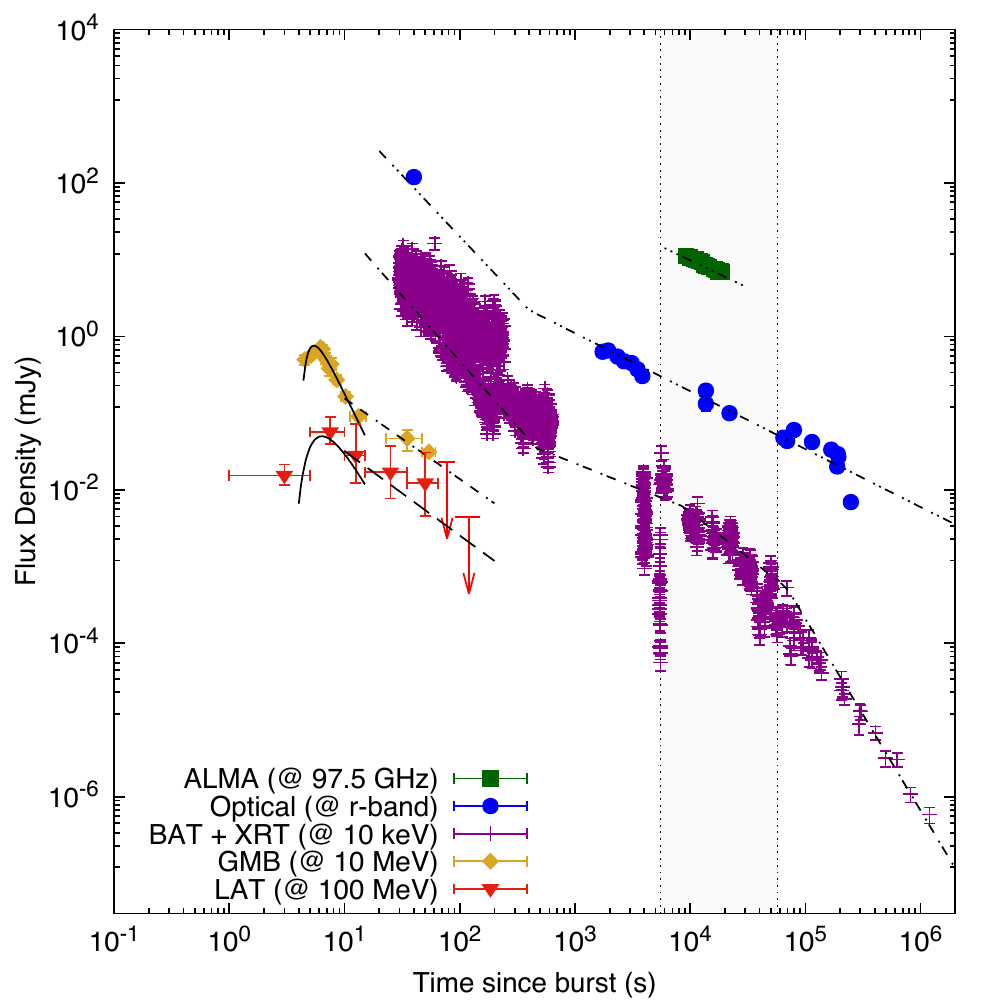}}\\
\resizebox*{0.83\textwidth}{0.4\textheight}
{\includegraphics{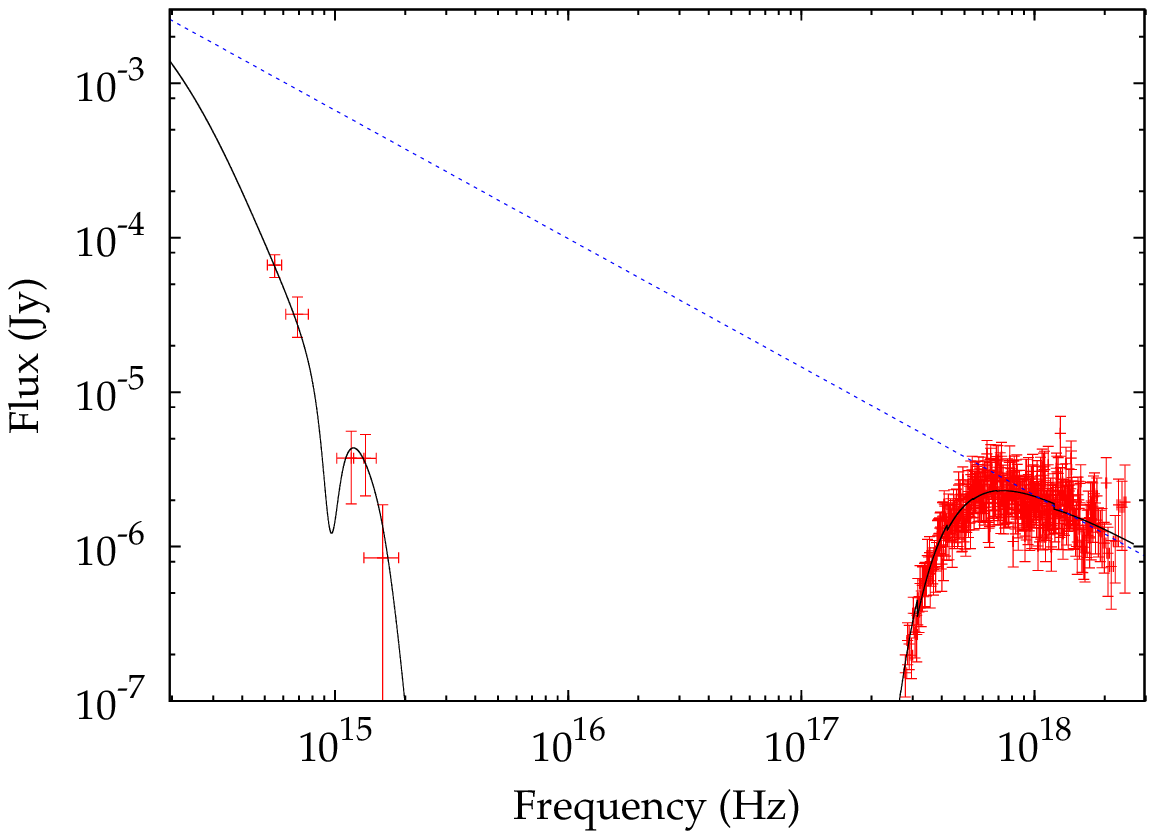}}\\
}
\caption{Top:  Light curves and fits of the multi-wavelength observation of GRB 190114C with the synchrotron forward-shock and SSC reverse-shock models.   Bottom:  The broadband SED of the X-ray and optical (UVOT) observations during the period of  5539 - 57216 s. The solid black line is the best-fit curve from XSPEC.  The shaded period in the upper panel corresponds to the spectrum on the lower panel.}
\label{grb190114c}
\end{figure}

Figure \ref{grb190114c} shows the LAT, GBM, X-ray, optical and radio light curves (upper panel) and the broadband SED of the X-ray and optical (UVOT) observations during the period of  5539 - 57216 s (lower panel) of GRB 190114C with the best-fit curves. The shaded period in the upper panel corresponds to the spectrum on the lower panel.  The best-fit values of the temporal PL indexes obtained through the Chi-square $\chi^2$ minimization function are reported in Table \ref{table3}.  In order to obtain the best-fit values of the spectral PL indexes, we analyze the broadband SED of GRB 190114C taking into account the available X-ray and optical data, and the values reported of LAT, GBM and radio bands.\\ 
During the first 70 s, the observations are almost covered by the LAT and GBM instruments with only one optical (r-band) data point. The LAT collaboration reported a spectral PL index above 100 MeV of  $\beta_{\rm LAT}=1.98\pm0.06$ \citep{2019GCN.23709....1D}. Analyzing the LAT spectrum, \cite{2019arXiv190107505W} reported PL indexes of  $\beta_{\rm LAT}=1.06\pm0.30$ and $1.10\pm0.31$ for two time intervals $\sim$ 6 - 7 s and 11 - 14 s, respectively. Analyzing the PL component of the GBM data, \cite{2019arXiv190201861R} reported a spectral index of $\beta_{\rm GBM}=0.81\pm0.08$. From 70 to 400 s, X-rays dominate the observations with one optical data point in the white band.\\
During the time interval from 5539 to 57216 s,  the optical (UVOT) and X-ray (XRT) available data are quasi-simultaneous, as shown in the lower panel in Figure \ref{grb190114c}.  From  X-ray to optical data, the SED is modelled with a SPL with PL index $\beta_{\rm X}=0.83\pm 0.04$.  The blue dashed line is the best-fit curve obtained from XSPEC. During this period, \cite{2019arXiv190407261L} described the SED of the radio data at 0.2 days; VLA at radio cm-band and ALMA at mm-band.  Using a BPL model  they found a value of spectral index of $\beta_{\rm R}=0.3\pm 0.2$ below a break of $24\pm 4$ GHz. In addition, authors found that the GROND K-band and ALMA observations were consistent with a SPL at 0.16 days. For the period of time longer than 57216 s, it is not possible to analyze the multi-wavelength observations because  there is no quasi-simultaneous available data.   The best-fit values of the temporal and spectral PL indexes of the LAT, GBM, X-ray, optical and radio fluxes are reported in Table \ref{table3}.\\
\subsection{Synchrotron forward-shock model and analysis of the long-lived multi-wavelength observations}
\subsubsection{Light Curves in a stratified stellar-wind like medium}
Taking into consideration a  Wolf-Rayet (WR) star as progenitor with typical values of a mass-loss rate of $\dot{M}\simeq 10^{-6}\,{\rm M_\odot\,yr^{-1}}$ and a constant wind velocity  of $\rm v_W\simeq 10^8\, {\rm cm\, s^{-1}}$, the density of the stratified stellar-wind like medium is given by  $\rho(r)=A\, r^{-2}$, where  $A=\frac{\dot{M}}{4\pi \rm v_W}= A_{\star}\,\, (5\times 10^{11})\, {\rm g\, cm^{-1}}$ with $A_{\star}$ a parameter of stellar wind density \citep{2000ApJ...543...66P, 2000A&A...362..295V, 2005A&A...442..587V, 2004ApJ...606..369C, 1998MNRAS.298...87D, 2000ApJ...536..195C}.   Using  the typical timescales together with the maximum power emitted by relativistic electrons, the characteristic (for $p\geq 2$) and cooling energy breaks and the maximum flux evolve as $\epsilon^{\rm syn}_{\rm m, f}\propto t^{-\frac32}$,  $\epsilon^{\rm syn}_{\rm c, f}\propto t^{\frac12}$ and $F^{\rm syn}_{\rm max, f}\propto t^{-\frac12}$, respectively. The subscript {\rm f} refers throughout this  manuscript  to  the  forward shock. 
The synchrotron breaks and the maximum flux are functions of $\varepsilon_{\rm e,f}$, $\varepsilon_{\rm B,f}$, $E$ and $A$.    The terms $\varepsilon_{\rm e,f}$ and $\varepsilon_{\rm B,f}$ refer to the microphysical parameters given to accelerate electrons and to amplify the magnetic field, respectively,   $E$ is the equivalent kinetic energy given by the isotropic energy $E_{\rm \gamma, iso}$ and the efficiency $\eta$ to convert  the kinetic to gamma-ray energy, $\xi$ is a constant parameter which lies in the range of $0.4 <\xi< 0.78$ \citep{1998ApJ...493L..31P, 2000ApJ...536..195C}.    Given the synchrotron spectra for fast- and slow-cooling regime together with the synchrotron spectral breaks and the maximum flux,  the synchrotron light curves in the fast (slow)- cooling regime  are
{\small
\begin{eqnarray}
\label{fcsyn_t-w}
F^{\rm syn}_{\rm \nu, f}\propto \cases{ 
t^{-\frac{2}{3}}\,(t^{0}) \epsilon_\gamma^{\frac13},\hspace{1.4cm} \epsilon_\gamma<\epsilon^{\rm syn}_{\rm c,f} (\epsilon^{\rm syn}_{\rm m,f}), \cr
t^{-\frac{1}{4}}\epsilon_\gamma^{-\frac12}(t^{-\frac{3p-1}{4}}\epsilon_\gamma^{-\frac{p-1}{2}} )  ,\,\, \hspace{0.1cm}       \epsilon^{\rm syn}_{\rm c,f} (\epsilon^{\rm syn}_{\rm m,f})   <\epsilon_\gamma<\epsilon^{\rm syn}_{\rm m,f} (\epsilon^{\rm syn}_{\rm c,f})\hspace{.2cm} \cr
t^{-\frac{3p-2}{4}}\,(t^{-\frac{3p-2}{4}})\epsilon_\gamma^{-\frac{p}{2}},\,\, \epsilon^{\rm syn}_{\rm m,f} (\epsilon^{\rm syn}_{\rm c,f}) <\epsilon_\gamma < \epsilon^{\rm syn}_{\rm max,f}  \,, \cr
}
\end{eqnarray}
}
where $\epsilon_\gamma$ is the energy at which the flux is detected.   Given the evolution of the bulk Lorentz factor in the stellar wind-like medium $\Gamma\propto t^{-\frac14}$,  the maximum synchrotron energy in this case evolves as $\epsilon^{\rm syn}_{\rm max,f}\propto t^{-\frac14}$.
\subsubsection{Light Curves in a uniform ISM-like medium}
The dynamics of the forward shocks for a relativistic outflow interacting  with a homogeneous medium (n) is usually  analyzed through the deceleration timescale and the equivalent kinetic energy evolved in the shock \citep[e.g., see][]{1998ApJ...497L..17S, 1995ApJ...455L.143S, 2000ApJ...532..286K,1999A&AS..138..537S}.   Taking into account  the typical timescales  together with the maximum power emitted by the electron population, the synchrotron spectral breaks and the maximum flux evolve as  $\epsilon^{\rm syn}_{\rm m,f}\propto t^{-\frac32}$, $\epsilon^{\rm syn}_{\rm c,f}\propto t^{-\frac12}$ and $F^{\rm syn}_{\rm max,f}\propto t^0$, respectively \citep{1998ApJ...497L..17S}.
Given the synchrotron spectra for fast- and slow-cooling regime  together with the synchrotron spectral breaks and the maximum flux,  the synchrotron light curves in the fast (slow)-cooling regime are
{\small
\begin{eqnarray}
\label{fcsyn_t-h}
F^{\rm syn}_{\rm \nu, f}\propto \cases{ 
t^{\frac{1}{6}}\,  (t^{\frac12}) \epsilon_\gamma^{\frac13},\hspace{1.7cm} \epsilon_\gamma<\epsilon^{\rm syn}_{\rm c,f} (\epsilon^{\rm syn}_{\rm m,f}), \cr
t^{-\frac{1}{4}}\epsilon_\gamma^{-\frac12}(t^{-\frac{3p-3}{4}}\epsilon_\gamma^{-\frac{p-1}{2}})\,\, \hspace{0.1cm}       \epsilon^{\rm syn}_{\rm c,f} (\epsilon^{\rm syn}_{\rm m,f})  <\epsilon_\gamma<\epsilon^{\rm syn}_{\rm m,f} (\epsilon^{\rm syn}_{\rm c,f})\hspace{.3cm} \cr
t^{-\frac{3p-2}{4}}\,(t^{-\frac{3p-2}{4}})\epsilon_\gamma^{-\frac{p}{2}},\,\,\,\,  \epsilon^{\rm syn}_{\rm m,f} (\epsilon^{\rm syn}_{\rm c,f})<\epsilon_\gamma < \epsilon^{\rm syn}_{\rm max,f}\,, \cr
}
\end{eqnarray}
}
where $\epsilon_\gamma$ is the energy at which the flux is detected.  Given the evolution of the bulk Lorentz factor $\Gamma\propto t^{-\frac38}$ in the forward shock,  the maximum synchrotron energy evolves as $\epsilon^{\rm syn}_{\rm max,f} \propto t^{-\frac38}$. 
\subsubsection{Analysis of long-lived multi-wavelength observations}
Given the spectral and temporal indexes of the LAT, GBM, X-ray, optical and radio bands,  it can be observed from Table \ref{table3} that the evolution of synchrotron emission can be separated into four distinct periods.\\
During the first period ($t\lesssim 400\, {\rm s}$),  the temporal decays of the optical and X-ray observations are equal and are steeper ($\Delta\alpha\sim 0.4$) than the ones of the LAT and GBM light curves. During this period the spectral indexes of the LAT and GBM observations are consistent each one within the uncertainties. It is worth noting that the temporal PL index of the X-ray light curve cannot be associated with the end of prompt emission that is larger than 2.5. We conclude that both the LAT and GBM observations  evolve in the third PL segment and the optical and X-ray fluxes evolve in the second PL segment of the slow-cooling regime in  the stratified stellar-wind like medium  for $p=2.2\pm0.3$.\\
During the second and third periods ($ 400\lesssim t\lesssim 10^5$),  the X-ray flux presents a chromatic break at $\sim 10^4 s$. During this transition, the temporal PL index varied from $0.57\pm0.09$ to $1.09\pm0.11$ while the spectral index remained unchanged.  The temporal PL index  after the break is consistent with the afterglow model evolving in a uniform IMS-like medium, while the temporal index before the break is associated to the ``plateau" phase. It is worth mentioning that during this shallow-to-normal transition as found in a large fraction of GRBs, the spectral index does not vary. During this period, the spectral analysis presented in this work reveals that the optical and X-ray observations are consistent with a SPL.
Moreover, the temporal PL indexes of radio (ALMA) and optical observations are consistent each other, and the spectral analysis  reported by \cite{2019arXiv190407261L}  indicated that these observations are consistent with a SPL. Similarly, their analysis  reported that the radio observations between VLA and ALMA are consistent with a BPL with a break at 24 GHz.  Therefore, we conclude that X-ray, optical and radio (ALMA) fluxes  evolve in the second PL segment between the cutoff and characteristic energy breaks, and the radio (VLA) evolves in first PL segment of the slow-cooling regime in the uniform ISM-like medium for $p=2.2\pm0.3$.\\
\\
During the four period ($t\gtrsim 10^5$), the temporal index in the X-ray flux is consistent with the jet break. 
\\
The temporal and spectral theoretical indexes obtained by the evolution of the standard synchrotron model in  the stratified stellar-wind like medium and  in the uniform ISM-like medium are reported in Table \ref{table3}.  Theoretical and observational spectral and temporal indexes are in agreement. The best explanation for this behavior is that the synchrotron radiation undergoes  a phase transition from a stratified stellar-wind like to uniform ISM-like  medium around $\sim$ 400 s\\ 
\subsection{The SSC reverse-shock model and Analysis of the short-lasting bright LAT peak}
\subsubsection{SSC model in the stratified stellar-wind like medium}
The quantities of synchrotron reverse-shock model  such as the spectral breaks, the fluxes and the light curves that describe the optical flashes  are introduced in \cite{2000ApJ...536..195C}.  In the thick-shell case ($\Gamma <\Gamma_c$) where the deceleration time is assumed to be smaller than the duration of the prompt phase and then the outflow is decelerated by the reverse shock are derived in \citep{2005ApJ...628..315Z}.   The term $\Gamma_c$ is the critical Lorentz factor.    The relationship among the characteristic energy breaks and maximum fluxes in the forward and reverse shock were derived in \citep{2005ApJ...628..315Z}.\\
The quantities of the SSC reverse-shock model  as the spectral breaks, the fluxes and the light curves have been widely explored \citep[e. g. see,][]{2001ApJ...546L..33W,  2001ApJ...556.1010W, 2012ApJ...755...12V, 2016ApJ...818..190F}. In the thick-shell case, the  SSC light curve  at the shock crossing time ($t_{\rm d}$) was showed in \cite{2016ApJ...818..190F}.  At $t < t_{\rm d}$,  the SSC  emission increases as $\propto t^{1/2}$ reaching at the shock crossing time the maximum value  of $F_{\rm \nu, r} \sim F_{\rm \nu, max, r} \left( \frac{\epsilon_{\rm LAT}}{\epsilon^{\rm ssc}_{\rm c, r}}\right)^{-\frac12}$  where the energy range observed by LAT instrument ($\epsilon_{\rm LAT}$) is constrained by the characteristic break ($\epsilon_{\rm LAT} < \epsilon^{\rm ssc}_{\rm m,r}$).  After at $t  > t_{\rm d}$,  the LAT flux initially evolves  as $\propto t^{-\frac{p+1}{2}}$,  later as $\propto t^{-\frac52}$ and finally as $\propto t^{-\frac{p+4}{2}}$ induced by  the  angular time delay effect \citep{2003ApJ...597..455K, 2000ApJ...541L..51K}.  The shock crossing time can be estimated as $t_d\sim\left(\Gamma/\Gamma_c\right)^{-4} T_{90}$ \citep{2007ApJ...655..973K}.\\
\subsubsection{Analysis of the LAT/GBM-peak observations}
In order to model the Fermi LAT/GBM data, the function given by equation (\ref{funcion_lat}) was used \citep{2017ApJ...848...15F}.  The best-fit values of $t_0=2.61\pm0.51$ s and $3.09\pm0.23$ s indicate the onset of the reverse shock as suggested by  \cite{2006Natur.442..172V}.   The values of the temporal decay indexes of $\alpha_{\gamma, pk}=2.65\pm0.19$ and $2.96\pm0.19$ are consistent with the decay slope of the synchrotron/SSC reverse-shock emission from high latitudes (due to the curvature effect) \citep{2003ApJ...595..950Z, 2019ApJ...871..123F, 2017ApJ...848...94F}. The values of bulk Lorentz factor and the parameter of the stellar wind density can be constrained through the deceleration time  {\small $t_{\rm dec}\propto \left(1+z \right)\,\xi^{-2}\,E\,A^{-1}\,\Gamma^{-4}$} with the LAT/GBM-peak flux at $\sim$ 6 - 7 s and the critical Lorentz factor in the thick-shell regime $\Gamma > \Gamma_{\rm c}$ \citep{2003ApJ...595..950Z}.   In the thick-shell regime, the shock crossing time is $t_d\sim\left(\Gamma/\Gamma_c\right)^{-4} T_{90}\simeq 6- 7$ s \citep{2007ApJ...655..973K}, which is much shorter than the duration of the main burst.  The peak of the LAT and GBM  fluxes will be modelled  with $F_{\rm \nu, r} \sim F_{\rm \nu, max, r} \left( \frac{\epsilon_{\rm \gamma}}{\epsilon^{\rm ssc}_{\rm c, r}}\right)^{-\frac12}$ \citep{2003ApJ...595..950Z,2016ApJ...831...22F} and the value of the spectral index of electrons ${\rm p}=2.2\pm 0.3$ found with multi-wavelength observations  and synchrotron forward-shock model will be used.  We want to emphasize that the synchrotron emission from the reverse shock is usually invoked to describe early optical afterglows \citep{2000ApJ...545..807K, 2003ApJ...597..455K, 2016ApJ...818..190F}, so the SSC emission used in this work is required to describe the LAT/GBM-peak observations.\\
\subsection{Transition from a stratified stellar-wind like to uniform ISM-like medium}
As indicated in subsection 3.2.1,  the progenitor of GRB 190114C can be associated with the core collapse of a WR star, indicating that the circumburst medium close to the progenitor is principally composed by the stratified stellar wind of the WR.  At a distance away from the parent a uniform medium is expected. Therefore, a transition phase between the stratified to uniform medium is expected at a distance larger than $\gtrsim 10^{-2}$ pc \citep{1977ApJ...218..377W, 1975ApJ...200L.107C, 2006ApJ...647.1269F}. \citeauthor{1977ApJ...218..377W} studied this phase considering a four-region structure which are (i) the unshocked stratified stellar-wind like medium with density $\rho(r)$, (ii) a quasi-isobaric zone consisting of the stellar wind mixed with a small fraction of interstellar gas, (iii) a dense-thin shell formed by most of ISM and (iv) the unshocked ambient ISM (see Figure 1 in \citealp{2006ApJ...643.1036P}).\\
Taking into consideration an adiabatic expansion, two strong shocks are formed, the outer and inner shocks. The outer termination (forward) shock radius can be estimated as 
{\small
\be
R_{FS, W}= 1.2\times 10^{19}\,  {\rm cm}\,\,\dot{M}^\frac15_{-6}\,\rm v_{W,8}^\frac25\,n^{-\frac15}\,t_{\star,5}^{-\frac35}\,,
\ee
}
where $t_\star$ is the lifetime of the WR.\\
The inner (reverse) shock radius for which the transition from stratified to uniform medium occurs \citep[$R_{\rm tr}$;][]{2006ApJ...643.1036P} is obtained by equaling the pressures in regions (ii) and (iii) \citep[e.g. see, ][]{2006ApJ...643.1036P, 1996ApJ...469..171G}
{\small
\be
P_{\rm (ii)}=P_{\rm (iii)}= 1.4\times 10^{-11}\,{\rm dynes\, cm^{-2}}\,\,\dot{M}^\frac25_{-6}\,\rm v_{W,8}^\frac45\,n^{-\frac{3}{5}}\,t_{\star,5}^{\frac45}\,.
\ee
}
The distance from the progenitor to the wind-to-homogeneous transition  is given by
{\small
\be
R_{\rm tr}\equiv R_{RS,W}=  5.1\times10^{18} {\rm cm} \,\,\dot{M}^\frac{3}{10}_{-6}\,\rm v_{W,8}^\frac{1}{10}\,n^{-\frac{3}{10}}\,t_{\star,5}^{\frac25}\,.
\ee
}
The density of the stellar wind medium  at $r=R_{\rm tr}$ can be written as
{\small
\be
\rho(R_{\rm tr})=1.8\times 10^{-27}\, {\rm g\,cm^{-3}}\,\,  R^{-2}_{\rm tr}\,\dot{M}_{-6}\, \rm v_{W,8}^{-1}\,,
\ee
}
which corresponds to a particle number density of  $\sim 10^{-3}\, {\rm cm^{-3}}$.
\section{Results and Discussion}
\begin{figure}[h!]
{ \centering
\resizebox*{0.9\textwidth}{0.5\textheight}
{\includegraphics{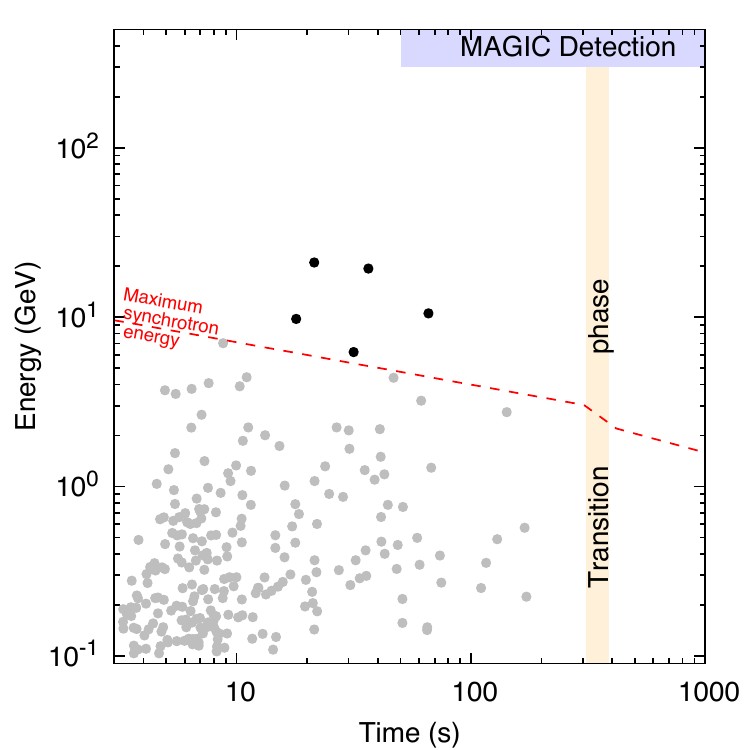}}
}
\caption{All the photons with energies $> 100$ MeV and probabilities $>90$\% of being associated with  GRB 190114C.  The red dashed line is the maximum photon energies released by synchrotron forward-shock model in a stratified stellar-wind medium and a uniform ISM-like medium. The yellow region represents the transition phase from a stratified to uniform medium and, the purple region the interval and the energy range of VHE photons reported by the MAGIC collaboration.
Photons with energy above the maximum synchrotron energy are in black and below are in gray.}
\label{photons_MeV}
\end{figure}

\begin{figure}[h!]
{ \centering
\resizebox*{1.\textwidth}{0.6\textheight}
{\includegraphics{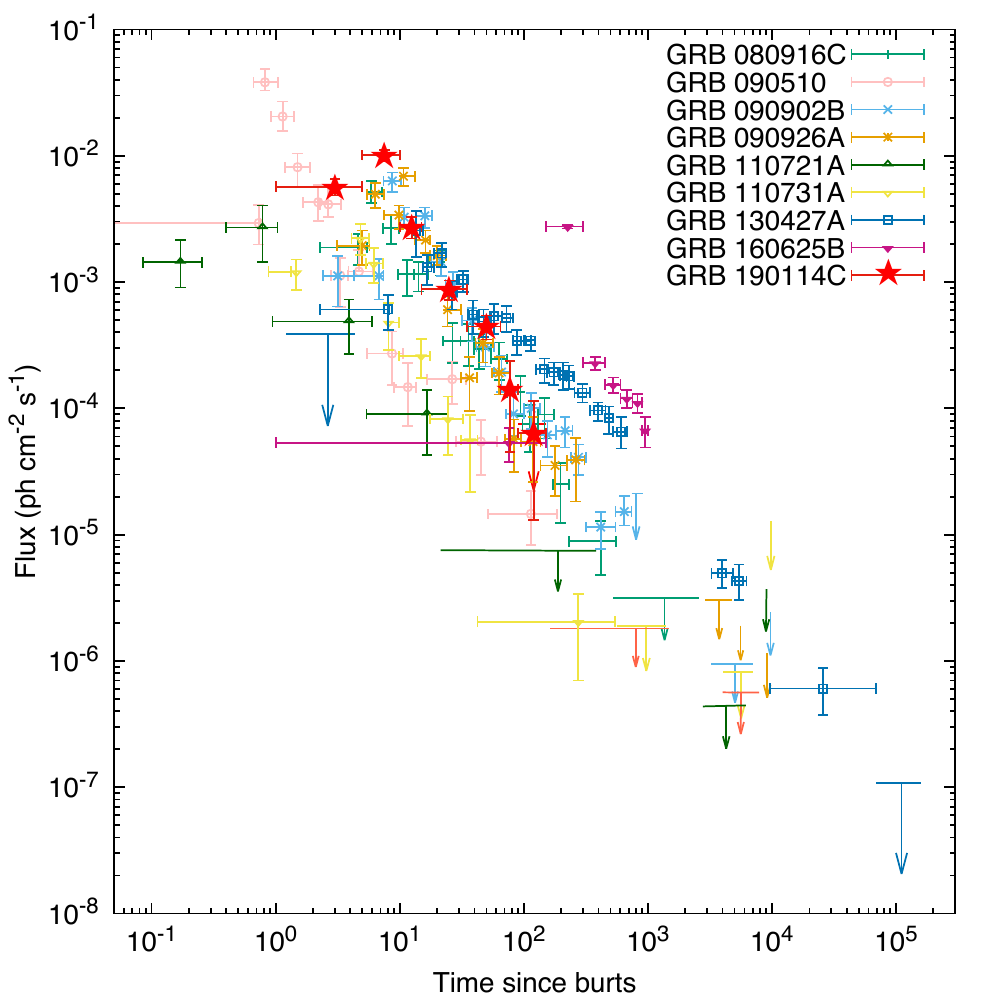}}
}
\caption{Comparison of the Fermi-LAT photon flux light curve from GRB 190114C (red filled stars) with those LAT-detected burst with short-lasting bright peaks  and long-lived emissions. For the LAT-detected burst  data are taken from \cite{2013ApJS..209...11A,2013ApJ...763...71A,  2014Sci...343...42A, 2017ApJ...848...94F}.}
\label{all_GRBs}
\end{figure}

We show that temporal and spectral analysis of the long-lived multi-wavelength observations of GRB 190114C is consistent with the closure relations of the synchrotron forward-shock model and the short-lasting LAT and GBM peaks with SSC reverse-shock model.  The LAT and GBM observations favor the emission originated from the forward and reverse shocks in a stratified stellar-wind like medium, and the X-ray and optical observations are consistent with the emission from forward shocks in both a stratified stellar-wind and a uniform ISM-like medium. The radio observations are consistent with the synchrotron emission radiated in a uniform ISM-like medium.  The transition from the stratified  to  uniform medium is found to be around $\sim$ 400 s after the GBM trigger. Now,  we obtain the electron spectral index, the microphysical parameters and the circumburst densities for which our model is satisfied. The photon energies of each PL segment at $\epsilon_\gamma=$ 97.5 GHz, 1 eV, 10 keV, 10 MeV and 100 MeV  are considered to describe the radio, optical, X-ray, GBM and LAT fluxes, respectively. We use the synchrotron light curves in the slow-cooling regime evolving in a stratified stellar-wind like medium (eqs. \ref{fcsyn_t-w}) before $\lesssim$ 400 s  and in a uniform ISM-like medium (eqs. \ref{fcsyn_t-h}) after $\gtrsim$ 400 s.  The values reported of the observed quantities such as the redshift $z=0.42$, the equivalent isotropic energy $3\times 10^{53}\,{\rm erg}$ and the duration of the prompt emission $T_{90}=116\,{\rm s}$  are required. In order to compute the luminosity distance, the values of cosmological parameters derived in \cite{2018arXiv180706209P} are used (Hubble constant $H_0=(67.4\pm 0.5)\,{\rm km\,s^{-1}\,Mpc^{-1}}$ and the matter density parameter $\Omega_{\rm m}=0.315\pm 0.007$).   The equivalent kinetic energy is obtained using  the isotropic energy  and the efficiency to convert the kinetic to photons of  $\eta=0.15$ \citep{2015MNRAS.454.1073B}. The value of the parameter $\xi=0.6$ was chosen taking into account the range of values  reported in the literature \citep{1998ApJ...493L..31P, 2000ApJ...536..195C}.\\ 
To find the best-fit values of the parameters that reproduce the multi-wavelength observations of GRB 190114C,  we use the Bayesian statistical technique  based on the Markov-chain Monte Carlo (MCMC) method \cite[see][]{2019ApJ...871..200F, 2019arXiv190407732F,2019arXiv190600502F}. The MCMC code computes the synchrotron forward-shock and the SSC reverse-shock models using, in general, a set of seven parameters, \{$A_\star$, $n$, $\epsilon_{\rm B,f}$, $\epsilon_{\rm e,f}$, $\epsilon_{\rm B,r}$, $\epsilon_{\rm e,r}$ and $p$\}. In particular, we use in each electromagnetic band only five parameters. For instance, the parameter  \{$n$\} is not used for the LAT and GBM observations, the parameters \{$\epsilon_{\rm B,r}$ and $\epsilon_{\rm e,r}$\} are not used for radio, optical and X-ray observations and the microphysical parameters \{$\epsilon_{\rm e,f}$ and $\epsilon_{\rm B,f}$\} are used to fit the radio observations.   A total of 16000 samples with 4000 tuning steps were run.   The best-fit value of each parameter for LAT, GBM, X-ray, optical and radio observations is reported in Table \ref{table4}.  The obtained values are 
typical with those reported by other luminous GRBs  \citep{2010ApJ...716.1178A, 2013ApJ...763...71A, 2014Sci...343...42A, 2015ApJ...804..105F, 2016ApJ...818..190F, 2016ApJ...831...22F, 2017ApJ...848...94F}.  Given the values of the observed quantities and the best-fit values reported in Table \ref{table4}, the results are discussed as follows.\\
\\
Taking into account the evolution of the maximum photon energy radiated by synchrotron emission from forward shock in both a stratified stellar-wind and a uniform ISM-like medium and the best-fit values of both densities,  we plot in Figure \ref{photons_MeV}  all photons with energies larger than $> 100$ MeV detected by Fermi LAT and associated to GRB 190114C. In addition, this figure shows in a yellow region the transition from the stratified to uniform medium, and the interval and the energy range of VHE photons (purple region) reported by the MAGIC collaboration \citep{2019GCN.23701....1M}.   Photons with energies above the maximum photon energy radiated by synchrotron emission (synchrotron limit) are in black and below are in gray.  In this figure is shown that the standard synchrotron forward-shock model can hardy explain all photons, therefore this model has to be varied or some additional processes to synchrotron in the forward shocks such as SSC emission, photo-hadronic interactions \citep[e.g.][]{2014MNRAS.437.2187F, 2015MNRAS.450.2784F}  and proton synchrotron radiation \citep[e.g.][]{2010ApJ...724L.109R} has to be evoked to interpret these VHE photons.  We want to emphasize that the LAT photons below the maximum synchrotron energy (the red dashed line) can be interpreted in the synchrotron forward-shock framework and beyond the synchrotron limit some additional mechanisms must be present to explain the VHE LAT photons. It is worth noting that a combination of synchrotron and SSC emission originated in the forward shock works well to explain the LAT photons \citep[e.g., see][]{2015MNRAS.454.1073B}.\\
\\
The best-fit values of the microphysical parameters found in forward- and reverse-shock regions are different. The microphysical parameter associated to the magnetic field in the reverse shock lies in the range of the expected values for the reverse shock to be formed and leads to an estimate of the magnetization parameter which is defined as the ratio of Poynting flux to matter energy flux $\sigma=\frac{L_{\rm pf}}{L_{\rm kn}}\simeq \frac{B^2_r}{4\pi \rho(r)\,\Gamma^2}\simeq 8\epsilon_{\rm B, r}\simeq0.8$ \citep{2005ApJ...628..315Z, 2002A&A...387..714D}.  This value indicates that the outflow is magnetized. In a different situation (e. g. $\sigma\gg$1),  particle acceleration would be hardly efficient and the LAT and GBM emissions  from the reverse shock would have been suppressed \citep{2004A&A...424..477F}.  Considering the microphysical parameter associated to the magnetic field in the reverse-shock region, we found that the strength of magnetic field in this region is stronger that the magnetic field in the forward-shock region ($\simeq 20$ times). It suggests that the jet composition of GRB 190114C could be Poynting  dominated.  \cite{2005ApJ...628..315Z} described the emission generated in the reverse shock from an outflow with an arbitrary value of the magnetization parameter. They found that the Poynting energy is transferred to the medium only until the reverse shock has disappeared. Given the timescale of the reverse shock associated to the short-lasting LAT and GBM peaks ($<100$ s),  the shallow decay segment observed in the X-ray light curve of GRB 190114C  might be interpreted as the late transferring of the Poynting energy to the uniform medium.  This result agrees with the linear polarization reported in radio \citep{2019arXiv190407261L} during the ``plateau" phase. These results agree with  some authors who claim that Poynting flux-dominated models with  a moderate degree of magnetization can explain the LAT observations in several powerful GRBs \citep{2014NatPh..10..351U,2011ApJ...726...90Z}, and in particular the properties exhibited in the light curve of GRB 190114C.\\
\\
Using the synchrotron reverse-shock model \citep{2003ApJ...597..455K, 2000ApJ...545..807K} and the best-fit values found, the self-absorption, the characteristic and cutoff energy breaks of $4.5\times10^{-8}\,{\rm eV}$, $0.5\,{\rm eV}$ and $8.1\times 10^{-3}\,{\rm eV}$, respectively, indicate that the synchrotron radiation evolves in the fast-cooling regime. Therefore, an optical bright flash with a maximum flux (at the peak) of  $F_{\nu, r}\sim F_{\rm max, r} \left(\frac{\epsilon_\gamma}{\epsilon_{\rm c, r}}\right)^{-\frac12} \sim 10^4\,{\rm mJy}$ in temporal coincidence with the LAT bright peaks similar to that reported for GRB 130427A is expected \citep[see,][]{2003ApJ...597..455K, 2016ApJ...831...22F}. The maximum flux and the spectral  break of the cutoff energy  are calculated with the best-fit parameters reported in Table \ref{table4}  for $\epsilon_\gamma=1\,{\rm eV}$.    Given that the self-absorption energy break is smaller than the cutoff and characteristic ones, the synchrotron emission is in the weak self-absorption regime and hence, a thermal component from the reverse shock cannot be expected \citep{2003ApJ...597..455K}. Taking into consideration that the outflow composition is Poynting dominated and  the synchrotron emission from the reverse shock is stronger than the radiation originated from the forward shock,  then polarization is expected in different wavelength bands.\\
\\
Using the best-fit values we calculate the theoretical fluxes at the maximum fluxes reported by the LAT and GBM instruments. We find  that the synchrotron emission from the forward-shock region is $\sim$ 3 times smaller than the SSC one from the reverse-shock. Once the LAT flux decreases, the synchrotron emission from forward shock begins dominating. Therefore,  the SSC emission from the reverse shock can only explain the short-lasting LAT peak and the high-energy photons associated temporally with it and not the high-energy photons detected at different time interval ($\gtrsim 10\, {\rm s}$). \\
\\
The spectral and temporal analysis of the forward and reverse shocks at the beginning of the afterglow phase together the best-fit value of the circumburst density  lead to an estimate of the initial bulk Lorentz factor, the critical Lorentz factor and the shock crossing time  $\Gamma\simeq$600, $\Gamma_{\rm c}\simeq270$ and $t_d\simeq 4\,{\rm s}$, respectively. The value of the initial bulk Lorentz factor lies in the range of values reported for the luminous LAT-detected GBRs \citep{2012ApJ...755...12V}. This value is consistent with the evolution of reverse shock in the thick-shell case and the duration of the short-lasting LAT and GBM peaks.\\
\\
The best-fit values found after modelling the LAT, GBM, X-ray, optical and radio observations with reverse and forward shocks indicate that the high-energy photons were originated in external shocks as was previously suggested for others GRBs \citep{2009MNRAS.400L..75K,2010MNRAS.409..226K,2009MNRAS.396.1163Z,2010MNRAS.403..926G,2011ApJ...733...22H,2014MNRAS.443.3578N,2016ApJ...831...22F,2017ApJ...848...15F}. It is worth highlighting that the values found of $t_0$ are in the range of the first high-energy photons detected by Fermi LAT. \\
\\
Given the best-fit values of the wind-like and homogeneous medium, the deceleration radius and the bulk Lorentz factor at the transition from the stratified to uniform medium is  $R_{\rm tr}\simeq 2.3\times10^{17}$ cm and $\Gamma_{\rm tr}\simeq220$, respectively, which agree with the breaks in the X-ray and optical light curves. In comparison with other bursts that exhibited this transition \citep[GRB 050319, 081109A and GRB 160626B;][]{2007ApJ...664L...5K, 2009MNRAS.400.1829J, 2017ApJ...848...15F}, the value obtained for GRB 190114C corresponds to the nearest value to the progenitor. \\
\\
With the best-fit values, we find that the characteristic and cutoff energy breaks of the synchrotron emission in the uniform medium at $6\times 10^3\, (6\times 10^4)\,{\rm s}$ as indicated with dotted lines in the upper panel are 93.2 (5.1) GHz and $166.5$ (27.6) keV, respectively. It indicates that during this time interval, X-ray, optical and radio fluxes evolve in the second PL segment, as shown in Figure \ref{grb190114c}.  {\rm The dotted lines mark the period for which the energy breaks were calculated.}   At 0.2 days,  the characteristic and cutoff energy breaks are 22.1 GHZ and 77.6 keV, respectively.  This result is consistent with the radio observations reported by \cite{2019arXiv190407261L}: i) the optical and radio (ALMA) observations evolved in the similar PL segment and, ii) the break energy of $24\pm 4$ GHz found in the radio spectrum  between  VLA and ALMA data. In this case this energy break is explained with the characteristic energy calculated in our model. \\  
\\
The Fermi-LAT photon flux light curve of GRB 190114C presented similar features to other bright LAT-detected bursts, as shown in Figure \ref{all_GRBs}.  For instance, the equivalent isotropic energy of these bursts was measured to be larger than $>10^{53}\,{\rm erg}$, \footnote{GRB 080916C \citep{2009Sci...323.1688A}, GRB 090510 \citep{2010ApJ...716.1178A}, GRB 090902B \citep{2009ApJ...706L.138A}, GRB 090926A \citep{2011ApJ...729..114A} GRB 110721A \citep{2013ApJ...763...71A,2017ApJ...848...94F}, GRB 110731A \citep{2013ApJ...763...71A}, GRB 130427A \citep{2014Sci...343...42A} and GRB 160625B \citep{2017ApJ...848...15F}} they exhibited long-lived emission lasting much longer than the prompt phase and had a short-lasting bright peak located at the beginning of the long-lived emission  \citep{2009MNRAS.400L..75K, 2010ApJ...718L..63P, 2010MNRAS.409..226K, 2013ApJ...763...71A, 2015ApJ...804..105F, 2016ApJ...818..190F, 2017ApJ...848...15F, 2016ApJ...831...22F, 2017ApJ...848...94F}.  All of them presented high-energy photons ($\geq$ 100 MeV), which arrived delayed alongside the onset of the prompt phase. In addition to exhibiting the previous features, GRB 160625B showed the wind-to-uniform transition. These bursts have been interpreted in the framework of external shocks. The best-fit parameters found for GRB 190114C lie in the range of the values reported in these bursts $0.01\leq\epsilon_{\rm e, f}\leq0.1$, $10^{-5}\leq\epsilon_{\rm B, f}\leq 10^{-3}$ and $2.15\leq p \leq 2.4$. Figure \ref{all_GRBs} shows that GRB 190114C (red filled stars) is one of the brightest during the first $\sim$ 100 s and, given that it is the second closest one, VHE photons are expected from this burst. 
\section{Conclusions}
We have obtained the Fermi LAT light curve around the reported position of GRB 190114C and showed that it exhibits similar features to the LAT-detected bursts.   The first photon detected by the LAT instrument had an energy of 571.4 MeV, arriving at $\sim$ 2.7 s late with respect to first low-energy photon reported by GBM. The time arrival of this energetic photon is consistent with the starting times of the LAT  ($t_0=2.61\pm0.51$ s) and GBM ($t_0=3.09\pm0.23$ s) emissions.  The highest-energy photons of 10 , 21, 6, 19 and  11 GeV  detected by the LAT instrument at 18, 21, 32, 36 and 65 s, respectively,  after the GBM trigger can be hardly interpreted in the standard synchrotron forward-shock model and some additional mechanisms must be present to interpret the VHE LAT photons. We want to emphasize that the MAGIC-detected photons cannot either be interpreted in the standard synchrotron forward-shock model. The other LAT photons can be explained well by synchrotron emission from the forward shock.   The LAT and GBM light curves exhibited a short-lasting bright peak and a long-lived extended emission.  The temporal and spectral indexes of the long-lived extended component are consistent with synchrotron forward-shock model and the  short-lasting bright peaks with SSC reverse-shock model. Given the best-fit values, a bright optical flash produced by synchrotron reverse-shock is expected.\\   
The X-ray and optical light curves are consistent with a BPL function with  a break at $\sim$  400 s. Using the closure relations and the synchrotron forward-shock model among the LAT, GBM, X-ray, optical and radio observations we claim that this break corresponded to a transition phase between a stratified stellar-wind like and uniform ISM-like medium.\\ 
With the values of best-fit values of the stratified and uniform medium, we infer that high-energy observed photons are produced in the deceleration phase of the outflow and  a different mechanism of the standard synchrotron model such as SSC emission, photo-hadronic interactions and proton synchrotron radiation from forward shocks has to be invoked to interpret these VHE photons. Given the values of the microphysical parameters, we claim that the outflow is endowed with magnetic fields.\\
The best-fit values of the microphysical parameters and the derived value of $\sigma$-parameter  indicates that an outflow with arbitrary magnetization could explain the features exhibited in the light curves of GRB 190114C (the short-lasting peaks, the ``plateau" phase, etc).  Taking into consideration that the ejecta must  be magnetized  and  the synchrotron emission from the reverse shock is stronger than the radiation originated in the forward shock,  then polarization in distinct wavelengths is expected.\\
\acknowledgements
We thank Peter Veres, Alexander A. Kann, Michelle Hui, Eleonora Troja, Alan Watson, Fabio De Colle and Diego Lopez-Camara for useful discussions. NF  acknowledges  financial  support  from UNAM-DGAPA-PAPIIT  through  grant  IA102019. RBD acknowledges support from the National Science Foundation under Grant 1816694 BBZ acknowledges support from National Thousand Young Talents program of China and National Key Research and Development Program of China (2018YFA0404204) and The National Natural Science Foundation of China (Grant No. 11833003).

\clearpage
%

\begin{thebibliography}{106}%
\makeatletter
\providecommand \@ifxundefined [1]{%
 \@ifx{#1\undefined}
}%
\providecommand \@ifnum [1]{%
 \ifnum #1\expandafter \@firstoftwo
 \else \expandafter \@secondoftwo
 \fi
}%
\providecommand \@ifx [1]{%
 \ifx #1\expandafter \@firstoftwo
 \else \expandafter \@secondoftwo
 \fi
}%
\providecommand \natexlab [1]{#1}%
\providecommand \enquote  [1]{``#1''}%
\providecommand \bibnamefont  [1]{#1}%
\providecommand \bibfnamefont [1]{#1}%
\providecommand \citenamefont [1]{#1}%
\providecommand \href@noop [0]{\@secondoftwo}%
\providecommand \href [0]{\begingroup \@sanitize@url \@href}%
\providecommand \@href[1]{\@@startlink{#1}\@@href}%
\providecommand \@@href[1]{\endgroup#1\@@endlink}%
\providecommand \@sanitize@url [0]{\catcode `\\12\catcode `\$12\catcode
  `\&12\catcode `\#12\catcode `\^12\catcode `\_12\catcode `\%12\relax}%
\providecommand \@@startlink[1]{}%
\providecommand \@@endlink[0]{}%
\providecommand \url  [0]{\begingroup\@sanitize@url \@url }%
\providecommand \@url [1]{\endgroup\@href {#1}{\urlprefix }}%
\providecommand \urlprefix  [0]{URL }%
\providecommand \Eprint [0]{\href }%
\providecommand \doibase [0]{http://dx.doi.org/}%
\providecommand \selectlanguage [0]{\@gobble}%
\providecommand \bibinfo  [0]{\@secondoftwo}%
\providecommand \bibfield  [0]{\@secondoftwo}%
\providecommand \translation [1]{[#1]}%
\providecommand \BibitemOpen [0]{}%
\providecommand \bibitemStop [0]{}%
\providecommand \bibitemNoStop [0]{.\EOS\space}%
\providecommand \EOS [0]{\spacefactor3000\relax}%
\providecommand \BibitemShut  [1]{\csname bibitem#1\endcsname}%
\let\auto@bib@innerbib\@empty
\bibitem [{\citenamefont {{Zhang}}\ and\ \citenamefont
  {{M{\'e}sz{\'a}ros}}(2004)}]{2004IJMPA..19.2385Z}%
  \BibitemOpen
  \bibfield  {author} {\bibinfo {author} {\bibfnamefont {B.}~\bibnamefont
  {{Zhang}}}\ and\ \bibinfo {author} {\bibfnamefont {P.}~\bibnamefont
  {{M{\'e}sz{\'a}ros}}},\ }\href {\doibase 10.1142/S0217751X0401746X}
  {\bibfield  {journal} {\bibinfo  {journal} {International Journal of Modern
  Physics A}\ }\textbf {\bibinfo {volume} {19}},\ \bibinfo {pages} {2385}
  (\bibinfo {year} {2004})},\ \Eprint
  {http://arxiv.org/abs/arXiv:astro-ph/0311321} {arXiv:astro-ph/0311321}
  \BibitemShut {NoStop}%
\bibitem [{\citenamefont {{Kumar}}\ and\ \citenamefont
  {{Zhang}}(2015)}]{2015PhR...561....1K}%
  \BibitemOpen
  \bibfield  {author} {\bibinfo {author} {\bibfnamefont {P.}~\bibnamefont
  {{Kumar}}}\ and\ \bibinfo {author} {\bibfnamefont {B.}~\bibnamefont
  {{Zhang}}},\ }\href {\doibase 10.1016/j.physrep.2014.09.008} {\bibfield
  {journal} {\bibinfo  {journal} {\physrep}\ }\textbf {\bibinfo {volume}
  {561}},\ \bibinfo {pages} {1} (\bibinfo {year} {2015})},\ \Eprint
  {http://arxiv.org/abs/1410.0679} {arXiv:1410.0679 [astro-ph.HE]} \BibitemShut
  {NoStop}%
\bibitem [{\citenamefont {{Giblin}}\ \emph {et~al.}(1999)\citenamefont
  {{Giblin}}, \citenamefont {{van Paradijs}}, \citenamefont {{Kouveliotou}},
  \citenamefont {{Connaughton}}, \citenamefont {{Wijers}}, \citenamefont
  {{Briggs}}, \citenamefont {{Preece}},\ and\ \citenamefont
  {{Fishman}}}]{1999ApJ...524L..47G}%
  \BibitemOpen
  \bibfield  {author} {\bibinfo {author} {\bibfnamefont {T.~W.}\ \bibnamefont
  {{Giblin}}}, \bibinfo {author} {\bibfnamefont {J.}~\bibnamefont {{van
  Paradijs}}}, \bibinfo {author} {\bibfnamefont {C.}~\bibnamefont
  {{Kouveliotou}}}, \bibinfo {author} {\bibfnamefont {V.}~\bibnamefont
  {{Connaughton}}}, \bibinfo {author} {\bibfnamefont {R.~A.~M.~J.}\
  \bibnamefont {{Wijers}}}, \bibinfo {author} {\bibfnamefont {M.~S.}\
  \bibnamefont {{Briggs}}}, \bibinfo {author} {\bibfnamefont {R.~D.}\
  \bibnamefont {{Preece}}}, \ and\ \bibinfo {author} {\bibfnamefont {G.~J.}\
  \bibnamefont {{Fishman}}},\ }\href {\doibase 10.1086/312285} {\bibfield
  {journal} {\bibinfo  {journal} {\apjl}\ }\textbf {\bibinfo {volume} {524}},\
  \bibinfo {pages} {L47} (\bibinfo {year} {1999})},\ \Eprint
  {http://arxiv.org/abs/astro-ph/9908139} {astro-ph/9908139} \BibitemShut
  {NoStop}%
\bibitem [{\citenamefont {{Kumar}}\ and\ \citenamefont
  {{Panaitescu}}(2000)}]{2000ApJ...541L..51K}%
  \BibitemOpen
  \bibfield  {author} {\bibinfo {author} {\bibfnamefont {P.}~\bibnamefont
  {{Kumar}}}\ and\ \bibinfo {author} {\bibfnamefont {A.}~\bibnamefont
  {{Panaitescu}}},\ }\href {\doibase 10.1086/312905} {\bibfield  {journal}
  {\bibinfo  {journal} {\apjl}\ }\textbf {\bibinfo {volume} {541}},\ \bibinfo
  {pages} {L51} (\bibinfo {year} {2000})},\ \Eprint
  {http://arxiv.org/abs/astro-ph/0006317} {astro-ph/0006317} \BibitemShut
  {NoStop}%
\bibitem [{\citenamefont {{Fraija}}\ \emph
  {et~al.}(2019{\natexlab{a}})\citenamefont {{Fraija}}, \citenamefont {{De
  Colle}}, \citenamefont {{Veres}}, \citenamefont {{Dichiara}}, \citenamefont
  {{Barniol Duran}}, \citenamefont {{Galvan-Gamez}},\ and\ \citenamefont
  {{Pedreira}}}]{2019ApJ...871..123F}%
  \BibitemOpen
  \bibfield  {author} {\bibinfo {author} {\bibfnamefont {N.}~\bibnamefont
  {{Fraija}}}, \bibinfo {author} {\bibfnamefont {F.}~\bibnamefont {{De
  Colle}}}, \bibinfo {author} {\bibfnamefont {P.}~\bibnamefont {{Veres}}},
  \bibinfo {author} {\bibfnamefont {S.}~\bibnamefont {{Dichiara}}}, \bibinfo
  {author} {\bibfnamefont {R.}~\bibnamefont {{Barniol Duran}}}, \bibinfo
  {author} {\bibfnamefont {A.}~\bibnamefont {{Galvan-Gamez}}}, \ and\ \bibinfo
  {author} {\bibfnamefont {A.~C.~C.~d.~E.~S.}\ \bibnamefont {{Pedreira}}},\
  }\href {\doibase 10.3847/1538-4357/aaf564} {\bibfield  {journal} {\bibinfo
  {journal} {\apj}\ }\textbf {\bibinfo {volume} {871}},\ \bibinfo {eid} {123}
  (\bibinfo {year} {2019}{\natexlab{a}})}\BibitemShut {NoStop}%
\bibitem [{\citenamefont {{Troja}}\ \emph {et~al.}(2017)\citenamefont
  {{Troja}}, \citenamefont {{Lipunov}}, \citenamefont {{Mundell}},\ and\
  \citenamefont {et~al.}}]{2017Natur.547..425T}%
  \BibitemOpen
  \bibfield  {author} {\bibinfo {author} {\bibfnamefont {E.}~\bibnamefont
  {{Troja}}}, \bibinfo {author} {\bibfnamefont {V.~M.}\ \bibnamefont
  {{Lipunov}}}, \bibinfo {author} {\bibfnamefont {C.~G.}\ \bibnamefont
  {{Mundell}}}, \ and\ \bibinfo {author} {\bibnamefont {et~al.}},\ }\href@noop
  {} {\bibfield  {journal} {\bibinfo  {journal} {\nat}\ }\textbf {\bibinfo
  {volume} {547}},\ \bibinfo {pages} {425} (\bibinfo {year}
  {2017})}\BibitemShut {NoStop}%
\bibitem [{\citenamefont {{Granot}}(2003)}]{2003ApJ...596L..17G}%
  \BibitemOpen
  \bibfield  {author} {\bibinfo {author} {\bibfnamefont {J.}~\bibnamefont
  {{Granot}}},\ }\href {\doibase 10.1086/379110} {\bibfield  {journal}
  {\bibinfo  {journal} {\apjl}\ }\textbf {\bibinfo {volume} {596}},\ \bibinfo
  {pages} {L17} (\bibinfo {year} {2003})},\ \Eprint
  {http://arxiv.org/abs/astro-ph/0306322} {astro-ph/0306322} \BibitemShut
  {NoStop}%
\bibitem [{\citenamefont {{Fraija}}\ \emph
  {et~al.}(2017{\natexlab{a}})\citenamefont {{Fraija}}, \citenamefont {{Lee}},
  \citenamefont {{Araya}}, \citenamefont {{Veres}}, \citenamefont {{Barniol
  Duran}},\ and\ \citenamefont {{Guiriec}}}]{2017ApJ...848...94F}%
  \BibitemOpen
  \bibfield  {author} {\bibinfo {author} {\bibfnamefont {N.}~\bibnamefont
  {{Fraija}}}, \bibinfo {author} {\bibfnamefont {W.~H.}\ \bibnamefont {{Lee}}},
  \bibinfo {author} {\bibfnamefont {M.}~\bibnamefont {{Araya}}}, \bibinfo
  {author} {\bibfnamefont {P.}~\bibnamefont {{Veres}}}, \bibinfo {author}
  {\bibfnamefont {R.}~\bibnamefont {{Barniol Duran}}}, \ and\ \bibinfo {author}
  {\bibfnamefont {S.}~\bibnamefont {{Guiriec}}},\ }\href {\doibase
  10.3847/1538-4357/aa8d65} {\bibfield  {journal} {\bibinfo  {journal} {\apj}\
  }\textbf {\bibinfo {volume} {848}},\ \bibinfo {eid} {94} (\bibinfo {year}
  {2017}{\natexlab{a}})},\ \Eprint {http://arxiv.org/abs/1709.06263}
  {arXiv:1709.06263 [astro-ph.HE]} \BibitemShut {NoStop}%
\bibitem [{\citenamefont {{Kobayashi}}\ and\ \citenamefont
  {{Zhang}}(2007)}]{2007ApJ...655..973K}%
  \BibitemOpen
  \bibfield  {author} {\bibinfo {author} {\bibfnamefont {S.}~\bibnamefont
  {{Kobayashi}}}\ and\ \bibinfo {author} {\bibfnamefont {B.}~\bibnamefont
  {{Zhang}}},\ }\href {\doibase 10.1086/510203} {\bibfield  {journal} {\bibinfo
   {journal} {\apj}\ }\textbf {\bibinfo {volume} {655}},\ \bibinfo {pages}
  {973} (\bibinfo {year} {2007})},\ \Eprint
  {http://arxiv.org/abs/arXiv:astro-ph/0608132} {arXiv:astro-ph/0608132}
  \BibitemShut {NoStop}%
\bibitem [{\citenamefont {{Fraija}}\ \emph
  {et~al.}(2016{\natexlab{a}})\citenamefont {{Fraija}}, \citenamefont {{Lee}},\
  and\ \citenamefont {{Veres}}}]{2016ApJ...818..190F}%
  \BibitemOpen
  \bibfield  {author} {\bibinfo {author} {\bibfnamefont {N.}~\bibnamefont
  {{Fraija}}}, \bibinfo {author} {\bibfnamefont {W.}~\bibnamefont {{Lee}}}, \
  and\ \bibinfo {author} {\bibfnamefont {P.}~\bibnamefont {{Veres}}},\ }\href
  {\doibase 10.3847/0004-637X/818/2/190} {\bibfield  {journal} {\bibinfo
  {journal} {\apj}\ }\textbf {\bibinfo {volume} {818}},\ \bibinfo {eid} {190}
  (\bibinfo {year} {2016}{\natexlab{a}})},\ \Eprint
  {http://arxiv.org/abs/1601.01264} {arXiv:1601.01264 [astro-ph.HE]}
  \BibitemShut {NoStop}%
\bibitem [{\citenamefont {{Fraija}}\ and\ \citenamefont
  {{Veres}}(2018)}]{2018ApJ...859...70F}%
  \BibitemOpen
  \bibfield  {author} {\bibinfo {author} {\bibfnamefont {N.}~\bibnamefont
  {{Fraija}}}\ and\ \bibinfo {author} {\bibfnamefont {P.}~\bibnamefont
  {{Veres}}},\ }\href {\doibase 10.3847/1538-4357/aabd79} {\bibfield  {journal}
  {\bibinfo  {journal} {\apj}\ }\textbf {\bibinfo {volume} {859}},\ \bibinfo
  {eid} {70} (\bibinfo {year} {2018})},\ \Eprint
  {http://arxiv.org/abs/1804.02449} {arXiv:1804.02449 [astro-ph.HE]}
  \BibitemShut {NoStop}%
\bibitem [{\citenamefont {{Becerra}}\ \emph
  {et~al.}(2019{\natexlab{a}})\citenamefont {{Becerra}}, \citenamefont
  {{Dichiara}}, \citenamefont {{Watson}}, \citenamefont {{Troja}},
  \citenamefont {{Fraija}}, \citenamefont {{Klotz}}, \citenamefont {{Butler}},
  \citenamefont {{Lee}}, \citenamefont {{Veres}},\ and\ \citenamefont
  {{Bloom}}}]{2019arXiv190405987B}%
  \BibitemOpen
  \bibfield  {author} {\bibinfo {author} {\bibfnamefont {R.~L.}\ \bibnamefont
  {{Becerra}}}, \bibinfo {author} {\bibfnamefont {S.}~\bibnamefont
  {{Dichiara}}}, \bibinfo {author} {\bibfnamefont {A.~M.}\ \bibnamefont
  {{Watson}}}, \bibinfo {author} {\bibfnamefont {E.}~\bibnamefont {{Troja}}},
  \bibinfo {author} {\bibfnamefont {N.~I.}\ \bibnamefont {{Fraija}}}, \bibinfo
  {author} {\bibfnamefont {A.}~\bibnamefont {{Klotz}}}, \bibinfo {author}
  {\bibfnamefont {N.~R.}\ \bibnamefont {{Butler}}}, \bibinfo {author}
  {\bibfnamefont {W.~H.}\ \bibnamefont {{Lee}}}, \bibinfo {author}
  {\bibfnamefont {P.}~\bibnamefont {{Veres}}}, \ and\ \bibinfo {author}
  {\bibfnamefont {J.~S.}\ \bibnamefont {{Bloom}}},\ }\href@noop {} {\bibfield
  {journal} {\bibinfo  {journal} {arXiv e-prints}\ ,\ \bibinfo {eid}
  {arXiv:1904.05987}} (\bibinfo {year} {2019}{\natexlab{a}})},\ \Eprint
  {http://arxiv.org/abs/1904.05987} {arXiv:1904.05987 [astro-ph.HE]}
  \BibitemShut {NoStop}%
\bibitem [{\citenamefont {{Kumar}}\ and\ \citenamefont {{Barniol
  Duran}}(2009)}]{2009MNRAS.400L..75K}%
  \BibitemOpen
  \bibfield  {author} {\bibinfo {author} {\bibfnamefont {P.}~\bibnamefont
  {{Kumar}}}\ and\ \bibinfo {author} {\bibfnamefont {R.}~\bibnamefont {{Barniol
  Duran}}},\ }\href {\doibase 10.1111/j.1745-3933.2009.00766.x} {\bibfield
  {journal} {\bibinfo  {journal} {\mnras}\ }\textbf {\bibinfo {volume} {400}},\
  \bibinfo {pages} {L75} (\bibinfo {year} {2009})},\ \Eprint
  {http://arxiv.org/abs/0905.2417} {arXiv:0905.2417 [astro-ph.HE]} \BibitemShut
  {NoStop}%
\bibitem [{\citenamefont {{Kumar}}\ and\ \citenamefont {{Barniol
  Duran}}(2010)}]{2010MNRAS.409..226K}%
  \BibitemOpen
  \bibfield  {author} {\bibinfo {author} {\bibfnamefont {P.}~\bibnamefont
  {{Kumar}}}\ and\ \bibinfo {author} {\bibfnamefont {R.}~\bibnamefont {{Barniol
  Duran}}},\ }\href {\doibase 10.1111/j.1365-2966.2010.17274.x} {\bibfield
  {journal} {\bibinfo  {journal} {\mnras}\ }\textbf {\bibinfo {volume} {409}},\
  \bibinfo {pages} {226} (\bibinfo {year} {2010})},\ \Eprint
  {http://arxiv.org/abs/0910.5726} {arXiv:0910.5726 [astro-ph.HE]} \BibitemShut
  {NoStop}%
\bibitem [{\citenamefont {{Ghisellini}}\ \emph {et~al.}(2010)\citenamefont
  {{Ghisellini}}, \citenamefont {{Ghirlanda}}, \citenamefont {{Nava}},\ and\
  \citenamefont {{Celotti}}}]{2010MNRAS.403..926G}%
  \BibitemOpen
  \bibfield  {author} {\bibinfo {author} {\bibfnamefont {G.}~\bibnamefont
  {{Ghisellini}}}, \bibinfo {author} {\bibfnamefont {G.}~\bibnamefont
  {{Ghirlanda}}}, \bibinfo {author} {\bibfnamefont {L.}~\bibnamefont {{Nava}}},
  \ and\ \bibinfo {author} {\bibfnamefont {A.}~\bibnamefont {{Celotti}}},\
  }\href {\doibase 10.1111/j.1365-2966.2009.16171.x} {\bibfield  {journal}
  {\bibinfo  {journal} {\mnras}\ }\textbf {\bibinfo {volume} {403}},\ \bibinfo
  {pages} {926} (\bibinfo {year} {2010})},\ \Eprint
  {http://arxiv.org/abs/0910.2459} {arXiv:0910.2459 [astro-ph.HE]} \BibitemShut
  {NoStop}%
\bibitem [{\citenamefont {{Nava}}\ \emph {et~al.}(2014)\citenamefont {{Nava}},
  \citenamefont {{Vianello}}, \citenamefont {{Omodei}}, \citenamefont
  {{Ghisellini}}, \citenamefont {{Ghirlanda}}, \citenamefont {{Celotti}},
  \citenamefont {{Longo}}, \citenamefont {{Desiante}},\ and\ \citenamefont
  {{Barniol Duran}}}]{2014MNRAS.443.3578N}%
  \BibitemOpen
  \bibfield  {author} {\bibinfo {author} {\bibfnamefont {L.}~\bibnamefont
  {{Nava}}}, \bibinfo {author} {\bibfnamefont {G.}~\bibnamefont {{Vianello}}},
  \bibinfo {author} {\bibfnamefont {N.}~\bibnamefont {{Omodei}}}, \bibinfo
  {author} {\bibfnamefont {G.}~\bibnamefont {{Ghisellini}}}, \bibinfo {author}
  {\bibfnamefont {G.}~\bibnamefont {{Ghirlanda}}}, \bibinfo {author}
  {\bibfnamefont {A.}~\bibnamefont {{Celotti}}}, \bibinfo {author}
  {\bibfnamefont {F.}~\bibnamefont {{Longo}}}, \bibinfo {author} {\bibfnamefont
  {R.}~\bibnamefont {{Desiante}}}, \ and\ \bibinfo {author} {\bibfnamefont
  {R.}~\bibnamefont {{Barniol Duran}}},\ }\href {\doibase
  10.1093/mnras/stu1451} {\bibfield  {journal} {\bibinfo  {journal} {\mnras}\
  }\textbf {\bibinfo {volume} {443}},\ \bibinfo {pages} {3578} (\bibinfo {year}
  {2014})},\ \Eprint {http://arxiv.org/abs/1406.6693} {arXiv:1406.6693
  [astro-ph.HE]} \BibitemShut {NoStop}%
\bibitem [{\citenamefont {{Zou}}\ \emph {et~al.}(2009)\citenamefont {{Zou}},
  \citenamefont {{Fan}},\ and\ \citenamefont {{Piran}}}]{2009MNRAS.396.1163Z}%
  \BibitemOpen
  \bibfield  {author} {\bibinfo {author} {\bibfnamefont {Y.-C.}\ \bibnamefont
  {{Zou}}}, \bibinfo {author} {\bibfnamefont {Y.-Z.}\ \bibnamefont {{Fan}}}, \
  and\ \bibinfo {author} {\bibfnamefont {T.}~\bibnamefont {{Piran}}},\ }\href
  {\doibase 10.1111/j.1365-2966.2009.14779.x} {\bibfield  {journal} {\bibinfo
  {journal} {\mnras}\ }\textbf {\bibinfo {volume} {396}},\ \bibinfo {pages}
  {1163} (\bibinfo {year} {2009})},\ \Eprint {http://arxiv.org/abs/0811.2997}
  {arXiv:0811.2997} \BibitemShut {NoStop}%
\bibitem [{\citenamefont {{Becerra}}\ \emph {et~al.}(2017)\citenamefont
  {{Becerra}}, \citenamefont {{Watson}}, \citenamefont {{Lee}}, \citenamefont
  {{Fraija}}, \citenamefont {{Butler}}, \citenamefont {{Bloom}}, \citenamefont
  {{Capone}}, \citenamefont {{Cucchiara}}, \citenamefont {{de Diego}},
  \citenamefont {{Fox}}, \citenamefont {{Gehrels}}, \citenamefont {{Georgiev}},
  \citenamefont {{Gonz{\'a}lez}}, \citenamefont {{Kutyrev}}, \citenamefont
  {{Littlejohns}}, \citenamefont {{Prochaska}}, \citenamefont {{Ramirez-Ruiz}},
  \citenamefont {{Richer}}, \citenamefont {{Rom{\'a}n-Z{\'u}{\~n}iga}},
  \citenamefont {{Toy}},\ and\ \citenamefont {{Troja}}}]{2017ApJ...837..116B}%
  \BibitemOpen
  \bibfield  {author} {\bibinfo {author} {\bibfnamefont {R.~L.}\ \bibnamefont
  {{Becerra}}}, \bibinfo {author} {\bibfnamefont {A.~M.}\ \bibnamefont
  {{Watson}}}, \bibinfo {author} {\bibfnamefont {W.~H.}\ \bibnamefont {{Lee}}},
  \bibinfo {author} {\bibfnamefont {N.}~\bibnamefont {{Fraija}}}, \bibinfo
  {author} {\bibfnamefont {N.~R.}\ \bibnamefont {{Butler}}}, \bibinfo {author}
  {\bibfnamefont {J.~S.}\ \bibnamefont {{Bloom}}}, \bibinfo {author}
  {\bibfnamefont {J.~I.}\ \bibnamefont {{Capone}}}, \bibinfo {author}
  {\bibfnamefont {A.}~\bibnamefont {{Cucchiara}}}, \bibinfo {author}
  {\bibfnamefont {J.~A.}\ \bibnamefont {{de Diego}}}, \bibinfo {author}
  {\bibfnamefont {O.~D.}\ \bibnamefont {{Fox}}}, \bibinfo {author}
  {\bibfnamefont {N.}~\bibnamefont {{Gehrels}}}, \bibinfo {author}
  {\bibfnamefont {L.~N.}\ \bibnamefont {{Georgiev}}}, \bibinfo {author}
  {\bibfnamefont {J.~J.}\ \bibnamefont {{Gonz{\'a}lez}}}, \bibinfo {author}
  {\bibfnamefont {A.~S.}\ \bibnamefont {{Kutyrev}}}, \bibinfo {author}
  {\bibfnamefont {O.~M.}\ \bibnamefont {{Littlejohns}}}, \bibinfo {author}
  {\bibfnamefont {J.~X.}\ \bibnamefont {{Prochaska}}}, \bibinfo {author}
  {\bibfnamefont {E.}~\bibnamefont {{Ramirez-Ruiz}}}, \bibinfo {author}
  {\bibfnamefont {M.~G.}\ \bibnamefont {{Richer}}}, \bibinfo {author}
  {\bibfnamefont {C.~G.}\ \bibnamefont {{Rom{\'a}n-Z{\'u}{\~n}iga}}}, \bibinfo
  {author} {\bibfnamefont {V.~L.}\ \bibnamefont {{Toy}}}, \ and\ \bibinfo
  {author} {\bibfnamefont {E.}~\bibnamefont {{Troja}}},\ }\href {\doibase
  10.3847/1538-4357/aa610f} {\bibfield  {journal} {\bibinfo  {journal} {\apj}\
  }\textbf {\bibinfo {volume} {837}},\ \bibinfo {eid} {116} (\bibinfo {year}
  {2017})},\ \Eprint {http://arxiv.org/abs/1702.04762} {arXiv:1702.04762
  [astro-ph.HE]} \BibitemShut {NoStop}%
\bibitem [{\citenamefont {{Fraija}}\ \emph
  {et~al.}(2019{\natexlab{b}})\citenamefont {{Fraija}}, \citenamefont
  {{Dichiara}}, \citenamefont {{Pedreira}}, \citenamefont {{Galvan-Gamez}},
  \citenamefont {{Becerra}}, \citenamefont {{Montalvo}}, \citenamefont
  {{Montero}}, \citenamefont {{Betancourt Kamenetskaia}},\ and\ \citenamefont
  {{Zhang}}}]{2019arXiv190513572F}%
  \BibitemOpen
  \bibfield  {author} {\bibinfo {author} {\bibfnamefont {N.}~\bibnamefont
  {{Fraija}}}, \bibinfo {author} {\bibfnamefont {S.}~\bibnamefont
  {{Dichiara}}}, \bibinfo {author} {\bibfnamefont {A.~C. C. d. E.~S.}\
  \bibnamefont {{Pedreira}}}, \bibinfo {author} {\bibfnamefont
  {A.}~\bibnamefont {{Galvan-Gamez}}}, \bibinfo {author} {\bibfnamefont
  {R.~L.}\ \bibnamefont {{Becerra}}}, \bibinfo {author} {\bibfnamefont
  {A.}~\bibnamefont {{Montalvo}}}, \bibinfo {author} {\bibfnamefont
  {J.}~\bibnamefont {{Montero}}}, \bibinfo {author} {\bibfnamefont
  {B.}~\bibnamefont {{Betancourt Kamenetskaia}}}, \ and\ \bibinfo {author}
  {\bibfnamefont {B.~B.}\ \bibnamefont {{Zhang}}},\ }\href@noop {} {\bibfield
  {journal} {\bibinfo  {journal} {arXiv e-prints}\ ,\ \bibinfo {eid}
  {arXiv:1905.13572}} (\bibinfo {year} {2019}{\natexlab{b}})},\ \Eprint
  {http://arxiv.org/abs/1905.13572} {arXiv:1905.13572 [astro-ph.HE]}
  \BibitemShut {NoStop}%
\bibitem [{\citenamefont {{Fraija}}(2015{\natexlab{a}})}]{2015ApJ...804..105F}%
  \BibitemOpen
  \bibfield  {author} {\bibinfo {author} {\bibfnamefont {N.}~\bibnamefont
  {{Fraija}}},\ }\href {\doibase 10.1088/0004-637X/804/2/105} {\bibfield
  {journal} {\bibinfo  {journal} {\apj}\ }\textbf {\bibinfo {volume} {804}},\
  \bibinfo {eid} {105} (\bibinfo {year} {2015}{\natexlab{a}})},\ \Eprint
  {http://arxiv.org/abs/1503.07449} {arXiv:1503.07449 [astro-ph.HE]}
  \BibitemShut {NoStop}%
\bibitem [{\citenamefont {{Piran}}\ and\ \citenamefont
  {{Nakar}}(2010)}]{2010ApJ...718L..63P}%
  \BibitemOpen
  \bibfield  {author} {\bibinfo {author} {\bibfnamefont {T.}~\bibnamefont
  {{Piran}}}\ and\ \bibinfo {author} {\bibfnamefont {E.}~\bibnamefont
  {{Nakar}}},\ }\href {\doibase 10.1088/2041-8205/718/2/L63} {\bibfield
  {journal} {\bibinfo  {journal} {\apjl}\ }\textbf {\bibinfo {volume} {718}},\
  \bibinfo {pages} {L63} (\bibinfo {year} {2010})},\ \Eprint
  {http://arxiv.org/abs/1003.5919} {arXiv:1003.5919 [astro-ph.HE]} \BibitemShut
  {NoStop}%
\bibitem [{\citenamefont {{Abdo}}\ \emph
  {et~al.}(2009{\natexlab{a}})\citenamefont {{Abdo}}, \citenamefont
  {{Ackermann}}, \citenamefont {{Ajello}}, \citenamefont {{Asano}},
  \citenamefont {{Atwood}}, \citenamefont {{Axelsson}}, \citenamefont
  {{Baldini}}, \citenamefont {{Ballet}}, \citenamefont {{Barbiellini}},
  \citenamefont {{Baring}}, \citenamefont {{Bastieri}}, \citenamefont
  {{Bechtol}}, \citenamefont {{Bellazzini}},\ and\ \citenamefont
  {et~al}}]{2009ApJ...706L.138A}%
  \BibitemOpen
  \bibfield  {author} {\bibinfo {author} {\bibfnamefont {A.~A.}\ \bibnamefont
  {{Abdo}}}, \bibinfo {author} {\bibfnamefont {M.}~\bibnamefont {{Ackermann}}},
  \bibinfo {author} {\bibfnamefont {M.}~\bibnamefont {{Ajello}}}, \bibinfo
  {author} {\bibfnamefont {K.}~\bibnamefont {{Asano}}}, \bibinfo {author}
  {\bibfnamefont {W.~B.}\ \bibnamefont {{Atwood}}}, \bibinfo {author}
  {\bibfnamefont {M.}~\bibnamefont {{Axelsson}}}, \bibinfo {author}
  {\bibfnamefont {L.}~\bibnamefont {{Baldini}}}, \bibinfo {author}
  {\bibfnamefont {J.}~\bibnamefont {{Ballet}}}, \bibinfo {author}
  {\bibfnamefont {G.}~\bibnamefont {{Barbiellini}}}, \bibinfo {author}
  {\bibfnamefont {M.~G.}\ \bibnamefont {{Baring}}}, \bibinfo {author}
  {\bibfnamefont {D.}~\bibnamefont {{Bastieri}}}, \bibinfo {author}
  {\bibfnamefont {K.}~\bibnamefont {{Bechtol}}}, \bibinfo {author}
  {\bibfnamefont {R.}~\bibnamefont {{Bellazzini}}}, \ and\ \bibinfo {author}
  {\bibnamefont {et~al}},\ }\href {\doibase 10.1088/0004-637X/706/1/L138}
  {\bibfield  {journal} {\bibinfo  {journal} {\apjl}\ }\textbf {\bibinfo
  {volume} {706}},\ \bibinfo {pages} {L138} (\bibinfo {year}
  {2009}{\natexlab{a}})},\ \Eprint {http://arxiv.org/abs/0909.2470}
  {arXiv:0909.2470 [astro-ph.HE]} \BibitemShut {NoStop}%
\bibitem [{\citenamefont {{Barniol Duran}}\ and\ \citenamefont
  {{Kumar}}(2011)}]{2011MNRAS.412..522B}%
  \BibitemOpen
  \bibfield  {author} {\bibinfo {author} {\bibfnamefont {R.}~\bibnamefont
  {{Barniol Duran}}}\ and\ \bibinfo {author} {\bibfnamefont {P.}~\bibnamefont
  {{Kumar}}},\ }\href {\doibase 10.1111/j.1365-2966.2010.17927.x} {\bibfield
  {journal} {\bibinfo  {journal} {\mnras}\ }\textbf {\bibinfo {volume} {412}},\
  \bibinfo {pages} {522} (\bibinfo {year} {2011})},\ \Eprint
  {http://arxiv.org/abs/1003.5916} {arXiv:1003.5916 [astro-ph.HE]} \BibitemShut
  {NoStop}%
\bibitem [{\citenamefont {{Gropp}}(2019)}]{2019GCN.23688....1G}%
  \BibitemOpen
  \bibfield  {author} {\bibinfo {author} {\bibfnamefont {J.~D. e.~a.}\
  \bibnamefont {{Gropp}}},\ }\href@noop {} {\bibfield  {journal} {\bibinfo
  {journal} {GRB Coordinates Network, Circular Service, No.~23688}\ }\textbf
  {\bibinfo {volume} {23688}} (\bibinfo {year} {2019})}\BibitemShut {NoStop}%
\bibitem [{\citenamefont {{Kocevski}}(2019)}]{2019GCN.23709....1D}%
  \BibitemOpen
  \bibfield  {author} {\bibinfo {author} {\bibfnamefont {D.~e.~a.}\
  \bibnamefont {{Kocevski}}},\ }\href@noop {} {\bibfield  {journal} {\bibinfo
  {journal} {GRB Coordinates Network, Circular Service, No.~23709}\ }\textbf
  {\bibinfo {volume} {23709}} (\bibinfo {year} {2019})}\BibitemShut {NoStop}%
\bibitem [{\citenamefont {{Osborne}}(2019)}]{2019GCN.23704....1O}%
  \BibitemOpen
  \bibfield  {author} {\bibinfo {author} {\bibfnamefont {J.~P. e.~a.}\
  \bibnamefont {{Osborne}}},\ }\href@noop {} {\bibfield  {journal} {\bibinfo
  {journal} {GRB Coordinates Network, Circular Service, No.~23704}\ }\textbf
  {\bibinfo {volume} {23704}} (\bibinfo {year} {2019})}\BibitemShut {NoStop}%
\bibitem [{\citenamefont {{Siegel}}(2019)}]{2019GCN.23725....1S}%
  \BibitemOpen
  \bibfield  {author} {\bibinfo {author} {\bibfnamefont {M.~H. e.~a.}\
  \bibnamefont {{Siegel}}},\ }\href@noop {} {\bibfield  {journal} {\bibinfo
  {journal} {GRB Coordinates Network, Circular Service, No.~23725}\ }\textbf
  {\bibinfo {volume} {23725}} (\bibinfo {year} {2019})}\BibitemShut {NoStop}%
\bibitem [{\citenamefont {{Minaev}}\ and\ \citenamefont
  {{Pozanenko}}(2019)}]{2019GCN.23714....1M}%
  \BibitemOpen
  \bibfield  {author} {\bibinfo {author} {\bibfnamefont {P.}~\bibnamefont
  {{Minaev}}}\ and\ \bibinfo {author} {\bibfnamefont {A.}~\bibnamefont
  {{Pozanenko}}},\ }\href@noop {} {\bibfield  {journal} {\bibinfo  {journal}
  {GRB Coordinates Network, Circular Service, No.~23714, \#1 (2019)}\ }\textbf
  {\bibinfo {volume} {23714}} (\bibinfo {year} {2019})}\BibitemShut {NoStop}%
\bibitem [{\citenamefont {{Ursi}}\ \emph {et~al.}(2019)\citenamefont {{Ursi}},
  \citenamefont {{Tavani}}, \citenamefont {{Marisaldi}}, \citenamefont
  {{Parmiggiani}}, \citenamefont {{Longo}}, \citenamefont {{Argan}},
  \citenamefont {{Cardillo}}, \citenamefont {{Casentini}}, \citenamefont
  {{Evangelista}}, \citenamefont {{Piano}}, \citenamefont {{Lucarelli}},
  \citenamefont {{Pittori}}, \citenamefont {{Verrecchia}}, \citenamefont
  {{Bulgarelli}}, \citenamefont {{Fioretti}}, \citenamefont {{Fuschino}},
  \citenamefont {{Pilia}}, \citenamefont {{Trois}}, \citenamefont
  {{Donnarumma}},\ and\ \citenamefont {{Giuliani}}}]{2019GCN.23712....1U}%
  \BibitemOpen
  \bibfield  {author} {\bibinfo {author} {\bibfnamefont {A.}~\bibnamefont
  {{Ursi}}}, \bibinfo {author} {\bibfnamefont {M.}~\bibnamefont {{Tavani}}},
  \bibinfo {author} {\bibfnamefont {M.}~\bibnamefont {{Marisaldi}}}, \bibinfo
  {author} {\bibfnamefont {N.}~\bibnamefont {{Parmiggiani}}}, \bibinfo {author}
  {\bibfnamefont {F.}~\bibnamefont {{Longo}}}, \bibinfo {author} {\bibfnamefont
  {A.}~\bibnamefont {{Argan}}}, \bibinfo {author} {\bibfnamefont
  {M.}~\bibnamefont {{Cardillo}}}, \bibinfo {author} {\bibfnamefont
  {C.}~\bibnamefont {{Casentini}}}, \bibinfo {author} {\bibfnamefont
  {Y.}~\bibnamefont {{Evangelista}}}, \bibinfo {author} {\bibfnamefont
  {G.}~\bibnamefont {{Piano}}}, \bibinfo {author} {\bibfnamefont
  {F.}~\bibnamefont {{Lucarelli}}}, \bibinfo {author} {\bibfnamefont
  {C.}~\bibnamefont {{Pittori}}}, \bibinfo {author} {\bibfnamefont
  {F.}~\bibnamefont {{Verrecchia}}}, \bibinfo {author} {\bibfnamefont
  {A.}~\bibnamefont {{Bulgarelli}}}, \bibinfo {author} {\bibfnamefont
  {V.}~\bibnamefont {{Fioretti}}}, \bibinfo {author} {\bibfnamefont
  {F.}~\bibnamefont {{Fuschino}}}, \bibinfo {author} {\bibfnamefont
  {M.}~\bibnamefont {{Pilia}}}, \bibinfo {author} {\bibfnamefont
  {A.}~\bibnamefont {{Trois}}}, \bibinfo {author} {\bibfnamefont
  {I.}~\bibnamefont {{Donnarumma}}}, \ and\ \bibinfo {author} {\bibfnamefont
  {A.}~\bibnamefont {{Giuliani}}},\ }\href@noop {} {\bibfield  {journal}
  {\bibinfo  {journal} {GRB Coordinates Network, Circular Service, No.~23712,
  \#1 (2019)}\ }\textbf {\bibinfo {volume} {23712}} (\bibinfo {year}
  {2019})}\BibitemShut {NoStop}%
\bibitem [{\citenamefont {{Xiao}}\ \emph {et~al.}(2019)\citenamefont {{Xiao}},
  \citenamefont {{Li}}, \citenamefont {{Li}}, \citenamefont {{Li}},
  \citenamefont {{Liao}}, \citenamefont {{Xiong}}, \citenamefont {{Liu}},
  \citenamefont {{Li}}, \citenamefont {{Li}}, \citenamefont {{Chang}},
  \citenamefont {{Lu}}, \citenamefont {{Zhao}}, \citenamefont {{Zhang}},
  \citenamefont {{Zhang}}, \citenamefont {{Zou}}, \citenamefont {{Jin}},
  \citenamefont {{Zhang}}, \citenamefont {{Li}}, \citenamefont {{Lu}},
  \citenamefont {{Song}}, \citenamefont {{Wu}}, \citenamefont {{Xu}},\ and\
  \citenamefont {{Zhang}}}]{2019GCN.23716....1X}%
  \BibitemOpen
  \bibfield  {author} {\bibinfo {author} {\bibfnamefont {S.}~\bibnamefont
  {{Xiao}}}, \bibinfo {author} {\bibfnamefont {C.~K.}\ \bibnamefont {{Li}}},
  \bibinfo {author} {\bibfnamefont {X.~B.}\ \bibnamefont {{Li}}}, \bibinfo
  {author} {\bibfnamefont {G.}~\bibnamefont {{Li}}}, \bibinfo {author}
  {\bibfnamefont {J.~Y.}\ \bibnamefont {{Liao}}}, \bibinfo {author}
  {\bibfnamefont {S.~L.}\ \bibnamefont {{Xiong}}}, \bibinfo {author}
  {\bibfnamefont {C.~Z.}\ \bibnamefont {{Liu}}}, \bibinfo {author}
  {\bibfnamefont {X.~F.}\ \bibnamefont {{Li}}}, \bibinfo {author}
  {\bibfnamefont {Z.~W.}\ \bibnamefont {{Li}}}, \bibinfo {author}
  {\bibfnamefont {Z.}~\bibnamefont {{Chang}}}, \bibinfo {author} {\bibfnamefont
  {X.~F.}\ \bibnamefont {{Lu}}}, \bibinfo {author} {\bibfnamefont {J.~L.}\
  \bibnamefont {{Zhao}}}, \bibinfo {author} {\bibfnamefont {A.~M.}\
  \bibnamefont {{Zhang}}}, \bibinfo {author} {\bibfnamefont {Y.~F.}\
  \bibnamefont {{Zhang}}}, \bibinfo {author} {\bibfnamefont {C.~L.}\
  \bibnamefont {{Zou}}}, \bibinfo {author} {\bibfnamefont {Y.~J.}\ \bibnamefont
  {{Jin}}}, \bibinfo {author} {\bibfnamefont {Z.}~\bibnamefont {{Zhang}}},
  \bibinfo {author} {\bibfnamefont {T.~P.}\ \bibnamefont {{Li}}}, \bibinfo
  {author} {\bibfnamefont {F.~J.}\ \bibnamefont {{Lu}}}, \bibinfo {author}
  {\bibfnamefont {L.~M.}\ \bibnamefont {{Song}}}, \bibinfo {author}
  {\bibfnamefont {M.}~\bibnamefont {{Wu}}}, \bibinfo {author} {\bibfnamefont
  {Y.~P.}\ \bibnamefont {{Xu}}}, \ and\ \bibinfo {author} {\bibfnamefont
  {S.~N.}\ \bibnamefont {{Zhang}}},\ }\href@noop {} {\bibfield  {journal}
  {\bibinfo  {journal} {GRB Coordinates Network, Circular Service, No.~23716,
  \#1 (2019)}\ }\textbf {\bibinfo {volume} {23716}} (\bibinfo {year}
  {2019})}\BibitemShut {NoStop}%
\bibitem [{\citenamefont {{Frederiks}}\ \emph {et~al.}(2019)\citenamefont
  {{Frederiks}}, \citenamefont {{Golenetskii}}, \citenamefont {{Aptekar}},
  \citenamefont {{Kozlova}}, \citenamefont {{Lysenko}}, \citenamefont
  {{Svinkin}}, \citenamefont {{Tsvetkova}}, \citenamefont {{Ulanov}},\ and\
  \citenamefont {{Cline}}}]{2019GCN.23737....1F}%
  \BibitemOpen
  \bibfield  {author} {\bibinfo {author} {\bibfnamefont {D.}~\bibnamefont
  {{Frederiks}}}, \bibinfo {author} {\bibfnamefont {S.}~\bibnamefont
  {{Golenetskii}}}, \bibinfo {author} {\bibfnamefont {R.}~\bibnamefont
  {{Aptekar}}}, \bibinfo {author} {\bibfnamefont {A.}~\bibnamefont
  {{Kozlova}}}, \bibinfo {author} {\bibfnamefont {A.}~\bibnamefont
  {{Lysenko}}}, \bibinfo {author} {\bibfnamefont {D.}~\bibnamefont
  {{Svinkin}}}, \bibinfo {author} {\bibfnamefont {A.}~\bibnamefont
  {{Tsvetkova}}}, \bibinfo {author} {\bibfnamefont {M.}~\bibnamefont
  {{Ulanov}}}, \ and\ \bibinfo {author} {\bibfnamefont {T.}~\bibnamefont
  {{Cline}}},\ }\href@noop {} {\bibfield  {journal} {\bibinfo  {journal} {GRB
  Coordinates Network, Circular Service, No.~23737, \#1 (2019)}\ }\textbf
  {\bibinfo {volume} {23737}} (\bibinfo {year} {2019})}\BibitemShut {NoStop}%
\bibitem [{\citenamefont {{Laskar}}\ \emph {et~al.}(2019)\citenamefont
  {{Laskar}}, \citenamefont {{Alexander}}, \citenamefont {{Gill}},
  \citenamefont {{Granot}}, \citenamefont {{Berger}}, \citenamefont
  {{Mundell}}, \citenamefont {{Barniol-Duran}}, \citenamefont {{Bolmer}},
  \citenamefont {{Duffell}},\ and\ \citenamefont {{van
  Eerten}}}]{2019arXiv190407261L}%
  \BibitemOpen
  \bibfield  {author} {\bibinfo {author} {\bibfnamefont {T.}~\bibnamefont
  {{Laskar}}}, \bibinfo {author} {\bibfnamefont {K.~D.}\ \bibnamefont
  {{Alexander}}}, \bibinfo {author} {\bibfnamefont {R.}~\bibnamefont {{Gill}}},
  \bibinfo {author} {\bibfnamefont {J.}~\bibnamefont {{Granot}}}, \bibinfo
  {author} {\bibfnamefont {E.}~\bibnamefont {{Berger}}}, \bibinfo {author}
  {\bibfnamefont {C.~G.}\ \bibnamefont {{Mundell}}}, \bibinfo {author}
  {\bibfnamefont {R.}~\bibnamefont {{Barniol-Duran}}}, \bibinfo {author}
  {\bibfnamefont {J.}~\bibnamefont {{Bolmer}}}, \bibinfo {author}
  {\bibfnamefont {P.}~\bibnamefont {{Duffell}}}, \ and\ \bibinfo {author}
  {\bibfnamefont {H.}~\bibnamefont {{van Eerten}}},\ }\href@noop {} {\bibfield
  {journal} {\bibinfo  {journal} {arXiv e-prints}\ ,\ \bibinfo {eid}
  {arXiv:1904.07261}} (\bibinfo {year} {2019})},\ \Eprint
  {http://arxiv.org/abs/1904.07261} {arXiv:1904.07261 [astro-ph.HE]}
  \BibitemShut {NoStop}%
\bibitem [{\citenamefont {{Tyurina}}(2019)}]{2019GCN.23690....1T}%
  \BibitemOpen
  \bibfield  {author} {\bibinfo {author} {\bibfnamefont {N.~e.~a.}\
  \bibnamefont {{Tyurina}}},\ }\href@noop {} {\bibfield  {journal} {\bibinfo
  {journal} {GRB Coordinates Network, Circular Service, No.~23690}\ }\textbf
  {\bibinfo {volume} {23690}} (\bibinfo {year} {2019})}\BibitemShut {NoStop}%
\bibitem [{\citenamefont {{Lipunov }}(2019)}]{2019GCN.23693....1L}%
  \BibitemOpen
  \bibfield  {author} {\bibinfo {author} {\bibfnamefont {V.~e.~a.}\
  \bibnamefont {{Lipunov }}},\ }\href@noop {} {\bibfield  {journal} {\bibinfo
  {journal} {GRB Coordinates Network, Circular Service, No.~23693}\ }\textbf
  {\bibinfo {volume} {23693}} (\bibinfo {year} {2019})}\BibitemShut {NoStop}%
\bibitem [{\citenamefont {{Selsing }}(2019)}]{2019GCN.23695....1S}%
  \BibitemOpen
  \bibfield  {author} {\bibinfo {author} {\bibfnamefont {J.~e.~a.}\
  \bibnamefont {{Selsing }}},\ }\href@noop {} {\bibfield  {journal} {\bibinfo
  {journal} {GRB Coordinates Network, Circular Service, No.~23695}\ }\textbf
  {\bibinfo {volume} {23695}} (\bibinfo {year} {2019})}\BibitemShut {NoStop}%
\bibitem [{\citenamefont {{Izzo}}(2019)}]{2019GCN.23699....1L}%
  \BibitemOpen
  \bibfield  {author} {\bibinfo {author} {\bibfnamefont {L.~e.~a.}\
  \bibnamefont {{Izzo}}},\ }\href@noop {} {\bibfield  {journal} {\bibinfo
  {journal} {GRB Coordinates Network, Circular Service, No.~23699}\ }\textbf
  {\bibinfo {volume} {23699}} (\bibinfo {year} {2019})}\BibitemShut {NoStop}%
\bibitem [{\citenamefont {{Mirzoyan}}(2019)}]{2019GCN.23701....1M}%
  \BibitemOpen
  \bibfield  {author} {\bibinfo {author} {\bibfnamefont {R.~e.~a.}\
  \bibnamefont {{Mirzoyan}}},\ }\href@noop {} {\bibfield  {journal} {\bibinfo
  {journal} {GRB Coordinates Network, Circular Service, No.~23701}\ }\textbf
  {\bibinfo {volume} {23701}} (\bibinfo {year} {2019})}\BibitemShut {NoStop}%
\bibitem [{\citenamefont {{Bolmer}}\ and\ \citenamefont
  {{Shady}}(2019)}]{2019GCN.23702....1B}%
  \BibitemOpen
  \bibfield  {author} {\bibinfo {author} {\bibfnamefont {J.}~\bibnamefont
  {{Bolmer}}}\ and\ \bibinfo {author} {\bibfnamefont {P.}~\bibnamefont
  {{Shady}}},\ }\href@noop {} {\bibfield  {journal} {\bibinfo  {journal} {GRB
  Coordinates Network, Circular Service, No.~23702}\ }\textbf {\bibinfo
  {volume} {23702}} (\bibinfo {year} {2019})}\BibitemShut {NoStop}%
\bibitem [{\citenamefont {{Im}}(2019{\natexlab{a}})}]{2019GCN.23717....1I}%
  \BibitemOpen
  \bibfield  {author} {\bibinfo {author} {\bibfnamefont {M.~e.~a.}\
  \bibnamefont {{Im}}},\ }\href@noop {} {\bibfield  {journal} {\bibinfo
  {journal} {GRB Coordinates Network, Circular Service, No.~23717}\ }\textbf
  {\bibinfo {volume} {23717}} (\bibinfo {year}
  {2019}{\natexlab{a}})}\BibitemShut {NoStop}%
\bibitem [{\citenamefont {{Alexander}}(2019)}]{2019GCN.23726....1K}%
  \BibitemOpen
  \bibfield  {author} {\bibinfo {author} {\bibfnamefont {K.~D. e.~a.}\
  \bibnamefont {{Alexander}}},\ }\href@noop {} {\bibfield  {journal} {\bibinfo
  {journal} {GRB Coordinates Network, Circular Service, No.~23726}\ }\textbf
  {\bibinfo {volume} {23726}} (\bibinfo {year} {2019})}\BibitemShut {NoStop}%
\bibitem [{\citenamefont {{D'Avanzo}}(2019)}]{2019GCN.23729....1D}%
  \BibitemOpen
  \bibfield  {author} {\bibinfo {author} {\bibfnamefont {P.~e.~a.}\
  \bibnamefont {{D'Avanzo}}},\ }\href@noop {} {\bibfield  {journal} {\bibinfo
  {journal} {GRB Coordinates Network, Circular Service, No.~23729}\ }\textbf
  {\bibinfo {volume} {23729}} (\bibinfo {year} {2019})}\BibitemShut {NoStop}%
\bibitem [{\citenamefont {{Kim}}\ and\ \citenamefont
  {{Im}}(2019{\natexlab{a}})}]{2019GCN.23732....1K}%
  \BibitemOpen
  \bibfield  {author} {\bibinfo {author} {\bibfnamefont {J.}~\bibnamefont
  {{Kim}}}\ and\ \bibinfo {author} {\bibfnamefont {M.}~\bibnamefont {{Im}}},\
  }\href@noop {} {\bibfield  {journal} {\bibinfo  {journal} {GRB Coordinates
  Network, Circular Service, No.~23732}\ }\textbf {\bibinfo {volume} {23732}}
  (\bibinfo {year} {2019}{\natexlab{a}})}\BibitemShut {NoStop}%
\bibitem [{\citenamefont {{Kumar}}(2019)}]{2019GCN.23733....1K}%
  \BibitemOpen
  \bibfield  {author} {\bibinfo {author} {\bibfnamefont {H.~e.~a.}\
  \bibnamefont {{Kumar}}},\ }\href@noop {} {\bibfield  {journal} {\bibinfo
  {journal} {GRB Coordinates Network, Circular Service, No.~23733}\ }\textbf
  {\bibinfo {volume} {23733}} (\bibinfo {year} {2019})}\BibitemShut {NoStop}%
\bibitem [{\citenamefont {{Kim}}\ and\ \citenamefont
  {{Im}}(2019{\natexlab{b}})}]{2019GCN.23734....1K}%
  \BibitemOpen
  \bibfield  {author} {\bibinfo {author} {\bibfnamefont {J.}~\bibnamefont
  {{Kim}}}\ and\ \bibinfo {author} {\bibfnamefont {M.}~\bibnamefont {{Im}}},\
  }\href@noop {} {\bibfield  {journal} {\bibinfo  {journal} {GRB Coordinates
  Network, Circular Service, No.~23734}\ }\textbf {\bibinfo {volume} {23734}}
  (\bibinfo {year} {2019}{\natexlab{b}})}\BibitemShut {NoStop}%
\bibitem [{\citenamefont {{Im}}(2019{\natexlab{b}})}]{2019GCN.23740....1I}%
  \BibitemOpen
  \bibfield  {author} {\bibinfo {author} {\bibfnamefont {M.~e.~a.}\
  \bibnamefont {{Im}}},\ }\href@noop {} {\bibfield  {journal} {\bibinfo
  {journal} {GRB Coordinates Network, Circular Service, No.~23740}\ }\textbf
  {\bibinfo {volume} {23740}} (\bibinfo {year}
  {2019}{\natexlab{b}})}\BibitemShut {NoStop}%
\bibitem [{\citenamefont {{Mazaeva}}(2019)}]{2019GCN.23741....1M}%
  \BibitemOpen
  \bibfield  {author} {\bibinfo {author} {\bibfnamefont {E.~e.~a.}\
  \bibnamefont {{Mazaeva}}},\ }\href@noop {} {\bibfield  {journal} {\bibinfo
  {journal} {GRB Coordinates Network, Circular Service, No.~23741}\ }\textbf
  {\bibinfo {volume} {23741}} (\bibinfo {year} {2019})}\BibitemShut {NoStop}%
\bibitem [{\citenamefont {{Wang}}\ \emph {et~al.}(2019)\citenamefont {{Wang}},
  \citenamefont {{Li}}, \citenamefont {{Moradi}},\ and\ \citenamefont
  {{Ruffini}}}]{2019arXiv190107505W}%
  \BibitemOpen
  \bibfield  {author} {\bibinfo {author} {\bibfnamefont {Y.}~\bibnamefont
  {{Wang}}}, \bibinfo {author} {\bibfnamefont {L.}~\bibnamefont {{Li}}},
  \bibinfo {author} {\bibfnamefont {R.}~\bibnamefont {{Moradi}}}, \ and\
  \bibinfo {author} {\bibfnamefont {R.}~\bibnamefont {{Ruffini}}},\ }\href@noop
  {} {\bibfield  {journal} {\bibinfo  {journal} {arXiv e-prints}\ } (\bibinfo
  {year} {2019})},\ \Eprint {http://arxiv.org/abs/1901.07505} {arXiv:1901.07505
  [astro-ph.HE]} \BibitemShut {NoStop}%
\bibitem [{\citenamefont {{Hamburg}}(2019)}]{2019GCN.23707....1H}%
  \BibitemOpen
  \bibfield  {author} {\bibinfo {author} {\bibfnamefont {R.~e.~a.}\
  \bibnamefont {{Hamburg}}},\ }\href@noop {} {\bibfield  {journal} {\bibinfo
  {journal} {GRB Coordinates Network, Circular Service, No.~23707}\ }\textbf
  {\bibinfo {volume} {23707}} (\bibinfo {year} {2019})}\BibitemShut {NoStop}%
\bibitem [{\citenamefont {{Arnaud}}(1996)}]{1996ASPC..101...17A}%
  \BibitemOpen
  \bibfield  {author} {\bibinfo {author} {\bibfnamefont {K.~A.}\ \bibnamefont
  {{Arnaud}}},\ }in\ \href@noop {} {\emph {\bibinfo {booktitle} {Astronomical
  Data Analysis Software and Systems V}}},\ \bibinfo {series} {Astronomical
  Society of the Pacific Conference Series}, Vol.\ \bibinfo {volume} {101},\
  \bibinfo {editor} {edited by\ \bibinfo {editor} {\bibfnamefont {G.~H.}\
  \bibnamefont {{Jacoby}}}\ and\ \bibinfo {editor} {\bibfnamefont
  {J.}~\bibnamefont {{Barnes}}}}\ (\bibinfo {year} {1996})\ p.~\bibinfo {pages}
  {17}\BibitemShut {NoStop}%
\bibitem [{\citenamefont {{Vestrand}}\ \emph {et~al.}(2006)\citenamefont
  {{Vestrand}}, \citenamefont {{Wren}}, \citenamefont {{Wozniak}},
  \citenamefont {{Aptekar}}, \citenamefont {{Golentskii}}, \citenamefont
  {{Pal'Shin}}, \citenamefont {{Sakamoto}}, \citenamefont {{White}},
  \citenamefont {{Evans}}, \citenamefont {{Casperson}},\ and\ \citenamefont
  {{Fenimore}}}]{2006Natur.442..172V}%
  \BibitemOpen
  \bibfield  {author} {\bibinfo {author} {\bibfnamefont {W.~T.}\ \bibnamefont
  {{Vestrand}}}, \bibinfo {author} {\bibfnamefont {J.~A.}\ \bibnamefont
  {{Wren}}}, \bibinfo {author} {\bibfnamefont {P.~R.}\ \bibnamefont
  {{Wozniak}}}, \bibinfo {author} {\bibfnamefont {R.}~\bibnamefont
  {{Aptekar}}}, \bibinfo {author} {\bibfnamefont {S.}~\bibnamefont
  {{Golentskii}}}, \bibinfo {author} {\bibfnamefont {V.}~\bibnamefont
  {{Pal'Shin}}}, \bibinfo {author} {\bibfnamefont {T.}~\bibnamefont
  {{Sakamoto}}}, \bibinfo {author} {\bibfnamefont {R.~R.}\ \bibnamefont
  {{White}}}, \bibinfo {author} {\bibfnamefont {S.}~\bibnamefont {{Evans}}},
  \bibinfo {author} {\bibfnamefont {D.}~\bibnamefont {{Casperson}}}, \ and\
  \bibinfo {author} {\bibfnamefont {E.}~\bibnamefont {{Fenimore}}},\ }\href
  {\doibase 10.1038/nature04913} {\bibfield  {journal} {\bibinfo  {journal}
  {\nat}\ }\textbf {\bibinfo {volume} {442}},\ \bibinfo {pages} {172} (\bibinfo
  {year} {2006})},\ \Eprint {http://arxiv.org/abs/astro-ph/0605472}
  {astro-ph/0605472} \BibitemShut {NoStop}%
\bibitem [{\citenamefont {{Ravasio}}\ \emph {et~al.}(2019)\citenamefont
  {{Ravasio}}, \citenamefont {{Oganesyan}}, \citenamefont {{Salafia}},
  \citenamefont {{Ghirlanda}}, \citenamefont {{Ghisellini}}, \citenamefont
  {{Branchesi}}, \citenamefont {{Campana}}, \citenamefont {{Covino}},\ and\
  \citenamefont {{Salvaterra}}}]{2019arXiv190201861R}%
  \BibitemOpen
  \bibfield  {author} {\bibinfo {author} {\bibfnamefont {M.~E.}\ \bibnamefont
  {{Ravasio}}}, \bibinfo {author} {\bibfnamefont {G.}~\bibnamefont
  {{Oganesyan}}}, \bibinfo {author} {\bibfnamefont {O.~S.}\ \bibnamefont
  {{Salafia}}}, \bibinfo {author} {\bibfnamefont {G.}~\bibnamefont
  {{Ghirlanda}}}, \bibinfo {author} {\bibfnamefont {G.}~\bibnamefont
  {{Ghisellini}}}, \bibinfo {author} {\bibfnamefont {M.}~\bibnamefont
  {{Branchesi}}}, \bibinfo {author} {\bibfnamefont {S.}~\bibnamefont
  {{Campana}}}, \bibinfo {author} {\bibfnamefont {S.}~\bibnamefont {{Covino}}},
  \ and\ \bibinfo {author} {\bibfnamefont {R.}~\bibnamefont {{Salvaterra}}},\
  }\href@noop {} {\bibfield  {journal} {\bibinfo  {journal} {arXiv e-prints}\ }
  (\bibinfo {year} {2019})},\ \Eprint {http://arxiv.org/abs/1902.01861}
  {arXiv:1902.01861 [astro-ph.HE]} \BibitemShut {NoStop}%
\bibitem [{\citenamefont {{Krimm}}(2019)}]{2019GCN.23724....1K}%
  \BibitemOpen
  \bibfield  {author} {\bibinfo {author} {\bibfnamefont {H.~A. e.~a.}\
  \bibnamefont {{Krimm}}},\ }\href@noop {} {\bibfield  {journal} {\bibinfo
  {journal} {GRB Coordinates Network, Circular Service, No.~23724}\ }\textbf
  {\bibinfo {volume} {23724}} (\bibinfo {year} {2019})}\BibitemShut {NoStop}%
\bibitem [{\citenamefont {{Zhang}}\ \emph {et~al.}(2006)\citenamefont
  {{Zhang}}, \citenamefont {{Fan}}, \citenamefont {{Dyks}}, \citenamefont
  {{Kobayashi}}, \citenamefont {{M{\'e}sz{\'a}ros}}, \citenamefont {{Burrows}},
  \citenamefont {{Nousek}},\ and\ \citenamefont
  {{Gehrels}}}]{2006ApJ...642..354Z}%
  \BibitemOpen
  \bibfield  {author} {\bibinfo {author} {\bibfnamefont {B.}~\bibnamefont
  {{Zhang}}}, \bibinfo {author} {\bibfnamefont {Y.~Z.}\ \bibnamefont {{Fan}}},
  \bibinfo {author} {\bibfnamefont {J.}~\bibnamefont {{Dyks}}}, \bibinfo
  {author} {\bibfnamefont {S.}~\bibnamefont {{Kobayashi}}}, \bibinfo {author}
  {\bibfnamefont {P.}~\bibnamefont {{M{\'e}sz{\'a}ros}}}, \bibinfo {author}
  {\bibfnamefont {D.~N.}\ \bibnamefont {{Burrows}}}, \bibinfo {author}
  {\bibfnamefont {J.~A.}\ \bibnamefont {{Nousek}}}, \ and\ \bibinfo {author}
  {\bibfnamefont {N.}~\bibnamefont {{Gehrels}}},\ }\href {\doibase
  10.1086/500723} {\bibfield  {journal} {\bibinfo  {journal} {\apj}\ }\textbf
  {\bibinfo {volume} {642}},\ \bibinfo {pages} {354} (\bibinfo {year}
  {2006})},\ \Eprint {http://arxiv.org/abs/astro-ph/0508321} {astro-ph/0508321}
  \BibitemShut {NoStop}%
\bibitem [{\citenamefont {{Stratta}}\ \emph {et~al.}(2018)\citenamefont
  {{Stratta}}, \citenamefont {{Dainotti}}, \citenamefont {{Dall'Osso}},
  \citenamefont {{Hernandez}},\ and\ \citenamefont {{De
  Cesare}}}]{2018ApJ...869..155S}%
  \BibitemOpen
  \bibfield  {author} {\bibinfo {author} {\bibfnamefont {G.}~\bibnamefont
  {{Stratta}}}, \bibinfo {author} {\bibfnamefont {M.~G.}\ \bibnamefont
  {{Dainotti}}}, \bibinfo {author} {\bibfnamefont {S.}~\bibnamefont
  {{Dall'Osso}}}, \bibinfo {author} {\bibfnamefont {X.}~\bibnamefont
  {{Hernandez}}}, \ and\ \bibinfo {author} {\bibfnamefont {G.}~\bibnamefont
  {{De Cesare}}},\ }\href {\doibase 10.3847/1538-4357/aadd8f} {\bibfield
  {journal} {\bibinfo  {journal} {\apj}\ }\textbf {\bibinfo {volume} {869}},\
  \bibinfo {eid} {155} (\bibinfo {year} {2018})},\ \Eprint
  {http://arxiv.org/abs/1804.08652} {arXiv:1804.08652 [astro-ph.HE]}
  \BibitemShut {NoStop}%
\bibitem [{\citenamefont {{Vaughan}}\ \emph {et~al.}(2006)\citenamefont
  {{Vaughan}}, \citenamefont {{Goad}}, \citenamefont {{Beardmore}},
  \citenamefont {{O'Brien}}, \citenamefont {{Osborne}}, \citenamefont {{Page}},
  \citenamefont {{Barthelmy}}, \citenamefont {{Burrows}}, \citenamefont
  {{Campana}}, \citenamefont {{Cannizzo}}, \citenamefont {{Capalbi}},
  \citenamefont {{Chincarini}}, \citenamefont {{Cummings}}, \citenamefont
  {{Cusumano}}, \citenamefont {{Giommi}}, \citenamefont {{Godet}},
  \citenamefont {{Hill}}, \citenamefont {{Kobayashi}}, \citenamefont {{Kumar}},
  \citenamefont {{La Parola}}, \citenamefont {{Levan}}, \citenamefont
  {{Mangano}}, \citenamefont {{M{\'e}sz{\'a}ros}}, \citenamefont {{Moretti}},
  \citenamefont {{Morris}}, \citenamefont {{Nousek}}, \citenamefont {{Pagani}},
  \citenamefont {{Palmer}}, \citenamefont {{Racusin}}, \citenamefont
  {{Romano}}, \citenamefont {{Tagliaferri}}, \citenamefont {{Zhang}},\ and\
  \citenamefont {{Gehrels}}}]{2006ApJ...638..920V}%
  \BibitemOpen
  \bibfield  {author} {\bibinfo {author} {\bibfnamefont {S.}~\bibnamefont
  {{Vaughan}}}, \bibinfo {author} {\bibfnamefont {M.~R.}\ \bibnamefont
  {{Goad}}}, \bibinfo {author} {\bibfnamefont {A.~P.}\ \bibnamefont
  {{Beardmore}}}, \bibinfo {author} {\bibfnamefont {P.~T.}\ \bibnamefont
  {{O'Brien}}}, \bibinfo {author} {\bibfnamefont {J.~P.}\ \bibnamefont
  {{Osborne}}}, \bibinfo {author} {\bibfnamefont {K.~L.}\ \bibnamefont
  {{Page}}}, \bibinfo {author} {\bibfnamefont {S.~D.}\ \bibnamefont
  {{Barthelmy}}}, \bibinfo {author} {\bibfnamefont {D.~N.}\ \bibnamefont
  {{Burrows}}}, \bibinfo {author} {\bibfnamefont {S.}~\bibnamefont
  {{Campana}}}, \bibinfo {author} {\bibfnamefont {J.~K.}\ \bibnamefont
  {{Cannizzo}}}, \bibinfo {author} {\bibfnamefont {M.}~\bibnamefont
  {{Capalbi}}}, \bibinfo {author} {\bibfnamefont {G.}~\bibnamefont
  {{Chincarini}}}, \bibinfo {author} {\bibfnamefont {J.~R.}\ \bibnamefont
  {{Cummings}}}, \bibinfo {author} {\bibfnamefont {G.}~\bibnamefont
  {{Cusumano}}}, \bibinfo {author} {\bibfnamefont {P.}~\bibnamefont
  {{Giommi}}}, \bibinfo {author} {\bibfnamefont {O.}~\bibnamefont {{Godet}}},
  \bibinfo {author} {\bibfnamefont {J.~E.}\ \bibnamefont {{Hill}}}, \bibinfo
  {author} {\bibfnamefont {S.}~\bibnamefont {{Kobayashi}}}, \bibinfo {author}
  {\bibfnamefont {P.}~\bibnamefont {{Kumar}}}, \bibinfo {author} {\bibfnamefont
  {V.}~\bibnamefont {{La Parola}}}, \bibinfo {author} {\bibfnamefont
  {A.}~\bibnamefont {{Levan}}}, \bibinfo {author} {\bibfnamefont
  {V.}~\bibnamefont {{Mangano}}}, \bibinfo {author} {\bibfnamefont
  {P.}~\bibnamefont {{M{\'e}sz{\'a}ros}}}, \bibinfo {author} {\bibfnamefont
  {A.}~\bibnamefont {{Moretti}}}, \bibinfo {author} {\bibfnamefont {D.~C.}\
  \bibnamefont {{Morris}}}, \bibinfo {author} {\bibfnamefont {J.~A.}\
  \bibnamefont {{Nousek}}}, \bibinfo {author} {\bibfnamefont {C.}~\bibnamefont
  {{Pagani}}}, \bibinfo {author} {\bibfnamefont {D.~M.}\ \bibnamefont
  {{Palmer}}}, \bibinfo {author} {\bibfnamefont {J.~L.}\ \bibnamefont
  {{Racusin}}}, \bibinfo {author} {\bibfnamefont {P.}~\bibnamefont {{Romano}}},
  \bibinfo {author} {\bibfnamefont {G.}~\bibnamefont {{Tagliaferri}}}, \bibinfo
  {author} {\bibfnamefont {B.}~\bibnamefont {{Zhang}}}, \ and\ \bibinfo
  {author} {\bibfnamefont {N.}~\bibnamefont {{Gehrels}}},\ }\href {\doibase
  10.1086/499069} {\bibfield  {journal} {\bibinfo  {journal} {\apj}\ }\textbf
  {\bibinfo {volume} {638}},\ \bibinfo {pages} {920} (\bibinfo {year}
  {2006})},\ \Eprint {http://arxiv.org/abs/astro-ph/0510677} {astro-ph/0510677}
  \BibitemShut {NoStop}%
\bibitem [{\citenamefont {{Ugarte Postigo}}(2019)}]{2019GCN.23692....1U}%
  \BibitemOpen
  \bibfield  {author} {\bibinfo {author} {\bibfnamefont {A.~e.~a.}\
  \bibnamefont {{Ugarte Postigo}}},\ }\href@noop {} {\bibfield  {journal}
  {\bibinfo  {journal} {GRB Coordinates Network, Circular Service, No.~23692}\
  }\textbf {\bibinfo {volume} {23692}} (\bibinfo {year} {2019})}\BibitemShut
  {NoStop}%
\bibitem [{\citenamefont {{Fukugita}}\ \emph {et~al.}(1996)\citenamefont
  {{Fukugita}}, \citenamefont {{Ichikawa}}, \citenamefont {{Gunn}},
  \citenamefont {{Doi}}, \citenamefont {{Shimasaku}},\ and\ \citenamefont
  {{Schneider}}}]{1996AJ....111.1748F}%
  \BibitemOpen
  \bibfield  {author} {\bibinfo {author} {\bibfnamefont {M.}~\bibnamefont
  {{Fukugita}}}, \bibinfo {author} {\bibfnamefont {T.}~\bibnamefont
  {{Ichikawa}}}, \bibinfo {author} {\bibfnamefont {J.~E.}\ \bibnamefont
  {{Gunn}}}, \bibinfo {author} {\bibfnamefont {M.}~\bibnamefont {{Doi}}},
  \bibinfo {author} {\bibfnamefont {K.}~\bibnamefont {{Shimasaku}}}, \ and\
  \bibinfo {author} {\bibfnamefont {D.~P.}\ \bibnamefont {{Schneider}}},\
  }\href {\doibase 10.1086/117915} {\bibfield  {journal} {\bibinfo  {journal}
  {\aj}\ }\textbf {\bibinfo {volume} {111}},\ \bibinfo {pages} {1748} (\bibinfo
  {year} {1996})}\BibitemShut {NoStop}%
\bibitem [{\citenamefont {{Becerra}}\ \emph
  {et~al.}(2019{\natexlab{b}})\citenamefont {{Becerra}}, \citenamefont
  {{Watson}}, \citenamefont {{Fraija}}, \citenamefont {{Butler}}, \citenamefont
  {{Lee}}, \citenamefont {{Troja}}, \citenamefont {{Rom{\'a}n-Z{\'u}{\~n}iga}},
  \citenamefont {{Kutyrev}}, \citenamefont {{{\'A}lvarez Nu{\~n}ez}},
  \citenamefont {{{\'A}ngeles}}, \citenamefont {{Chapa}}, \citenamefont
  {{Cuevas}}, \citenamefont {{Farah}}, \citenamefont {{Fuentes-Fern{\'a}ndez}},
  \citenamefont {{Figueroa}}, \citenamefont {{Langarica}}, \citenamefont
  {{Quir{\'o}s}}, \citenamefont {{Ru{\'\i}z-D{\'\i}az-Soto}},\ and\
  \citenamefont {{Tinoco}}}]{2019arXiv190106051B}%
  \BibitemOpen
  \bibfield  {author} {\bibinfo {author} {\bibfnamefont {R.~L.}\ \bibnamefont
  {{Becerra}}}, \bibinfo {author} {\bibfnamefont {A.~M.}\ \bibnamefont
  {{Watson}}}, \bibinfo {author} {\bibfnamefont {N.}~\bibnamefont {{Fraija}}},
  \bibinfo {author} {\bibfnamefont {N.~R.}\ \bibnamefont {{Butler}}}, \bibinfo
  {author} {\bibfnamefont {W.~H.}\ \bibnamefont {{Lee}}}, \bibinfo {author}
  {\bibfnamefont {E.}~\bibnamefont {{Troja}}}, \bibinfo {author} {\bibfnamefont
  {C.~G.}\ \bibnamefont {{Rom{\'a}n-Z{\'u}{\~n}iga}}}, \bibinfo {author}
  {\bibfnamefont {A.~S.}\ \bibnamefont {{Kutyrev}}}, \bibinfo {author}
  {\bibfnamefont {L.~C.}\ \bibnamefont {{{\'A}lvarez Nu{\~n}ez}}}, \bibinfo
  {author} {\bibfnamefont {F.}~\bibnamefont {{{\'A}ngeles}}}, \bibinfo {author}
  {\bibfnamefont {O.}~\bibnamefont {{Chapa}}}, \bibinfo {author} {\bibfnamefont
  {S.}~\bibnamefont {{Cuevas}}}, \bibinfo {author} {\bibfnamefont {A.~S.}\
  \bibnamefont {{Farah}}}, \bibinfo {author} {\bibfnamefont {J.}~\bibnamefont
  {{Fuentes-Fern{\'a}ndez}}}, \bibinfo {author} {\bibfnamefont
  {L.}~\bibnamefont {{Figueroa}}}, \bibinfo {author} {\bibfnamefont
  {R.}~\bibnamefont {{Langarica}}}, \bibinfo {author} {\bibfnamefont
  {F.}~\bibnamefont {{Quir{\'o}s}}}, \bibinfo {author} {\bibfnamefont
  {J.}~\bibnamefont {{Ru{\'\i}z-D{\'\i}az-Soto}}}, \ and\ \bibinfo {author}
  {\bibfnamefont {C.~G. T. S.~J.}\ \bibnamefont {{Tinoco}}},\ }\href@noop {}
  {\bibfield  {journal} {\bibinfo  {journal} {arXiv e-prints}\ ,\ \bibinfo
  {eid} {arXiv:1901.06051}} (\bibinfo {year} {2019}{\natexlab{b}})},\ \Eprint
  {http://arxiv.org/abs/1901.06051} {arXiv:1901.06051 [astro-ph.HE]}
  \BibitemShut {NoStop}%
\bibitem [{\citenamefont {{Burenin}}\ \emph {et~al.}(2019)\citenamefont
  {{Burenin}}, \citenamefont {{Bikmaev}}, \citenamefont {{Irtuganov}},\ and\
  \citenamefont {{et al.Malesani}}}]{2019GCN.23766....1R}%
  \BibitemOpen
  \bibfield  {author} {\bibinfo {author} {\bibfnamefont {R.}~\bibnamefont
  {{Burenin}}}, \bibinfo {author} {\bibfnamefont {I.}~\bibnamefont
  {{Bikmaev}}}, \bibinfo {author} {\bibfnamefont {E.}~\bibnamefont
  {{Irtuganov}}}, \ and\ \bibinfo {author} {\bibnamefont {{et al.Malesani}}},\
  }\href@noop {} {\bibfield  {journal} {\bibinfo  {journal} {GRB Coordinates
  Network, Circular Service, No.~23766, \#1 (2019/January-21)}\ }\textbf
  {\bibinfo {volume} {23766}} (\bibinfo {year} {2019})}\BibitemShut {NoStop}%
\bibitem [{\citenamefont {{Melandri}}\ \emph {et~al.}(2019)\citenamefont
  {{Melandri}}, \citenamefont {{Izzo}}, \citenamefont {{D'Avanzo}},
  \citenamefont {{Malesani}}, \citenamefont {{Valle}}, \citenamefont {{Pian}},
  \citenamefont {{Tanvir}}, \citenamefont {{Ragosta}}, \citenamefont
  {{Olivares}}, \citenamefont {{Carini}}, \citenamefont {{Palazzi}},
  \citenamefont {{Piranomonte}}, \citenamefont {{Jonker}}, \citenamefont
  {{Rossi}}, \citenamefont {{Kann}}, \citenamefont {{Hartmann}}, \citenamefont
  {{Inserra}}, \citenamefont {{Kankare}}, \citenamefont {{Maguire}},
  \citenamefont {{Smartt}}, \citenamefont {{Yaron}}, \citenamefont {{Young}},\
  and\ \citenamefont {{Manulis}}}]{2019GCN.23983....1M}%
  \BibitemOpen
  \bibfield  {author} {\bibinfo {author} {\bibfnamefont {A.}~\bibnamefont
  {{Melandri}}}, \bibinfo {author} {\bibfnamefont {L.}~\bibnamefont {{Izzo}}},
  \bibinfo {author} {\bibfnamefont {P.}~\bibnamefont {{D'Avanzo}}}, \bibinfo
  {author} {\bibfnamefont {D.}~\bibnamefont {{Malesani}}}, \bibinfo {author}
  {\bibfnamefont {M.~D.}\ \bibnamefont {{Valle}}}, \bibinfo {author}
  {\bibfnamefont {E.}~\bibnamefont {{Pian}}}, \bibinfo {author} {\bibfnamefont
  {N.~R.}\ \bibnamefont {{Tanvir}}}, \bibinfo {author} {\bibfnamefont
  {F.}~\bibnamefont {{Ragosta}}}, \bibinfo {author} {\bibfnamefont
  {F.}~\bibnamefont {{Olivares}}}, \bibinfo {author} {\bibfnamefont
  {R.}~\bibnamefont {{Carini}}}, \bibinfo {author} {\bibfnamefont
  {E.}~\bibnamefont {{Palazzi}}}, \bibinfo {author} {\bibfnamefont
  {S.}~\bibnamefont {{Piranomonte}}}, \bibinfo {author} {\bibfnamefont
  {P.}~\bibnamefont {{Jonker}}}, \bibinfo {author} {\bibfnamefont
  {A.}~\bibnamefont {{Rossi}}}, \bibinfo {author} {\bibfnamefont {D.~A.}\
  \bibnamefont {{Kann}}}, \bibinfo {author} {\bibfnamefont {D.}~\bibnamefont
  {{Hartmann}}}, \bibinfo {author} {\bibfnamefont {C.}~\bibnamefont
  {{Inserra}}}, \bibinfo {author} {\bibfnamefont {E.}~\bibnamefont
  {{Kankare}}}, \bibinfo {author} {\bibfnamefont {K.}~\bibnamefont
  {{Maguire}}}, \bibinfo {author} {\bibfnamefont {S.~J.}\ \bibnamefont
  {{Smartt}}}, \bibinfo {author} {\bibfnamefont {O.}~\bibnamefont {{Yaron}}},
  \bibinfo {author} {\bibfnamefont {D.~R.}\ \bibnamefont {{Young}}}, \ and\
  \bibinfo {author} {\bibfnamefont {I.}~\bibnamefont {{Manulis}}},\ }\href@noop
  {} {\bibfield  {journal} {\bibinfo  {journal} {GRB Coordinates Network,
  Circular Service, No.~23983, \#1 (2019/March-0)}\ }\textbf {\bibinfo {volume}
  {23983}} (\bibinfo {year} {2019})}\BibitemShut {NoStop}%
\bibitem [{\citenamefont {{Panaitescu}}\ and\ \citenamefont
  {{Kumar}}(2000)}]{2000ApJ...543...66P}%
  \BibitemOpen
  \bibfield  {author} {\bibinfo {author} {\bibfnamefont {A.}~\bibnamefont
  {{Panaitescu}}}\ and\ \bibinfo {author} {\bibfnamefont {P.}~\bibnamefont
  {{Kumar}}},\ }\href {\doibase 10.1086/317090} {\bibfield  {journal} {\bibinfo
   {journal} {\apj}\ }\textbf {\bibinfo {volume} {543}},\ \bibinfo {pages} {66}
  (\bibinfo {year} {2000})},\ \Eprint {http://arxiv.org/abs/astro-ph/0003246}
  {astro-ph/0003246} \BibitemShut {NoStop}%
\bibitem [{\citenamefont {{Vink}}\ \emph {et~al.}(2000)\citenamefont {{Vink}},
  \citenamefont {{de Koter}},\ and\ \citenamefont
  {{Lamers}}}]{2000A&A...362..295V}%
  \BibitemOpen
  \bibfield  {author} {\bibinfo {author} {\bibfnamefont {J.~S.}\ \bibnamefont
  {{Vink}}}, \bibinfo {author} {\bibfnamefont {A.}~\bibnamefont {{de Koter}}},
  \ and\ \bibinfo {author} {\bibfnamefont {H.~J.~G.~L.~M.}\ \bibnamefont
  {{Lamers}}},\ }\href@noop {} {\bibfield  {journal} {\bibinfo  {journal}
  {\aap}\ }\textbf {\bibinfo {volume} {362}},\ \bibinfo {pages} {295} (\bibinfo
  {year} {2000})},\ \Eprint {http://arxiv.org/abs/astro-ph/0008183}
  {astro-ph/0008183} \BibitemShut {NoStop}%
\bibitem [{\citenamefont {{Vink}}\ and\ \citenamefont {{de
  Koter}}(2005)}]{2005A&A...442..587V}%
  \BibitemOpen
  \bibfield  {author} {\bibinfo {author} {\bibfnamefont {J.~S.}\ \bibnamefont
  {{Vink}}}\ and\ \bibinfo {author} {\bibfnamefont {A.}~\bibnamefont {{de
  Koter}}},\ }\href {\doibase 10.1051/0004-6361:20052862} {\bibfield  {journal}
  {\bibinfo  {journal} {\aap}\ }\textbf {\bibinfo {volume} {442}},\ \bibinfo
  {pages} {587} (\bibinfo {year} {2005})},\ \Eprint
  {http://arxiv.org/abs/astro-ph/0507352} {astro-ph/0507352} \BibitemShut
  {NoStop}%
\bibitem [{\citenamefont {{Chevalier}}\ \emph {et~al.}(2004)\citenamefont
  {{Chevalier}}, \citenamefont {{Li}},\ and\ \citenamefont
  {{Fransson}}}]{2004ApJ...606..369C}%
  \BibitemOpen
  \bibfield  {author} {\bibinfo {author} {\bibfnamefont {R.~A.}\ \bibnamefont
  {{Chevalier}}}, \bibinfo {author} {\bibfnamefont {Z.-Y.}\ \bibnamefont
  {{Li}}}, \ and\ \bibinfo {author} {\bibfnamefont {C.}~\bibnamefont
  {{Fransson}}},\ }\href {\doibase 10.1086/382867} {\bibfield  {journal}
  {\bibinfo  {journal} {\apj}\ }\textbf {\bibinfo {volume} {606}},\ \bibinfo
  {pages} {369} (\bibinfo {year} {2004})},\ \Eprint
  {http://arxiv.org/abs/astro-ph/0311326} {astro-ph/0311326} \BibitemShut
  {NoStop}%
\bibitem [{\citenamefont {{Dai}}\ and\ \citenamefont
  {{Lu}}(1998)}]{1998MNRAS.298...87D}%
  \BibitemOpen
  \bibfield  {author} {\bibinfo {author} {\bibfnamefont {Z.~G.}\ \bibnamefont
  {{Dai}}}\ and\ \bibinfo {author} {\bibfnamefont {T.}~\bibnamefont {{Lu}}},\
  }\href {\doibase 10.1046/j.1365-8711.1998.01681.x} {\bibfield  {journal}
  {\bibinfo  {journal} {\mnras}\ }\textbf {\bibinfo {volume} {298}},\ \bibinfo
  {pages} {87} (\bibinfo {year} {1998})},\ \Eprint
  {http://arxiv.org/abs/astro-ph/9806305} {astro-ph/9806305} \BibitemShut
  {NoStop}%
\bibitem [{\citenamefont {{Chevalier}}\ and\ \citenamefont
  {{Li}}(2000)}]{2000ApJ...536..195C}%
  \BibitemOpen
  \bibfield  {author} {\bibinfo {author} {\bibfnamefont {R.~A.}\ \bibnamefont
  {{Chevalier}}}\ and\ \bibinfo {author} {\bibfnamefont {Z.-Y.}\ \bibnamefont
  {{Li}}},\ }\href {\doibase 10.1086/308914} {\bibfield  {journal} {\bibinfo
  {journal} {\apj}\ }\textbf {\bibinfo {volume} {536}},\ \bibinfo {pages} {195}
  (\bibinfo {year} {2000})},\ \Eprint
  {http://arxiv.org/abs/arXiv:astro-ph/9908272} {arXiv:astro-ph/9908272}
  \BibitemShut {NoStop}%
\bibitem [{\citenamefont {{Panaitescu}}\ and\ \citenamefont
  {{M{\'e}sz{\'a}ros}}(1998)}]{1998ApJ...493L..31P}%
  \BibitemOpen
  \bibfield  {author} {\bibinfo {author} {\bibfnamefont {A.}~\bibnamefont
  {{Panaitescu}}}\ and\ \bibinfo {author} {\bibfnamefont {P.}~\bibnamefont
  {{M{\'e}sz{\'a}ros}}},\ }\href {\doibase 10.1086/311127} {\bibfield
  {journal} {\bibinfo  {journal} {\apjl}\ }\textbf {\bibinfo {volume} {493}},\
  \bibinfo {pages} {L31} (\bibinfo {year} {1998})},\ \Eprint
  {http://arxiv.org/abs/astro-ph/9709284} {astro-ph/9709284} \BibitemShut
  {NoStop}%
\bibitem [{\citenamefont {{Sari}}\ \emph {et~al.}(1998)\citenamefont {{Sari}},
  \citenamefont {{Piran}},\ and\ \citenamefont
  {{Narayan}}}]{1998ApJ...497L..17S}%
  \BibitemOpen
  \bibfield  {author} {\bibinfo {author} {\bibfnamefont {R.}~\bibnamefont
  {{Sari}}}, \bibinfo {author} {\bibfnamefont {T.}~\bibnamefont {{Piran}}}, \
  and\ \bibinfo {author} {\bibfnamefont {R.}~\bibnamefont {{Narayan}}},\ }\href
  {\doibase 10.1086/311269} {\bibfield  {journal} {\bibinfo  {journal} {\apjl}\
  }\textbf {\bibinfo {volume} {497}},\ \bibinfo {pages} {L17} (\bibinfo {year}
  {1998})},\ \Eprint {http://arxiv.org/abs/arXiv:astro-ph/9712005}
  {arXiv:astro-ph/9712005} \BibitemShut {NoStop}%
\bibitem [{\citenamefont {{Sari}}\ and\ \citenamefont
  {{Piran}}(1995)}]{1995ApJ...455L.143S}%
  \BibitemOpen
  \bibfield  {author} {\bibinfo {author} {\bibfnamefont {R.}~\bibnamefont
  {{Sari}}}\ and\ \bibinfo {author} {\bibfnamefont {T.}~\bibnamefont
  {{Piran}}},\ }\href {\doibase 10.1086/309835} {\bibfield  {journal} {\bibinfo
   {journal} {\apjl}\ }\textbf {\bibinfo {volume} {455}},\ \bibinfo {pages}
  {L143} (\bibinfo {year} {1995})},\ \Eprint
  {http://arxiv.org/abs/astro-ph/9508081} {astro-ph/9508081} \BibitemShut
  {NoStop}%
\bibitem [{\citenamefont {{Kumar}}\ and\ \citenamefont
  {{Piran}}(2000)}]{2000ApJ...532..286K}%
  \BibitemOpen
  \bibfield  {author} {\bibinfo {author} {\bibfnamefont {P.}~\bibnamefont
  {{Kumar}}}\ and\ \bibinfo {author} {\bibfnamefont {T.}~\bibnamefont
  {{Piran}}},\ }\href {\doibase 10.1086/308537} {\bibfield  {journal} {\bibinfo
   {journal} {\apj}\ }\textbf {\bibinfo {volume} {532}},\ \bibinfo {pages}
  {286} (\bibinfo {year} {2000})},\ \Eprint
  {http://arxiv.org/abs/astro-ph/9906002} {astro-ph/9906002} \BibitemShut
  {NoStop}%
\bibitem [{\citenamefont {{Sari}}\ and\ \citenamefont
  {{Piran}}(1999)}]{1999A&AS..138..537S}%
  \BibitemOpen
  \bibfield  {author} {\bibinfo {author} {\bibfnamefont {R.}~\bibnamefont
  {{Sari}}}\ and\ \bibinfo {author} {\bibfnamefont {T.}~\bibnamefont
  {{Piran}}},\ }\href {\doibase 10.1051/aas:1999342} {\bibfield  {journal}
  {\bibinfo  {journal} {\aaps}\ }\textbf {\bibinfo {volume} {138}},\ \bibinfo
  {pages} {537} (\bibinfo {year} {1999})},\ \Eprint
  {http://arxiv.org/abs/astro-ph/9901105} {astro-ph/9901105} \BibitemShut
  {NoStop}%
\bibitem [{\citenamefont {{Zhang}}\ and\ \citenamefont
  {{Kobayashi}}(2005)}]{2005ApJ...628..315Z}%
  \BibitemOpen
  \bibfield  {author} {\bibinfo {author} {\bibfnamefont {B.}~\bibnamefont
  {{Zhang}}}\ and\ \bibinfo {author} {\bibfnamefont {S.}~\bibnamefont
  {{Kobayashi}}},\ }\href {\doibase 10.1086/429787} {\bibfield  {journal}
  {\bibinfo  {journal} {\apj}\ }\textbf {\bibinfo {volume} {628}},\ \bibinfo
  {pages} {315} (\bibinfo {year} {2005})},\ \Eprint
  {http://arxiv.org/abs/astro-ph/0404140} {astro-ph/0404140} \BibitemShut
  {NoStop}%
\bibitem [{\citenamefont {{Wang}}\ \emph
  {et~al.}(2001{\natexlab{a}})\citenamefont {{Wang}}, \citenamefont {{Dai}},\
  and\ \citenamefont {{Lu}}}]{2001ApJ...546L..33W}%
  \BibitemOpen
  \bibfield  {author} {\bibinfo {author} {\bibfnamefont {X.~Y.}\ \bibnamefont
  {{Wang}}}, \bibinfo {author} {\bibfnamefont {Z.~G.}\ \bibnamefont {{Dai}}}, \
  and\ \bibinfo {author} {\bibfnamefont {T.}~\bibnamefont {{Lu}}},\ }\href
  {\doibase 10.1086/318064} {\bibfield  {journal} {\bibinfo  {journal} {\apjl}\
  }\textbf {\bibinfo {volume} {546}},\ \bibinfo {pages} {L33} (\bibinfo {year}
  {2001}{\natexlab{a}})},\ \Eprint {http://arxiv.org/abs/astro-ph/0010320}
  {astro-ph/0010320} \BibitemShut {NoStop}%
\bibitem [{\citenamefont {{Wang}}\ \emph
  {et~al.}(2001{\natexlab{b}})\citenamefont {{Wang}}, \citenamefont {{Dai}},\
  and\ \citenamefont {{Lu}}}]{2001ApJ...556.1010W}%
  \BibitemOpen
  \bibfield  {author} {\bibinfo {author} {\bibfnamefont {X.~Y.}\ \bibnamefont
  {{Wang}}}, \bibinfo {author} {\bibfnamefont {Z.~G.}\ \bibnamefont {{Dai}}}, \
  and\ \bibinfo {author} {\bibfnamefont {T.}~\bibnamefont {{Lu}}},\ }\href
  {\doibase 10.1086/321608} {\bibfield  {journal} {\bibinfo  {journal} {\apj}\
  }\textbf {\bibinfo {volume} {556}},\ \bibinfo {pages} {1010} (\bibinfo {year}
  {2001}{\natexlab{b}})},\ \Eprint {http://arxiv.org/abs/astro-ph/0104128}
  {astro-ph/0104128} \BibitemShut {NoStop}%
\bibitem [{\citenamefont {{Veres}}\ and\ \citenamefont
  {{M{\'e}sz{\'a}ros}}(2012)}]{2012ApJ...755...12V}%
  \BibitemOpen
  \bibfield  {author} {\bibinfo {author} {\bibfnamefont {P.}~\bibnamefont
  {{Veres}}}\ and\ \bibinfo {author} {\bibfnamefont {P.}~\bibnamefont
  {{M{\'e}sz{\'a}ros}}},\ }\href {\doibase 10.1088/0004-637X/755/1/12}
  {\bibfield  {journal} {\bibinfo  {journal} {\apj}\ }\textbf {\bibinfo
  {volume} {755}},\ \bibinfo {eid} {12} (\bibinfo {year} {2012})},\ \Eprint
  {http://arxiv.org/abs/1202.2821} {arXiv:1202.2821 [astro-ph.HE]} \BibitemShut
  {NoStop}%
\bibitem [{\citenamefont {{Kobayashi}}\ and\ \citenamefont
  {{Zhang}}(2003)}]{2003ApJ...597..455K}%
  \BibitemOpen
  \bibfield  {author} {\bibinfo {author} {\bibfnamefont {S.}~\bibnamefont
  {{Kobayashi}}}\ and\ \bibinfo {author} {\bibfnamefont {B.}~\bibnamefont
  {{Zhang}}},\ }\href {\doibase 10.1086/378283} {\bibfield  {journal} {\bibinfo
   {journal} {\apj}\ }\textbf {\bibinfo {volume} {597}},\ \bibinfo {pages}
  {455} (\bibinfo {year} {2003})},\ \Eprint
  {http://arxiv.org/abs/astro-ph/0304086} {astro-ph/0304086} \BibitemShut
  {NoStop}%
\bibitem [{\citenamefont {{Fraija}}\ \emph
  {et~al.}(2017{\natexlab{b}})\citenamefont {{Fraija}}, \citenamefont
  {{Veres}}, \citenamefont {{Zhang}}, \citenamefont {{Barniol Duran}},
  \citenamefont {{Becerra}}, \citenamefont {{Zhang}}, \citenamefont {{Lee}},
  \citenamefont {{Watson}}, \citenamefont {{Ordaz-Salazar}},\ and\
  \citenamefont {{Galvan-Gamez}}}]{2017ApJ...848...15F}%
  \BibitemOpen
  \bibfield  {author} {\bibinfo {author} {\bibfnamefont {N.}~\bibnamefont
  {{Fraija}}}, \bibinfo {author} {\bibfnamefont {P.}~\bibnamefont {{Veres}}},
  \bibinfo {author} {\bibfnamefont {B.~B.}\ \bibnamefont {{Zhang}}}, \bibinfo
  {author} {\bibfnamefont {R.}~\bibnamefont {{Barniol Duran}}}, \bibinfo
  {author} {\bibfnamefont {R.~L.}\ \bibnamefont {{Becerra}}}, \bibinfo {author}
  {\bibfnamefont {B.}~\bibnamefont {{Zhang}}}, \bibinfo {author} {\bibfnamefont
  {W.~H.}\ \bibnamefont {{Lee}}}, \bibinfo {author} {\bibfnamefont {A.~M.}\
  \bibnamefont {{Watson}}}, \bibinfo {author} {\bibfnamefont {C.}~\bibnamefont
  {{Ordaz-Salazar}}}, \ and\ \bibinfo {author} {\bibfnamefont {A.}~\bibnamefont
  {{Galvan-Gamez}}},\ }\href {\doibase 10.3847/1538-4357/aa8a72} {\bibfield
  {journal} {\bibinfo  {journal} {\apj}\ }\textbf {\bibinfo {volume} {848}},\
  \bibinfo {eid} {15} (\bibinfo {year} {2017}{\natexlab{b}})},\ \Eprint
  {http://arxiv.org/abs/1705.09311} {arXiv:1705.09311 [astro-ph.HE]}
  \BibitemShut {NoStop}%
\bibitem [{\citenamefont {{Zhang}}\ \emph {et~al.}(2003)\citenamefont
  {{Zhang}}, \citenamefont {{Kobayashi}},\ and\ \citenamefont
  {{M{\'e}sz{\'a}ros}}}]{2003ApJ...595..950Z}%
  \BibitemOpen
  \bibfield  {author} {\bibinfo {author} {\bibfnamefont {B.}~\bibnamefont
  {{Zhang}}}, \bibinfo {author} {\bibfnamefont {S.}~\bibnamefont
  {{Kobayashi}}}, \ and\ \bibinfo {author} {\bibfnamefont {P.}~\bibnamefont
  {{M{\'e}sz{\'a}ros}}},\ }\href {\doibase 10.1086/377363} {\bibfield
  {journal} {\bibinfo  {journal} {\apj}\ }\textbf {\bibinfo {volume} {595}},\
  \bibinfo {pages} {950} (\bibinfo {year} {2003})},\ \Eprint
  {http://arxiv.org/abs/astro-ph/0302525} {astro-ph/0302525} \BibitemShut
  {NoStop}%
\bibitem [{\citenamefont {{Fraija}}\ \emph
  {et~al.}(2016{\natexlab{b}})\citenamefont {{Fraija}}, \citenamefont {{Lee}},
  \citenamefont {{Veres}},\ and\ \citenamefont {{Barniol
  Duran}}}]{2016ApJ...831...22F}%
  \BibitemOpen
  \bibfield  {author} {\bibinfo {author} {\bibfnamefont {N.}~\bibnamefont
  {{Fraija}}}, \bibinfo {author} {\bibfnamefont {W.~H.}\ \bibnamefont {{Lee}}},
  \bibinfo {author} {\bibfnamefont {P.}~\bibnamefont {{Veres}}}, \ and\
  \bibinfo {author} {\bibfnamefont {R.}~\bibnamefont {{Barniol Duran}}},\
  }\href {\doibase 10.3847/0004-637X/831/1/22} {\bibfield  {journal} {\bibinfo
  {journal} {\apj}\ }\textbf {\bibinfo {volume} {831}},\ \bibinfo {eid} {22}
  (\bibinfo {year} {2016}{\natexlab{b}})}\BibitemShut {NoStop}%
\bibitem [{\citenamefont {{Kobayashi}}(2000)}]{2000ApJ...545..807K}%
  \BibitemOpen
  \bibfield  {author} {\bibinfo {author} {\bibfnamefont {S.}~\bibnamefont
  {{Kobayashi}}},\ }\href {\doibase 10.1086/317869} {\bibfield  {journal}
  {\bibinfo  {journal} {\apj}\ }\textbf {\bibinfo {volume} {545}},\ \bibinfo
  {pages} {807} (\bibinfo {year} {2000})},\ \Eprint
  {http://arxiv.org/abs/astro-ph/0009319} {astro-ph/0009319} \BibitemShut
  {NoStop}%
\bibitem [{\citenamefont {{Weaver}}\ \emph {et~al.}(1977)\citenamefont
  {{Weaver}}, \citenamefont {{McCray}}, \citenamefont {{Castor}}, \citenamefont
  {{Shapiro}},\ and\ \citenamefont {{Moore}}}]{1977ApJ...218..377W}%
  \BibitemOpen
  \bibfield  {author} {\bibinfo {author} {\bibfnamefont {R.}~\bibnamefont
  {{Weaver}}}, \bibinfo {author} {\bibfnamefont {R.}~\bibnamefont {{McCray}}},
  \bibinfo {author} {\bibfnamefont {J.}~\bibnamefont {{Castor}}}, \bibinfo
  {author} {\bibfnamefont {P.}~\bibnamefont {{Shapiro}}}, \ and\ \bibinfo
  {author} {\bibfnamefont {R.}~\bibnamefont {{Moore}}},\ }\href {\doibase
  10.1086/155692} {\bibfield  {journal} {\bibinfo  {journal} {\apj}\ }\textbf
  {\bibinfo {volume} {218}},\ \bibinfo {pages} {377} (\bibinfo {year}
  {1977})}\BibitemShut {NoStop}%
\bibitem [{\citenamefont {{Castor}}\ \emph {et~al.}(1975)\citenamefont
  {{Castor}}, \citenamefont {{McCray}},\ and\ \citenamefont
  {{Weaver}}}]{1975ApJ...200L.107C}%
  \BibitemOpen
  \bibfield  {author} {\bibinfo {author} {\bibfnamefont {J.}~\bibnamefont
  {{Castor}}}, \bibinfo {author} {\bibfnamefont {R.}~\bibnamefont {{McCray}}},
  \ and\ \bibinfo {author} {\bibfnamefont {R.}~\bibnamefont {{Weaver}}},\
  }\href {\doibase 10.1086/181908} {\bibfield  {journal} {\bibinfo  {journal}
  {\apjl}\ }\textbf {\bibinfo {volume} {200}},\ \bibinfo {pages} {L107}
  (\bibinfo {year} {1975})}\BibitemShut {NoStop}%
\bibitem [{\citenamefont {{Fryer}}\ \emph {et~al.}(2006)\citenamefont
  {{Fryer}}, \citenamefont {{Rockefeller}},\ and\ \citenamefont
  {{Young}}}]{2006ApJ...647.1269F}%
  \BibitemOpen
  \bibfield  {author} {\bibinfo {author} {\bibfnamefont {C.~L.}\ \bibnamefont
  {{Fryer}}}, \bibinfo {author} {\bibfnamefont {G.}~\bibnamefont
  {{Rockefeller}}}, \ and\ \bibinfo {author} {\bibfnamefont {P.~A.}\
  \bibnamefont {{Young}}},\ }\href {\doibase 10.1086/505590} {\bibfield
  {journal} {\bibinfo  {journal} {\apj}\ }\textbf {\bibinfo {volume} {647}},\
  \bibinfo {pages} {1269} (\bibinfo {year} {2006})},\ \Eprint
  {http://arxiv.org/abs/astro-ph/0604432} {astro-ph/0604432} \BibitemShut
  {NoStop}%
\bibitem [{\citenamefont {{Pe'er}}\ and\ \citenamefont
  {{Wijers}}(2006)}]{2006ApJ...643.1036P}%
  \BibitemOpen
  \bibfield  {author} {\bibinfo {author} {\bibfnamefont {A.}~\bibnamefont
  {{Pe'er}}}\ and\ \bibinfo {author} {\bibfnamefont {R.~A.~M.~J.}\ \bibnamefont
  {{Wijers}}},\ }\href {\doibase 10.1086/500969} {\bibfield  {journal}
  {\bibinfo  {journal} {\apj}\ }\textbf {\bibinfo {volume} {643}},\ \bibinfo
  {pages} {1036} (\bibinfo {year} {2006})},\ \Eprint
  {http://arxiv.org/abs/astro-ph/0511508} {astro-ph/0511508} \BibitemShut
  {NoStop}%
\bibitem [{\citenamefont {{Garcia-Segura}}\ and\ \citenamefont
  {{Franco}}(1996)}]{1996ApJ...469..171G}%
  \BibitemOpen
  \bibfield  {author} {\bibinfo {author} {\bibfnamefont {G.}~\bibnamefont
  {{Garcia-Segura}}}\ and\ \bibinfo {author} {\bibfnamefont {J.}~\bibnamefont
  {{Franco}}},\ }\href {\doibase 10.1086/177769} {\bibfield  {journal}
  {\bibinfo  {journal} {\apj}\ }\textbf {\bibinfo {volume} {469}},\ \bibinfo
  {pages} {171} (\bibinfo {year} {1996})}\BibitemShut {NoStop}%
\bibitem [{\citenamefont {{Planck Collaboration}}\ \emph
  {et~al.}(2018)\citenamefont {{Planck Collaboration}}, \citenamefont
  {{Aghanim}}, \citenamefont {{Akrami}}, \citenamefont {{Ashdown}},
  \citenamefont {{Aumont}}, \citenamefont {{Baccigalupi}}, \citenamefont
  {{Ballardini}},\ and\ \citenamefont {{Banday}}}]{2018arXiv180706209P}%
  \BibitemOpen
  \bibfield  {author} {\bibinfo {author} {\bibnamefont {{Planck
  Collaboration}}}, \bibinfo {author} {\bibfnamefont {N.}~\bibnamefont
  {{Aghanim}}}, \bibinfo {author} {\bibfnamefont {Y.}~\bibnamefont {{Akrami}}},
  \bibinfo {author} {\bibfnamefont {M.}~\bibnamefont {{Ashdown}}}, \bibinfo
  {author} {\bibfnamefont {J.}~\bibnamefont {{Aumont}}}, \bibinfo {author}
  {\bibfnamefont {C.}~\bibnamefont {{Baccigalupi}}}, \bibinfo {author}
  {\bibfnamefont {M.}~\bibnamefont {{Ballardini}}}, \ and\ \bibinfo {author}
  {\bibfnamefont {A.~J. e.~a.}\ \bibnamefont {{Banday}}},\ }\href@noop {}
  {\bibfield  {journal} {\bibinfo  {journal} {arXiv e-prints}\ ,\ \bibinfo
  {eid} {arXiv:1807.06209}} (\bibinfo {year} {2018})},\ \Eprint
  {http://arxiv.org/abs/1807.06209} {arXiv:1807.06209 [astro-ph.CO]}
  \BibitemShut {NoStop}%
\bibitem [{\citenamefont {{Beniamini}}\ \emph {et~al.}(2015)\citenamefont
  {{Beniamini}}, \citenamefont {{Nava}}, \citenamefont {{Duran}},\ and\
  \citenamefont {{Piran}}}]{2015MNRAS.454.1073B}%
  \BibitemOpen
  \bibfield  {author} {\bibinfo {author} {\bibfnamefont {P.}~\bibnamefont
  {{Beniamini}}}, \bibinfo {author} {\bibfnamefont {L.}~\bibnamefont {{Nava}}},
  \bibinfo {author} {\bibfnamefont {R.~B.}\ \bibnamefont {{Duran}}}, \ and\
  \bibinfo {author} {\bibfnamefont {T.}~\bibnamefont {{Piran}}},\ }\href
  {\doibase 10.1093/mnras/stv2033} {\bibfield  {journal} {\bibinfo  {journal}
  {\mnras}\ }\textbf {\bibinfo {volume} {454}},\ \bibinfo {pages} {1073}
  (\bibinfo {year} {2015})},\ \Eprint {http://arxiv.org/abs/1504.04833}
  {arXiv:1504.04833 [astro-ph.HE]} \BibitemShut {NoStop}%
\bibitem [{\citenamefont {{Fraija}}\ \emph
  {et~al.}(2019{\natexlab{c}})\citenamefont {{Fraija}}, \citenamefont
  {{Pedreira}},\ and\ \citenamefont {{Veres}}}]{2019ApJ...871..200F}%
  \BibitemOpen
  \bibfield  {author} {\bibinfo {author} {\bibfnamefont {N.}~\bibnamefont
  {{Fraija}}}, \bibinfo {author} {\bibfnamefont {A.~C.~C.~d.~E.~S.}\
  \bibnamefont {{Pedreira}}}, \ and\ \bibinfo {author} {\bibfnamefont
  {P.}~\bibnamefont {{Veres}}},\ }\href {\doibase 10.3847/1538-4357/aaf80e}
  {\bibfield  {journal} {\bibinfo  {journal} {\apj}\ }\textbf {\bibinfo
  {volume} {871}},\ \bibinfo {eid} {200} (\bibinfo {year}
  {2019}{\natexlab{c}})}\BibitemShut {NoStop}%
\bibitem [{\citenamefont {{Fraija}}\ \emph
  {et~al.}(2019{\natexlab{d}})\citenamefont {{Fraija}}, \citenamefont
  {{Lopez-Camara}}, \citenamefont {{Pedreira}}, \citenamefont {{Betancourt
  Kamenetskaia}}, \citenamefont {{Veres}},\ and\ \citenamefont
  {{Dichiara}}}]{2019arXiv190407732F}%
  \BibitemOpen
  \bibfield  {author} {\bibinfo {author} {\bibfnamefont {N.}~\bibnamefont
  {{Fraija}}}, \bibinfo {author} {\bibfnamefont {D.}~\bibnamefont
  {{Lopez-Camara}}}, \bibinfo {author} {\bibfnamefont {A.~C. C. d. E.~S.}\
  \bibnamefont {{Pedreira}}}, \bibinfo {author} {\bibfnamefont
  {B.}~\bibnamefont {{Betancourt Kamenetskaia}}}, \bibinfo {author}
  {\bibfnamefont {P.}~\bibnamefont {{Veres}}}, \ and\ \bibinfo {author}
  {\bibfnamefont {S.}~\bibnamefont {{Dichiara}}},\ }\href@noop {} {\bibfield
  {journal} {\bibinfo  {journal} {arXiv e-prints}\ ,\ \bibinfo {eid}
  {arXiv:1904.07732}} (\bibinfo {year} {2019}{\natexlab{d}})},\ \Eprint
  {http://arxiv.org/abs/1904.07732} {arXiv:1904.07732 [astro-ph.HE]}
  \BibitemShut {NoStop}%
\bibitem [{\citenamefont {{Fraija}}\ \emph
  {et~al.}(2019{\natexlab{e}})\citenamefont {{Fraija}}, \citenamefont {{De
  Colle}}, \citenamefont {{Veres}}, \citenamefont {{Dichiara}}, \citenamefont
  {{Barniol Duran}}, \citenamefont {{Pedreira}}, \citenamefont
  {{Galvan-Gamez}},\ and\ \citenamefont {{Betancourt
  Kamenetskaia}}}]{2019arXiv190600502F}%
  \BibitemOpen
  \bibfield  {author} {\bibinfo {author} {\bibfnamefont {N.}~\bibnamefont
  {{Fraija}}}, \bibinfo {author} {\bibfnamefont {F.}~\bibnamefont {{De
  Colle}}}, \bibinfo {author} {\bibfnamefont {P.}~\bibnamefont {{Veres}}},
  \bibinfo {author} {\bibfnamefont {S.}~\bibnamefont {{Dichiara}}}, \bibinfo
  {author} {\bibfnamefont {R.}~\bibnamefont {{Barniol Duran}}}, \bibinfo
  {author} {\bibfnamefont {A.~C. C. d. E.~S.}\ \bibnamefont {{Pedreira}}},
  \bibinfo {author} {\bibfnamefont {A.}~\bibnamefont {{Galvan-Gamez}}}, \ and\
  \bibinfo {author} {\bibfnamefont {B.}~\bibnamefont {{Betancourt
  Kamenetskaia}}},\ }\href@noop {} {\bibfield  {journal} {\bibinfo  {journal}
  {arXiv e-prints}\ ,\ \bibinfo {eid} {arXiv:1906.00502}} (\bibinfo {year}
  {2019}{\natexlab{e}})},\ \Eprint {http://arxiv.org/abs/1906.00502}
  {arXiv:1906.00502 [astro-ph.HE]} \BibitemShut {NoStop}%
\bibitem [{\citenamefont {{Ackermann}}\ and\ \citenamefont
  {et~al.}(2010)}]{2010ApJ...716.1178A}%
  \BibitemOpen
  \bibfield  {author} {\bibinfo {author} {\bibfnamefont {M.}~\bibnamefont
  {{Ackermann}}}\ and\ \bibinfo {author} {\bibnamefont {et~al.}},\ }\href
  {\doibase 10.1088/0004-637X/716/2/1178} {\bibfield  {journal} {\bibinfo
  {journal} {\apj}\ }\textbf {\bibinfo {volume} {716}},\ \bibinfo {pages}
  {1178} (\bibinfo {year} {2010})},\ \Eprint {http://arxiv.org/abs/1005.2141}
  {arXiv:1005.2141 [astro-ph.HE]} \BibitemShut {NoStop}%
\bibitem [{\citenamefont {{Ackermann}}\ and\ \citenamefont
  {et~al.}(2013)}]{2013ApJ...763...71A}%
  \BibitemOpen
  \bibfield  {author} {\bibinfo {author} {\bibfnamefont {M.}~\bibnamefont
  {{Ackermann}}}\ and\ \bibinfo {author} {\bibnamefont {et~al.}},\ }\href
  {\doibase 10.1088/0004-637X/763/2/71} {\bibfield  {journal} {\bibinfo
  {journal} {\apj}\ }\textbf {\bibinfo {volume} {763}},\ \bibinfo {eid} {71}
  (\bibinfo {year} {2013})},\ \Eprint {http://arxiv.org/abs/1212.0973}
  {arXiv:1212.0973 [astro-ph.HE]} \BibitemShut {NoStop}%
\bibitem [{\citenamefont {{Ackermann}}\ \emph {et~al.}(2014)\citenamefont
  {{Ackermann}}, \citenamefont {{Ajello}}, \citenamefont {{Asano}},
  \citenamefont {{Atwood}}, \citenamefont {{Axelsson}}, \citenamefont
  {{Baldini}}, \citenamefont {{Ballet}}, \citenamefont {{Barbiellini}},
  \citenamefont {{Baring}},\ and\ \citenamefont
  {et~al.}}]{2014Sci...343...42A}%
  \BibitemOpen
  \bibfield  {author} {\bibinfo {author} {\bibfnamefont {M.}~\bibnamefont
  {{Ackermann}}}, \bibinfo {author} {\bibfnamefont {M.}~\bibnamefont
  {{Ajello}}}, \bibinfo {author} {\bibfnamefont {K.}~\bibnamefont {{Asano}}},
  \bibinfo {author} {\bibfnamefont {W.~B.}\ \bibnamefont {{Atwood}}}, \bibinfo
  {author} {\bibfnamefont {M.}~\bibnamefont {{Axelsson}}}, \bibinfo {author}
  {\bibfnamefont {L.}~\bibnamefont {{Baldini}}}, \bibinfo {author}
  {\bibfnamefont {J.}~\bibnamefont {{Ballet}}}, \bibinfo {author}
  {\bibfnamefont {G.}~\bibnamefont {{Barbiellini}}}, \bibinfo {author}
  {\bibfnamefont {M.~G.}\ \bibnamefont {{Baring}}}, \ and\ \bibinfo {author}
  {\bibnamefont {et~al.}},\ }\href {\doibase 10.1126/science.1242353}
  {\bibfield  {journal} {\bibinfo  {journal} {Science}\ }\textbf {\bibinfo
  {volume} {343}},\ \bibinfo {pages} {42} (\bibinfo {year} {2014})}\BibitemShut
  {NoStop}%
\bibitem [{\citenamefont {{Fraija}}(2014)}]{2014MNRAS.437.2187F}%
  \BibitemOpen
  \bibfield  {author} {\bibinfo {author} {\bibfnamefont {N.}~\bibnamefont
  {{Fraija}}},\ }\href {\doibase 10.1093/mnras/stt2036} {\bibfield  {journal}
  {\bibinfo  {journal} {\mnras}\ }\textbf {\bibinfo {volume} {437}},\ \bibinfo
  {pages} {2187} (\bibinfo {year} {2014})},\ \Eprint
  {http://arxiv.org/abs/1310.7061} {arXiv:1310.7061 [astro-ph.HE]} \BibitemShut
  {NoStop}%
\bibitem [{\citenamefont {{Fraija}}(2015{\natexlab{b}})}]{2015MNRAS.450.2784F}%
  \BibitemOpen
  \bibfield  {author} {\bibinfo {author} {\bibfnamefont {N.}~\bibnamefont
  {{Fraija}}},\ }\href {\doibase 10.1093/mnras/stv737} {\bibfield  {journal}
  {\bibinfo  {journal} {\mnras}\ }\textbf {\bibinfo {volume} {450}},\ \bibinfo
  {pages} {2784} (\bibinfo {year} {2015}{\natexlab{b}})},\ \Eprint
  {http://arxiv.org/abs/1504.00328} {arXiv:1504.00328 [astro-ph.HE]}
  \BibitemShut {NoStop}%
\bibitem [{\citenamefont {{Razzaque}}(2010)}]{2010ApJ...724L.109R}%
  \BibitemOpen
  \bibfield  {author} {\bibinfo {author} {\bibfnamefont {S.}~\bibnamefont
  {{Razzaque}}},\ }\href {\doibase 10.1088/2041-8205/724/1/L109} {\bibfield
  {journal} {\bibinfo  {journal} {\apjl}\ }\textbf {\bibinfo {volume} {724}},\
  \bibinfo {pages} {L109} (\bibinfo {year} {2010})},\ \Eprint
  {http://arxiv.org/abs/1004.3330} {arXiv:1004.3330 [astro-ph.HE]} \BibitemShut
  {NoStop}%
\bibitem [{\citenamefont {{Drenkhahn}}(2002)}]{2002A&A...387..714D}%
  \BibitemOpen
  \bibfield  {author} {\bibinfo {author} {\bibfnamefont {G.}~\bibnamefont
  {{Drenkhahn}}},\ }\href {\doibase 10.1051/0004-6361:20020390} {\bibfield
  {journal} {\bibinfo  {journal} {\aap}\ }\textbf {\bibinfo {volume} {387}},\
  \bibinfo {pages} {714} (\bibinfo {year} {2002})},\ \Eprint
  {http://arxiv.org/abs/astro-ph/0112509} {astro-ph/0112509} \BibitemShut
  {NoStop}%
\bibitem [{\citenamefont {{Fan}}\ \emph {et~al.}(2004)\citenamefont {{Fan}},
  \citenamefont {{Wei}},\ and\ \citenamefont {{Wang}}}]{2004A&A...424..477F}%
  \BibitemOpen
  \bibfield  {author} {\bibinfo {author} {\bibfnamefont {Y.~Z.}\ \bibnamefont
  {{Fan}}}, \bibinfo {author} {\bibfnamefont {D.~M.}\ \bibnamefont {{Wei}}}, \
  and\ \bibinfo {author} {\bibfnamefont {C.~F.}\ \bibnamefont {{Wang}}},\
  }\href {\doibase 10.1051/0004-6361:20041115} {\bibfield  {journal} {\bibinfo
  {journal} {\aap}\ }\textbf {\bibinfo {volume} {424}},\ \bibinfo {pages} {477}
  (\bibinfo {year} {2004})},\ \Eprint
  {http://arxiv.org/abs/arXiv:astro-ph/0405392} {arXiv:astro-ph/0405392}
  \BibitemShut {NoStop}%
\bibitem [{\citenamefont {{Uhm}}\ and\ \citenamefont
  {{Zhang}}(2014)}]{2014NatPh..10..351U}%
  \BibitemOpen
  \bibfield  {author} {\bibinfo {author} {\bibfnamefont {Z.~L.}\ \bibnamefont
  {{Uhm}}}\ and\ \bibinfo {author} {\bibfnamefont {B.}~\bibnamefont
  {{Zhang}}},\ }\href {\doibase 10.1038/nphys2932} {\bibfield  {journal}
  {\bibinfo  {journal} {Nature Physics}\ }\textbf {\bibinfo {volume} {10}},\
  \bibinfo {pages} {351} (\bibinfo {year} {2014})},\ \Eprint
  {http://arxiv.org/abs/1303.2704} {arXiv:1303.2704 [astro-ph.HE]} \BibitemShut
  {NoStop}%
\bibitem [{\citenamefont {{Zhang}}\ and\ \citenamefont
  {{Yan}}(2011)}]{2011ApJ...726...90Z}%
  \BibitemOpen
  \bibfield  {author} {\bibinfo {author} {\bibfnamefont {B.}~\bibnamefont
  {{Zhang}}}\ and\ \bibinfo {author} {\bibfnamefont {H.}~\bibnamefont
  {{Yan}}},\ }\href {\doibase 10.1088/0004-637X/726/2/90} {\bibfield  {journal}
  {\bibinfo  {journal} {\apj}\ }\textbf {\bibinfo {volume} {726}},\ \bibinfo
  {eid} {90} (\bibinfo {year} {2011})},\ \Eprint
  {http://arxiv.org/abs/1011.1197} {arXiv:1011.1197 [astro-ph.HE]} \BibitemShut
  {NoStop}%
\bibitem [{\citenamefont {{He}}\ \emph {et~al.}(2011)\citenamefont {{He}},
  \citenamefont {{Wu}}, \citenamefont {{Toma}}, \citenamefont {{Wang}},\ and\
  \citenamefont {{M{\'e}sz{\'a}ros}}}]{2011ApJ...733...22H}%
  \BibitemOpen
  \bibfield  {author} {\bibinfo {author} {\bibfnamefont {H.-N.}\ \bibnamefont
  {{He}}}, \bibinfo {author} {\bibfnamefont {X.-F.}\ \bibnamefont {{Wu}}},
  \bibinfo {author} {\bibfnamefont {K.}~\bibnamefont {{Toma}}}, \bibinfo
  {author} {\bibfnamefont {X.-Y.}\ \bibnamefont {{Wang}}}, \ and\ \bibinfo
  {author} {\bibfnamefont {P.}~\bibnamefont {{M{\'e}sz{\'a}ros}}},\ }\href
  {\doibase 10.1088/0004-637X/733/1/22} {\bibfield  {journal} {\bibinfo
  {journal} {\apj}\ }\textbf {\bibinfo {volume} {733}},\ \bibinfo {eid} {22}
  (\bibinfo {year} {2011})},\ \Eprint {http://arxiv.org/abs/1009.1432}
  {arXiv:1009.1432 [astro-ph.HE]} \BibitemShut {NoStop}%
\bibitem [{\citenamefont {{Kamble}}\ \emph {et~al.}(2007)\citenamefont
  {{Kamble}}, \citenamefont {{Resmi}},\ and\ \citenamefont
  {{Misra}}}]{2007ApJ...664L...5K}%
  \BibitemOpen
  \bibfield  {author} {\bibinfo {author} {\bibfnamefont {A.}~\bibnamefont
  {{Kamble}}}, \bibinfo {author} {\bibfnamefont {L.}~\bibnamefont {{Resmi}}}, \
  and\ \bibinfo {author} {\bibfnamefont {K.}~\bibnamefont {{Misra}}},\ }\href
  {\doibase 10.1086/520533} {\bibfield  {journal} {\bibinfo  {journal} {\apjl}\
  }\textbf {\bibinfo {volume} {664}},\ \bibinfo {pages} {L5} (\bibinfo {year}
  {2007})},\ \Eprint {http://arxiv.org/abs/0709.3561} {arXiv:0709.3561}
  \BibitemShut {NoStop}%
\bibitem [{\citenamefont {{Jin}}\ \emph {et~al.}(2009)\citenamefont {{Jin}},
  \citenamefont {{Xu}}, \citenamefont {{Covino}}, \citenamefont {{D'Avanzo}},
  \citenamefont {{Antonelli}}, \citenamefont {{Fan}},\ and\ \citenamefont
  {{Wei}}}]{2009MNRAS.400.1829J}%
  \BibitemOpen
  \bibfield  {author} {\bibinfo {author} {\bibfnamefont {Z.~P.}\ \bibnamefont
  {{Jin}}}, \bibinfo {author} {\bibfnamefont {D.}~\bibnamefont {{Xu}}},
  \bibinfo {author} {\bibfnamefont {S.}~\bibnamefont {{Covino}}}, \bibinfo
  {author} {\bibfnamefont {P.}~\bibnamefont {{D'Avanzo}}}, \bibinfo {author}
  {\bibfnamefont {A.}~\bibnamefont {{Antonelli}}}, \bibinfo {author}
  {\bibfnamefont {Y.~Z.}\ \bibnamefont {{Fan}}}, \ and\ \bibinfo {author}
  {\bibfnamefont {D.~M.}\ \bibnamefont {{Wei}}},\ }\href {\doibase
  10.1111/j.1365-2966.2009.15555.x} {\bibfield  {journal} {\bibinfo  {journal}
  {\mnras}\ }\textbf {\bibinfo {volume} {400}},\ \bibinfo {pages} {1829}
  (\bibinfo {year} {2009})},\ \Eprint {http://arxiv.org/abs/0903.4476}
  {arXiv:0903.4476 [astro-ph.HE]} \BibitemShut {NoStop}%
\bibitem [{\citenamefont {{Abdo}}\ \emph
  {et~al.}(2009{\natexlab{b}})\citenamefont {{Abdo}}, \citenamefont
  {{Ackermann}}, \citenamefont {{Arimoto}}, \citenamefont {{Asano}},
  \citenamefont {{Atwood}}, \citenamefont {{Axelsson}}, \citenamefont
  {{Baldini}}, \citenamefont {{Ballet}}, \citenamefont {{Band}}, \citenamefont
  {{Barbiellini}},\ and\ \citenamefont {et~al.}}]{2009Sci...323.1688A}%
  \BibitemOpen
  \bibfield  {author} {\bibinfo {author} {\bibfnamefont {A.~A.}\ \bibnamefont
  {{Abdo}}}, \bibinfo {author} {\bibfnamefont {M.}~\bibnamefont {{Ackermann}}},
  \bibinfo {author} {\bibfnamefont {M.}~\bibnamefont {{Arimoto}}}, \bibinfo
  {author} {\bibfnamefont {K.}~\bibnamefont {{Asano}}}, \bibinfo {author}
  {\bibfnamefont {W.~B.}\ \bibnamefont {{Atwood}}}, \bibinfo {author}
  {\bibfnamefont {M.}~\bibnamefont {{Axelsson}}}, \bibinfo {author}
  {\bibfnamefont {L.}~\bibnamefont {{Baldini}}}, \bibinfo {author}
  {\bibfnamefont {J.}~\bibnamefont {{Ballet}}}, \bibinfo {author}
  {\bibfnamefont {D.~L.}\ \bibnamefont {{Band}}}, \bibinfo {author}
  {\bibfnamefont {G.}~\bibnamefont {{Barbiellini}}}, \ and\ \bibinfo {author}
  {\bibnamefont {et~al.}},\ }\href {\doibase 10.1126/science.1169101}
  {\bibfield  {journal} {\bibinfo  {journal} {Science}\ }\textbf {\bibinfo
  {volume} {323}},\ \bibinfo {pages} {1688} (\bibinfo {year}
  {2009}{\natexlab{b}})}\BibitemShut {NoStop}%
\bibitem [{\citenamefont {{Ackermann}}\ \emph {et~al.}(2011)\citenamefont
  {{Ackermann}}, \citenamefont {{Ajello}}, \citenamefont {{Asano}},
  \citenamefont {{Axelsson}}, \citenamefont {{Baldini}}, \citenamefont
  {{Ballet}}, \citenamefont {{Barbiellini}}, \citenamefont {{Baring}},
  \citenamefont {{Bastieri}}, \citenamefont {{Bechtol}},\ and\ \citenamefont
  {et~al}}]{2011ApJ...729..114A}%
  \BibitemOpen
  \bibfield  {author} {\bibinfo {author} {\bibfnamefont {M.}~\bibnamefont
  {{Ackermann}}}, \bibinfo {author} {\bibfnamefont {M.}~\bibnamefont
  {{Ajello}}}, \bibinfo {author} {\bibfnamefont {K.}~\bibnamefont {{Asano}}},
  \bibinfo {author} {\bibfnamefont {M.}~\bibnamefont {{Axelsson}}}, \bibinfo
  {author} {\bibfnamefont {L.}~\bibnamefont {{Baldini}}}, \bibinfo {author}
  {\bibfnamefont {J.}~\bibnamefont {{Ballet}}}, \bibinfo {author}
  {\bibfnamefont {G.}~\bibnamefont {{Barbiellini}}}, \bibinfo {author}
  {\bibfnamefont {M.~G.}\ \bibnamefont {{Baring}}}, \bibinfo {author}
  {\bibfnamefont {D.}~\bibnamefont {{Bastieri}}}, \bibinfo {author}
  {\bibfnamefont {K.}~\bibnamefont {{Bechtol}}}, \ and\ \bibinfo {author}
  {\bibnamefont {et~al}},\ }\href {\doibase 10.1088/0004-637X/729/2/114}
  {\bibfield  {journal} {\bibinfo  {journal} {\apj}\ }\textbf {\bibinfo
  {volume} {729}},\ \bibinfo {eid} {114} (\bibinfo {year} {2011})},\ \Eprint
  {http://arxiv.org/abs/1101.2082} {arXiv:1101.2082 [astro-ph.HE]} \BibitemShut
  {NoStop}%
\bibitem [{\citenamefont {{Ackermann}}\ \emph {et~al.}(2013)\citenamefont
  {{Ackermann}}, \citenamefont {{Ajello}}, \citenamefont {{Asano}},
  \citenamefont {{Axelsson}}, \citenamefont {{Baldini}}, \citenamefont
  {{Ballet}}, \citenamefont {{Barbiellini}}, \citenamefont {{Bastieri}},
  \citenamefont {{Bechtol}}, \citenamefont {{Bellazzini}}, \citenamefont
  {{Bhat}},\ and\ \citenamefont {et~al.}}]{2013ApJS..209...11A}%
  \BibitemOpen
  \bibfield  {author} {\bibinfo {author} {\bibfnamefont {M.}~\bibnamefont
  {{Ackermann}}}, \bibinfo {author} {\bibfnamefont {M.}~\bibnamefont
  {{Ajello}}}, \bibinfo {author} {\bibfnamefont {K.}~\bibnamefont {{Asano}}},
  \bibinfo {author} {\bibfnamefont {M.}~\bibnamefont {{Axelsson}}}, \bibinfo
  {author} {\bibfnamefont {L.}~\bibnamefont {{Baldini}}}, \bibinfo {author}
  {\bibfnamefont {J.}~\bibnamefont {{Ballet}}}, \bibinfo {author}
  {\bibfnamefont {G.}~\bibnamefont {{Barbiellini}}}, \bibinfo {author}
  {\bibfnamefont {D.}~\bibnamefont {{Bastieri}}}, \bibinfo {author}
  {\bibfnamefont {K.}~\bibnamefont {{Bechtol}}}, \bibinfo {author}
  {\bibfnamefont {R.}~\bibnamefont {{Bellazzini}}}, \bibinfo {author}
  {\bibfnamefont {P.~N.}\ \bibnamefont {{Bhat}}}, \ and\ \bibinfo {author}
  {\bibnamefont {et~al.}},\ }\href {\doibase 10.1088/0067-0049/209/1/11}
  {\bibfield  {journal} {\bibinfo  {journal} {\apjs}\ }\textbf {\bibinfo
  {volume} {209}},\ \bibinfo {eid} {11} (\bibinfo {year} {2013})},\ \Eprint
  {http://arxiv.org/abs/1303.2908} {arXiv:1303.2908 [astro-ph.HE]} \BibitemShut
  {NoStop}%
\end{thebibliography}
%


%

%
%

\clearpage

\begin{table}
\centering \renewcommand{\arraystretch}{2}\addtolength{\tabcolsep}{3pt}
\caption{Swift X-ray light curve of GRB 190114C  with the best-fit values of the temporal PL index with their respective $\chi^2$/ndf.}
\label{table1}
\begin{tabular}{ c c c c}
 \hline \hline
\scriptsize{X-rays} &\hspace{0.5cm}   \scriptsize{Period}  &\hspace{0.5cm}   \scriptsize{Index}    & \hspace{0.5cm} \scriptsize{ $\chi^2$/ndf} \\ 
\scriptsize{(PL function)} & \hspace{0.5cm}  & \hspace{0.5cm}  \scriptsize{($\alpha_{\rm X}$)}   &   \\ 
\hline \hline
\scriptsize{I}   	        & \hspace{0.5cm} \scriptsize{$\leq 400$ s}  &\hspace{0.5cm} \scriptsize{$1.59\pm0.12$}		&\hspace{0.5cm}  \scriptsize{$0.6$}\\		\scriptsize{II}   	        & \hspace{0.5cm} \scriptsize{$400 - 10^4$ s}  &\hspace{0.5cm} \scriptsize{$0.57\pm0.09$}		&\hspace{0.5cm}  \scriptsize{$0.81$}\\	\scriptsize{III}   	        & \hspace{0.5cm} \scriptsize{$10^4 - 10^5$ s}  &\hspace{0.5cm} \scriptsize{$1.09\pm0.11$}		&\hspace{0.5cm}  \scriptsize{$0.83$}\\		\scriptsize{IV}   	        & \hspace{0.5cm} \scriptsize{$\geq 10^5$ s}  &\hspace{0.5cm} \scriptsize{$2.54\pm0.14$}		&\hspace{0.5cm}  \scriptsize{$0.91$}\\								              \\
\\
\hline \hline
\end{tabular}
\end{table}
\vspace{1.5cm}
\begin{table}
\centering \renewcommand{\arraystretch}{2}\addtolength{\tabcolsep}{3pt}
\caption{Optical light curves of GRB 190114C in different filters with the best-fit values of the temporal PL index with their respective $\chi^2$/ndf.}
\label{table2}
\begin{tabular}{ c c c c c}
 \hline \hline
\scriptsize{Optical} &\hspace{0.5cm}   \scriptsize{Index}  &\hspace{0.5cm}   \scriptsize{Break time }  &\hspace{0.5cm}   \scriptsize{Index}    & \hspace{0.5cm} \scriptsize{ $\chi^2$/ndf} \\ 
\scriptsize{band} & \hspace{0.5cm}  \scriptsize{$\alpha_{\rm O}$}  & \hspace{0.5cm}  \scriptsize{$t_{\rm br}$(s)}   &\hspace{0.5cm}  \scriptsize{$\alpha_{\rm O}$}  &\hspace{0.5cm} \\ 
\hline \hline
\scriptsize{b}	        & \hspace{0.5cm} \scriptsize{-} & \hspace{0.5cm} \scriptsize{-}   &\hspace{0.5cm} \scriptsize{$0.8374\pm0.0064$}             &\hspace{0.5cm}  \scriptsize{$0.61$}             \\
\scriptsize{i}     	        & \hspace{0.5cm} \scriptsize{-} & \hspace{0.5cm} \scriptsize{-}  &\hspace{0.5cm} \scriptsize{$0.5835\pm0.0089$}		&\hspace{0.5cm} \scriptsize{$1.32$} 								                     \\
\scriptsize{r}   	        &\hspace{0.5cm} \scriptsize{-} & \hspace{0.5cm} \scriptsize{-}  &\hspace{0.5cm} \scriptsize{$0.7554\pm0.0073$}		&\hspace{0.5cm}  \scriptsize{$1.41$}							                     \\
\scriptsize{}   	        &\hspace{0.5cm} \scriptsize{$1.593\pm0.012$} & \hspace{0.5cm} \scriptsize{$8.1$}  &\hspace{0.5cm} \scriptsize{$0.7554\pm0.0034$}		&\hspace{0.5cm}  \scriptsize{$1.22$}							                     \\
\scriptsize{v}   	        & \hspace{0.5cm} \scriptsize{-} & \hspace{0.5cm} \scriptsize{-}  &\hspace{0.5cm} \scriptsize{$0.7828\pm0.0551$}		&\hspace{0.5cm}   \scriptsize{$0.41$}									                     \\
\scriptsize{White}   	& \hspace{0.5cm} \scriptsize{-}& \hspace{0.5cm} \scriptsize{-}  &\hspace{0.5cm} \scriptsize{$0.912\pm0.0719$}	        &\hspace{0.5cm}  \scriptsize{$1.45$}										                     \\
\scriptsize{}   	& \hspace{0.5cm} \scriptsize{$1.567\pm0.097$}&  \hspace{0.5cm} \scriptsize{$26.3$} &\hspace{0.5cm} \scriptsize{$0.911\pm0.081$}	        &\hspace{0.5cm}  \scriptsize{$1.72$}										                     \\
\hline \hline
\end{tabular}
\end{table}
\vspace{1.5cm}
\begin{table}
\centering \renewcommand{\arraystretch}{2}\addtolength{\tabcolsep}{-2pt}
\caption{The best-fit values of the spectral and temporal indexes using the LAT, X-ray and optical observational data.  In addition, the theoretical predictions of  the spectral and temporal indexes are calculated for $p=2.2\pm0.3$.  Values in round parenthesis are the chi-square minimization ($\chi^2$ / n.d.f.)}
\label{table3}
\begin{tabular}{ l c c c c c c c}
\hline
\hline
		                            &   									  \normalsize{Observation} & \normalsize{Theory} &  \normalsize{Observation} & \normalsize{Theory}  &  \normalsize{Observation} & \normalsize{Theory}         \\
		                           				                       	                & \scriptsize{($\leq$ 400 s)} &  \scriptsize{(Stratified medium)} &  \scriptsize{(400 - $10^4$ s)}\hspace{0.5cm} \scriptsize{($10^4$ - $10^5$ s)} & \scriptsize{(Uniform medium)}   & \scriptsize{($\geq 10^5$ s)}& \scriptsize{(Uniform medium)}     \\
		                           
\hline \hline
\scriptsize{\bf{LAT flux}} 	& 			 				                         &	  		 	         &	   & 	& &		 \\ 
\hline \hline  
 \scriptsize{$\alpha_{\rm LAT}$}	 \hspace{0.3cm}	&		\scriptsize{$1.10\pm0.15$}  &  \scriptsize{$1.15\pm0.22$}     &		\scriptsize{$ - $ }	&		\scriptsize{$ - $ } & 	\scriptsize{$ - $ }&	\scriptsize{$ - $ }\\
 \scriptsize{$\beta_{\rm LAT}$}	 \hspace{0.3cm}	&		\scriptsize{$1.10\pm0.31\footnote{This value was reported in \cite{2019arXiv190107505W}}$}       &   \scriptsize{$1.10\pm0.15$}   & 		\scriptsize{$ - $} & 		\scriptsize{$ - $} & 	\scriptsize{$ - $ }&	\scriptsize{$ - $ }	\\

\\
\hline \hline
\scriptsize{\bf{GBM flux}} 	& 			 				                         &	  		 	         &	   & 	&  &		 \\ 
\hline \hline  
 \scriptsize{$\alpha_{\rm GBM}$}	 \hspace{0.3cm}	&		\scriptsize{$1.05\pm0.13$}  &  \scriptsize{$1.15\pm0.22$}     &		\scriptsize{$ - $ }	&		\scriptsize{$ - $ } &	\scriptsize{$ - $ } &	\scriptsize{$ - $ }\\
 \scriptsize{$\beta_{\rm GBM}$}	 \hspace{0.3cm}	&		\scriptsize{$0.81\pm0.08\footnote{This value was reported in \cite{2019arXiv190201861R}}$}       &   \scriptsize{$1.10\pm0.15$}   & 		\scriptsize{$ - $} & 		\scriptsize{$ - $} &	\scriptsize{$ - $ } &	\scriptsize{$ - $ }	\\

\\
\hline \hline
\scriptsize{\bf{X-ray flux}}		         & 		I&    & II \hspace{1.5cm} III& II \hspace{1.5cm} III & IV                                                                              \\
\hline \hline
 \scriptsize{$\alpha_{\rm X}$}	\hspace{0.3cm}		 &			\scriptsize{$1.59\pm 0.12$} & \scriptsize{$1.40\pm0.22$} 	&\scriptsize{$0.57\pm0.09$}\hspace{0.3cm} \scriptsize{$1.09\pm0.11$}  &  \scriptsize{$(0.1-0.6)$}\hspace{0.3cm} \scriptsize{$1.15\pm0.22$} &	\scriptsize{$2.54\pm0.14$ } & 	\scriptsize{$2.2\pm0.3$ }\\
 \scriptsize{$\beta_{\rm X}$}       \hspace{0.3cm}	 &			\scriptsize{$-$} & \scriptsize{$-$}  & \scriptsize{$0.83\pm0.04$}  &   \scriptsize{$0.60\pm0.15$}	& 	\scriptsize{$ - $ }&	\scriptsize{$ - $ }\\
\\
\hline\hline
\scriptsize{\bf{Optical flux}}  		         &                                             &  &                                     \\
\hline\hline
  \scriptsize{$\alpha_{\rm O}$}   \hspace{0.3cm}	&	\scriptsize{$1.593\pm0.012$}  &\scriptsize{$1.40\pm0.22$}	&  \scriptsize{$ 0.755\pm0.003$} & \scriptsize{$0.90\pm0.22$} & 	\scriptsize{$ - $ }& 	\scriptsize{$ - $ }\\
\scriptsize{$\beta_{\rm O}$}        \hspace{0.3cm}	&	\scriptsize{$-$}  & \scriptsize{$-$}         & \scriptsize{$0.83\pm 0.04$} &  \scriptsize{$0.60\pm0.15$} &	\scriptsize{$ - $ } &	\scriptsize{$ - $ }	\\
\\
\hline\hline
\scriptsize{\bf{Radio flux}}  		         &                                                                                  \\
\hline\hline
  \scriptsize{$\alpha_{\rm R}$}   \hspace{0.3cm}	&	\scriptsize{$-$}  &\scriptsize{$-$}	&  \scriptsize{$ 0.71\pm0.01$} & \scriptsize{$0.90\pm0.22$} &	\scriptsize{$ - $ } &	\scriptsize{$ - $ }\\
\scriptsize{$\beta_{\rm R}$}        \hspace{0.3cm}	&	\scriptsize{$-$}  & \scriptsize{$-$}         & \scriptsize{$-(0.3\pm0.2)$\footnote{This value was reported in \cite{2019arXiv190407261L} below 24 GHz. Above this value, the radio mm-band and optical data can be described with a SPL.}} &  \scriptsize{$-0.33$\footnote{The value between radio mm-band and optical data is $0.60\pm0.15$. }} & 	\scriptsize{$ - $ }&	\scriptsize{$ - $ }
\\
\\
\hline
\end{tabular}
\end{table}
\vspace{1.5cm}
\newpage
\begin{table}
\centering \renewcommand{\arraystretch}{2}\addtolength{\tabcolsep}{3pt}
\caption{Median values of parameters found with symmetrical quantiles (15\%, 50\%, 85\%).\\ Our model was used to constrain the values of parameters.}
\label{table4}
\begin{tabular}{ l  c c c  c c}
\hline
\hline
{\large   Parameters}	& 	&	& {\large  Median}  & 		 				                           		 	        	   		 \\ 
                          	 	&  {\normalsize LAT (100 MeV)} & 	 {\normalsize GBM (10 MeV)} &  {\normalsize X-ray (10 keV)} & 	{\normalsize Optical (1 eV)} 	& 	{\normalsize Radio (97.5 GHz)}		                           		 	        	   		 \\ 

\hline \hline
\\
\small{$A_{\star}\,(10^{-2})$}	\hspace{1cm}& \small{$5.999^{+0.297}_{-0.295}$}	\hspace{0.7cm} &  \small{$6.149^{+0.298}_{-0.296}$} \hspace{0.7cm}&  \small{$6.101^{+0.099}_{-0.101}$}&  \small{$5.950^{+0.098}_{-0.099}$}&  \small{$6.000^{+0.100}_{-0.100}$}	 \\
\small{$n\,({\rm cm^{-3}})$}	\hspace{1cm}&     \small{$-$}	\hspace{0.7cm} &  \small{$-$} \hspace{0.7cm}&  \small{$1.060^{+0.102}_{-0.101}$} &  \small{$1.100^{+0.098}_{-0.096}$}&  \small{$1.084^{+0.099}_{-0.097}$}	 \\
\small{$\epsilon_{\rm B,f}\,(10^{-5.3})$}	\hspace{1cm}&     \small{$1.001^{+0.302}_{-0.298}$}	\hspace{0.7cm} &  \small{$1.200^{+0.301}_{-0.296}$} \hspace{0.7cm}&  \small{$1.148^{+0.304}_{-0.293}$} &  \small{$0.951^{+0.298}_{-0.301}$} &  \small{$0.993^{+0.228}_{-0.190}$}	 \\
\small{$\epsilon_{\rm e,f}\,(10^{-2})$}	\hspace{1cm}&     \small{$1.000^{+0.304}_{-0.303}$}	\hspace{0.7cm} &  \small{$1.150^{+0.294}_{-0.302}$} \hspace{0.7cm}&  \small{$1.140^{+0.604}_{-0.585}$} &  \small{$1.139^{+0.298}_{-0.297}$} &  \small{$1.095^{+0.156}_{-0.137}$}	 \\
\small{$\epsilon_{\rm B,r}\,(10^{-1})$}	\hspace{1cm}&     \small{$1.000^{+0.304}_{-0.303}$}	\hspace{0.7cm} &  \small{$0.999^{+0.298}_{-0.297}$} \hspace{0.7cm}&  \small{$-$} &  \small{$-$} &  \small{$-$}	 \\
\small{$\epsilon_{\rm e,r}\,(10^{-1})$}	\hspace{1cm}&     \small{$0.999^{+0.100}_{-0.099}$}	\hspace{0.7cm} &  \small{$1.150^{+0.104}_{-0.100}$} \hspace{0.7cm}&  \small{$-$} &  \small{$-$}&  \small{$-$}	 \\
\small{$p$}	\hspace{1cm}&     \small{$2.300^{+0.100}_{-0.099}$}	\hspace{0.7cm} &  \small{$2.202^{+0.098}_{-0.098}$} \hspace{0.7cm}&  \small{$2.250^{+0.098}_{-0.101}$} &  \small{$2.280^{+0.099}_{-0.101}$} &  \small{$2.296^{+0.010}_{-0.010}$}	 \\

\hline
\end{tabular}
\end{table}
\clearpage
\clearpage
\clearpage

\clearpage
\clearpage

\end{document}